\documentclass[14pt]{article}
\usepackage{epsf}
\usepackage{color}
\usepackage{framed}
\usepackage{amsmath,amsfonts,amsbsy,amssymb}
\usepackage{indentfirst}
\usepackage{mathtools}
\usepackage{arydshln} 
\usepackage{framed}

\newcommand{\nn}{\nonumber}

\def\be{\begin{equation}}
\def\ee{\end{equation}}
\def\bef{\begin{framed}}
\def\eef{\end{framed}}
\def\bse{\begin{subequations}}
\def\ese{\end{subequations}}
\def\bal{\begin{align}}
\def\ealn{\end{align}}
\def\tr{\text{tr}}

\def\bff{\begin{framed}}
\def\eff{\end{framed}}

\begin{document}

\begin{titlepage}

\def\slash#1{{\rlap{$#1$} \thinspace/}}

\begin{flushright} 

\end{flushright} 

\vspace{0cm}

\begin{Large}
\begin{center}

\vspace{-0.2cm}

{\bf 
A Unified Construction of 
Skyrme-type 
Non-linear sigma Models \\
 via  \\ The Higher Dimensional  Landau Models
}
\end{center}
\end{Large}

\vspace{0.5cm}

\begin{center}
{\bf Kazuki Hasebe}   \\ 
\vspace{0.5cm} 
\it{
National Institute of Technology, Sendai College,  
Ayashi, Sendai, 989-3128, Japan} \\ 

\vspace{0.5cm} 

{\sf
khasebe@sendai-nct.ac.jp} 

\vspace{1.0cm} 

{\today} 

\end{center}

\vspace{0.3cm}

\begin{abstract}
\noindent

\baselineskip=18pt

A curious correspondence has been known between Landau models and non-linear sigma models: Reinterpreting the base-manifolds of Landau models as  field-manifolds, the Landau models are transformed to non-linear sigma models with same global and local symmetries. With the idea of the dimensional hierarchy of  higher dimensional Landau models, we exploit this correspondence to present a systematic procedure for construction of non-linear sigma models in higher dimensions. We explicitly derive $O(2k+1)$ non-linear sigma models in $2k$ dimension based on the parent tensor gauge theories that originate from non-Abelian monopoles. The obtained non-linear sigma models turn out to be Skyrme-type non-linear sigma models with  $O(2k)$ local symmetry. Through a dimensional reduction of  Chern-Simons tensor field theories, we also derive Skyrme-type $O(2k)$ non-linear sigma models  in $2k-1$ dimension, which realize the original and other Skyrme models as their special cases.   As a unified description, we explore Skyrme-type  $O(d+1)$ non-linear sigma models and clarify their  basic properties, such as stability of soliton configurations, scale invariant solutions, and  field configurations with higher winding number.      

\end{abstract}

\end{titlepage}

\newpage 

\tableofcontents

\newpage 

\section{Introduction}

Non-linear sigma (NLS) models were originally introduced for a description of mesons in hadron physics around 1960 \cite{Skyrme-1958, Gursey-1960, Gursey-1961, Gellmann-Levy-1960, Skyrme-1961-1, 
Perring-Skyrme-1962}. Skyrme proposed his celebrated  NLS model with a higher derivative term \cite{Skyrme-1962} to describe  baryons as solitonic excitations of meson fluid.  We refer to such  non-linear sigma models with a higher derivative term
 as the Skyrme-type non-linear sigma model (S-NLS) in this paper. 
The Skyrmions, or more generally the NLS model topological solitons, accommodate  deep mathematical structure related to gauge theories.  In particular, relationship between the quaternionic projective non-linear sigma model and $SU(2)$ gauge theory  was  intensively investigated  around 1970  \cite{Lukierski-1979,  Gava-Jengo-Omero-1979-1, Gava-Jengo-Omero-1979-2, Balachandran-Stern-Trahern-1979,  Gursey-Jafarizadeh-Tze-1979,  Kafiev-1980, Kafiev-1981,   Gursey-Tze-1980,  Jafarizadeh-Snyder-Tze-1980,  Felsager-Leinas-1980}. The self-dual equations of higher dimensional gauge theories were also revealed  in 1980s \cite{Tchrakian-1980, Corrigan-Devchand-Fairlie-Nuyts-1983, Grossmanetal1984, Tchrakian-1985, Saclioglu1986, Bais-Batenburg-1986-1, Bais-Batenburg-1986-2, OSe-Tchrakian-1987, Ivanova-Popov-1993}.  
An explicit recipe for derivation of the Skyrmion field configuration from the $SU(2)$ instanton was proposed by  Atiyah and Manton \cite{Atiyah-Manton-1989, Atiyah-Manton-1993}, which   stimulated  recent studies about  connections of  topological solitons in different dimensions,   
\cite{Eto-Nitta-Ohashi-Tong-2005, Nitta-2012, Nitta-2013-1, Nitta-2013-2, Nitta-2014, Gudnason-Nitta-2014} and 
\cite{ 
Sutcliffe-2010,Sutcliffe-2011,Sutcliffe-2015, Nakamula-Sasaki-Takesue-2016, Nakamula-Sasaki-Takesue-2017, Takesue-2017, Loginov-2020}.  
Apart from such formal aspects, Skyrmions  now appear ubiquitously in many branches of theoretical physics \cite{Rho-Zahed-2017} and   are also observed in daily nanoscale magnetic experiments (see \cite{EverschorSitteet-al-2018} and references therein).

One of the most prominent early experiments about  Skyrmions, more precisely $O(3)$ NLS model solitons,    
is the NMR  Knight shift measurement of the spin texture in quantum Hall ferromagnets \cite{Barret-et-al-1995}. 
Besides of the quantum Hall ferromagnets,  we often come across the $O(3)$ NLS model solitons in various  contexts of the quantum Hall effect.  One  example is about anyonic excitations of the fractional quantum Hall effect.      The effective field theory of the fractional quantum Hall effect is the Chern-Simons topological field theory \cite{Girvin-MacDonald-1987,Zhang-Hansson-Kivelson-1988, Zhang-1992}.  The Chern-Simons statistical field coupled to  the $O(3)$ NLS model solitons  provides a  field theoretical description of  anyons \cite{WilczekZee1983, BowickKarabaliWijewardhana1986} and such anyons  are realized as fractionally charge excitations  of the fractional quantum Hall effect \cite{Arovas-Schrieffer-Wilczek-1984,Arovas-Schrieffer-Wilczek-1985}. 
Another important example is about their analogous mathematical structures.  
The Haldane's formulation of the quantum Hall effect \cite{Haldane-1983}  is based on    the $SO(3)$ Landau model  \cite{Wu-Yang-1976, Hasebe-2015} in the Dirac monopole background \cite{Dirac-1931}, in which the $\it{base\text{-}manifold}$  or physical space is given by $S^2$ and the gauge symmetry is  $U(1)$. Meanwhile in the $O(3)$ NLS model \cite{Belavin-Polyakov-1975, Faddeev-1974} or equivalently  the $\mathbb{C}P^1$ model \cite{Eichenherr-1978, Cremmer-Scherk-1978, Golo-Perelomov-1978}, 
the  $\it{target\text{-}manifold}$ manifold or the field-space is $S^2\simeq \mathbb{C}P^1$ and the hidden local symmetry is $U(1)$.\footnote{We used  $``SO(3)''$ for the Landau model, since the Landau model Hamiltonian is constructed by the  angular momentum operators of the $SO(3)$ group, while  $``O(3)''$ for the NLS model since  the NLS model Hamiltonian is invariant under the $O(3)$ transformation, $i.e.$,  $SO(3)$ rotations and $\mathbb{Z}_2$ reflection of the NLS field.}   
  One may find a curious correspondence between the Landau model and the NLS model:  The base-manifold $S^2$  of the Landau model is identical to the target-manifold of the $O(3)$ NLS model, and their  local  symmetries are  also given by $U(1)$. We will refer to this correspondence as the Landau/NLS model correspondence.  

The Landau/NLS model correspondence is not  a special property in 2D, but holds in 4D. In the 4D quantum Hall effect \cite{Zhang-Hu-2001}, the Landau model is given by the $SO(5)$ Landau model \cite{Yang-1978-2,  Hasebe-2020-1}  whose base-manifold  is $S^4$ and magnetic field background  is  given by the Yang's $SU(2)$  monopole  \cite{Yang-1978-1}. 
Meanwhile in the $O(5)$  NLS model or the $\mathbb{H}P^1$ model \cite{Lukierski-1979, Gava-Jengo-Omero-1979-1, Kafiev-1980, Kafiev-1981, Gursey-Tze-1980, Jafarizadeh-Snyder-Tze-1980,Felsager-Leinas-1980},  the field-manifold is  $S^4$ and the hidden local symmetry is $SU(2)$. 
Besides, anyonic excitations in the 4D quantum Hall effect are known to be membrane-like objects whose internal space is  $S^4$  described by the field-manifold of the $O(5)$ NLS model \cite{Wu-Zee-1988, BookBottTu}.  
The Landau/NLS model correspondence  is thus reasonably generalized from 2D to 4D. 
It may be natural to ask whether the Landau/NLS model correspondence can hold in even higher dimensions.   
Such correspondence indeed holds in arbitrary dimensions as suggested in \cite{Hasebe-2014-1}.   
Quantum Hall effect on arbitrary $d$-dimensional sphere   has been constructed in  \cite{Hasebe-2014-1, Hasebe-Kimura-2003, Hasebe-2017}\footnote{See \cite{Karabali-Nair-2006, Hasebe-2010} and references therein  about  early developments of the higher dimensional quantum Hall effect.} (see  \cite{Meng-2003, CoskunKurkcuogluToga2016} also),  and  
the mathematical set-up    is given by the $SO(d+1)$ Landau model   in the $SO(d)$ monopole background.  The excitations are  $(d-2)$-dimensionally extended anyonic  objects whose fractional statistics are well investigated in \cite{RavenelZee1985, NepomechieZee1987,  TzeNam1989, HorowitzSrednicki1990, Gursey-Tze-1996}.  
 The effective field theory is  a tensor-type topological field theory coupled to the $(d-2)$-brane with $S^{d}$ internal space, which is described by the field-manifold of   $O(d+1)$ NLS models \cite{Hasebe-2014-1, Hasebe-2017}.  Again,  the field-manifold of the NLS model is identical to the base-manifold of the quantum Hall effect.  
 Furthermore, it is widely known that  any $O(d+1)$ NLS models with field-manifold $S^d\simeq O(d+1)/O(d)$ possess the hidden local symmetry $O(d)$ \cite{Bando-Kugo-Yamawaki-1988, Bando-Kugo-Uehara-Yamawaki-Yanagida-1985,Ma-Harada-2016}.  
The Landau/NLS model correspondence thus actually holds in arbitrary  dimensions. 

\begin{figure}[tbph]
\center
\includegraphics*[width=130mm]{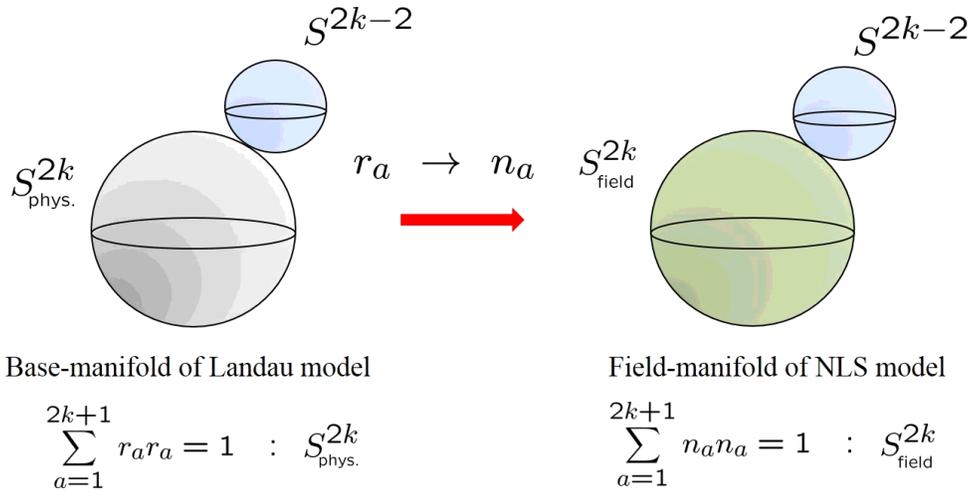}
\caption{The Landau/NLS model correspondence for $d=2k$. The differential topological structure  of the  $SO(2k+1)$ Landau model is same as  of the $O(2k+1)$ NLS model. The Landau model is transformed to the NLS model  under identification of  the base-manifold with the field-manifold.    }
\label{duality.fig}
\end{figure}

While NLS model solitons play crucial roles  in  the higher dimensional quantum Hall effect,  a systematic analysis of the $O(d+1)$ NLS model to host membrane excitations is still lacking. To be more precise,  there are  numerous possible NLS models with  field-manifold being   
$S^{d}$, but  there is no  criterion  to choose better models or  hopefully the best model among these models.     
A main purpose of this paper is to provide a  systematic procedure to construct appropriate NLS models  based on the Landau/NLS model correspondence [Fig.\ref{duality.fig}].  
For  the construction, we make use of the 
idea of the dimensional hierarchy  of the higher dimensional Landau models \cite{Hasebe-2014-1, Hasebe-Kimura-2003, Hasebe-2017}. Consequently, the obtained NLS models necessarily inherit  structures of the differential geometry of the Landau models.    
We also adopt the idea that was originally suggested by Tchrakian  \cite{Tchrakian-1980} and recently made manifest by Adam et al. \cite{Adam-Ferreira-Hora-Wereszczynski-Zakrzewski-2013} where a BPS equation 
is firstly given and   the Hamiltonian is  later derived so that the Hamiltonian may satisfy the BPS equation. 

The paper is organized as follows.   Sec.\ref{sec:diffgeolandau} reviews the differential geometry associated with  non-Abelian monopoles in the higher dimensional Landau models. 
In Sec.\ref{sec:skyrmeO4andO5NLS}, we reconsider geometric meanings  of the Skyrme's NLS field and the $O(5)$ S-NLS model in the light of the  Landau/NLS model correspondence.  We  present a systematic method for derivation of $O(2k+1)$ S-NLS models and explicitly construct the $O(7)$ NLS model and $O(2k+1)$ NLS model Hamiltonians  in  Sec.\ref{sec:evenhigherdim}. In Sec.\ref{sec:o2knls}, we construct $O(2k)$ S-NLS models  using the Chern-Simons term of pure gauge fields. We explore  general  $O(d+1)$ S-NLS models and analyze their basic properties  in Sec.\ref{sec:od+1nls}. Sec.\ref{sec:summdisc} is devoted to summary and discussions. 

\section{Differential Geometry of the Higher Dimensional Landau Model}\label{sec:diffgeolandau}

In this section,  we review the differential geometry  of the $SO(2k+1)$ Landau models and discuss extended objects that are  realized as the  $O(2k+1)$ NLS model solitons.  

\subsection{Non-Abelian monopole configuration of the $SO(2k+1)$ Landau model   }\label{subsec:difftoplandau2k+1}

The $SO(5)$ Landau model is formulated on $S^4$ embedded in $\mathbb{R}^5$ \cite{Zhang-Hu-2001,Yang-1978-2,Hasebe-2020-1}, and the background magnetic field is given by the 
Yang's $SU(2)$ monopole \cite{Yang-1978-1}  
\be
A = -\frac{1}{2r(r+r_{5})}\eta_{mn}^i  r_{n}\sigma_i  dr_{m},  ~~~~(m,n=1,2,3,4) 
\label{yangssu2mono}
\ee
where $\eta_{mn}^i\equiv \epsilon_{mni 4}+\delta_{m i}\delta_{n4}-\delta_{m4}\delta_{ni}$ denotes the 't Hooft symbol \cite{tHooft-1976}. The 1D reduction of the $SO(5)$ Landau model reproduces the $SO(4)$ Landau model \cite{Nair-Daemi-2004, Hasebe-2018}  on the $S^3$-equator  of $S^4$ \cite{Hasebe-2020-1, Hasebe-2014-2}. In Sec.\ref{sec:skyrmeO4andO5NLS}, we will consider the reverse  process  to derive the $O(5)$ S-NLS model from the Skyrme's field-manifold $S^3$. 

Generalizing the $SU(2) ~(\otimes SU(2)\simeq SO(4))$ to the $SO(2k)$ group \cite{Horvath-Palla-1978},\footnote{To be precise, $Spin(2k)$ group.}  the $SO(2k+1)$ Landau model is introduced on a base-manifold $S^{2k}$ in the $SO(2k)$ monopole background \cite{Hasebe-Kimura-2003, Hasebe-2014-1} [Table \ref{table:correslandau}]. 
\begin{table}
\center
   \begin{tabular}{|c|c|c|c|}\hline
      Landau model  &  $SO(3)$   &   $SO(5)$   & $SO(2k+1)$ \\ \hline
    Base-manifold           &  ${S}^2$     & ${S}^4$     &  $S^{2k}$  \\ \hline  
Global symmetry     &  $SO(3)\simeq SU(2)$      &   $SO(5)$     &       $SO(2k+1)$      \\ \hline 
Monopole gauge group      &  $SO(2)\simeq U(1)$       &  $SO(4)\simeq SU(2)~(\otimes SU(2))$    & $SO(2k)$ \\ \hline  
   Chern number   &  1st   &  2nd        & $k$th \\ \hline   
   Topological map    &  $\pi_1(U(1))\simeq \mathbb{Z}$   &  $\pi_3(SU(2))\simeq \mathbb{Z}$        & $\pi_{2k-1}(SO(2k))\simeq \mathbb{Z}$ \\ \hline   
    \end{tabular}       
\caption{  Geometric and topological features of the Landau models. The monopole gauge group $SO(2k)$ is chosen so that it is identical to the holonomy group of the base-manifold $S^{2k}\simeq SO(2k+1)/SO(2k)$ \cite{Hasebe-Kimura-2003}. In the $SO(5)$ Landau model, the holonomy of $S^4$ is $SO(4)\simeq SU(2)\otimes SU(2)$ and one of the two $SU(2)$s is adopted as the gauge group.      
}
\label{table:correslandau}
\end{table}
Notice that the gauge group is equal to the holonomy group of the basemanifold. 
The $SO(2k)$ monopole gauge field is represented as 
\be
A=\sum_{a=1}^{2k+1}A_a dr_a= -\frac{1}{r(r+r_{2k+1})}\sum_{m,n=1}^{2k}\sigma_{mn} r_{n}  dr_{m}, 
\label{fieldconfig}
\ee
or  
\be
A_m = -\frac{1}{r(r+r_{2k+1})}\sigma_{mn} r_{n}, ~~A_{2k+1}=0, ~~~(m,n=1,2,\cdots,2k) 
\label{fieldconfig2k}
\ee
which is regular except for the south pole.\footnote{
At $r_{2k+1}=0$, the $SO(2k)$ monopole configuration, (\ref{fieldconfig2k}) or (\ref{fieldconfigstrength}), is reduced to the meron configuration on $\mathbb{R}^{2k}$ \cite{Alfaro-Fubini-Furlan-1976}: 
\be
A_{\mu} =-\frac{1}{x^2}\sigma_{\mu\nu}x_\nu, ~~~F_{\mu\nu} =\frac{1}{x^2}\sigma_{\mu\nu}-\frac{1}{x^2}(x_\mu A_\nu -x_{\nu} A_{\mu} ), \label{configmeron}
\ee 
which satisfies the pure Yang-Mills field equation on $\mathbb{R}^{2k}$ \cite{O'Brien-Tchrakian-1987, Popov-1992}: 
\be
\frac{\partial}{\partial x_{\mu}}F_{\mu\nu}+i[A_{\mu}, F_{\mu\nu}]=0. 
\ee
} 
Here, $\sigma_{mn}$  are 
$Spin(2k)$ matrix generators:
\be
\sigma_{ij} =-i\frac{1}{4}[\gamma_i, \gamma_j], ~~~\sigma_{i,2k}=-\sigma_{2k,i}=\frac{1}{2}\gamma_i  
\ee
that satisfy 
\be
[\sigma_{mn}, \sigma_{pq}]=i(\delta_{mp}\sigma_{nq} -\delta_{mq}\sigma_{np} +\delta_{nq}\sigma_{mp} -\delta_{np}\sigma_{mq}    ). 
\label{so2kalgebra}
\ee
$\gamma_i$ $(i=1,2,\cdots,2k-1)$ stand for the $SO(2k-1)$ gamma matrices. 
The $SO(2k)$ monopole field strength is derived as  
\be
F=dA+i{A}^2 =\frac{1}{2}F_{ab}~dr_a\wedge dr_b, 
\ee
where  $F_{ab}=\partial_a A_b -\partial_b A_a +i[A_a, A_b]$ are 
\be 
F_{mn}=\frac{1}{r^2}\sigma_{mn}-\frac{1}{r^2}(r_{m}A_{n}-r_{n}A_{m}), ~~~F_{m ,{2k+1}} =-F_{2k+1, m}=\frac{1}{r^2}(r+r_{2k+1}) A_{m}. \label{fieldconfigstrength}
\ee
(\ref{fieldconfig2k}) and (\ref{fieldconfigstrength}) satisfy  
the field equations of motion of the pure Yang-Mills theory in $(2k+1)$D:\footnote{We will give an alternative verification of (\ref{EOMso2kmono}) in Appendix \ref{append:nonabemono}. 
 } 
\be
D_a F_{ab} =\partial_a F_{ab} +i[A_a, F_{ab}]=0. \label{EOMso2kmono}
\ee
One may need only the algebraic property of the $SO(2k)$ generators (\ref{so2kalgebra}) to verify (\ref{EOMso2kmono}),  and so  the monopole gauge field (\ref{fieldconfig2k}) of any $Spin(2k)$ representation realizes a  solution of the pure Yang-Mills field equation.  
The monopole configuration carries  unit Chern number. 
Indeed, substituting  (\ref{fieldconfigstrength}) into the $k$th Chern number 
\be
c_{k} =\frac{1}{k!(2\pi)^k }\int \tr(F^k), \label{defkthcsno}
\ee
we have 
\be
N_{2k}=\frac{1}{{A}(S^{2k}_{\text{phys.}})} \int_{S^{2k}_{\text{phys.}}} \frac{1}{(2k)!}\epsilon_{a_1 a_2 \cdots a_{2k+1}} r_{2a+1} dr_{a_1} dr_{a_2} \cdots dr_{a_{2k}} =1,  \label{kthchernunit}
\ee
with $A(S^{2k})$ being the area of $S^{2k}$: 
\be
{A}(S^{2k}) =\frac{2^{k+1}}{(2k-1)!!}\pi^{k}. 
\ee
(\ref{kthchernunit}) implies that the Chern number for the monopole configuration is accounted for by the   
  winding number (the Pontryagin index) from $S^{2k}_{\text{phys.}}$ to $S^{2k}_{\text{field}}$:  
\be
\pi_{2k}(S^{2k})\simeq \mathbb{Z}. 
\ee
Another expression of the $SO(2k)$ monopole  gauge field is 
\be
A'=-\frac{1}{r(r-r_{2k+1})}\bar{\sigma}_{mn} r_{n}  dr_{m},  \label{southgauge}
\ee
which is regular except for the north-pole.  
The two expressions of the monopole gauge fields, (\ref{fieldconfig}) and (\ref{southgauge}), are related by a gauge transformation  on the $S^{2k-1}$-equator of $S^{2k}$:  
\be
A'=g^{\dagger}Ag-ig^{\dagger}dg,  \label{gaugeons2k}
\ee
where  $g$  denotes a transition function of the form 
\begin{align}
g=\frac{1}{\sqrt{r^2-{r_{2k+1}}^2}}1_{2^{k-1}}  +i \frac{1}{\sqrt{r^2-{r_{2k+1}}^2}}\sum_{i=1}^{2k-1} r_{i}\gamma_i
 =\cos \theta ~1_{2^{k-1}} +i\sin \theta \sum_{i=1}^{2k-1} \hat{r}_i\gamma_i =e^{i\theta \sum_{i=1}^{2k-1}\hat{r}_i\gamma_i}.
 \label{so2kgroupele}
\end{align}
Here, 
\be
\hat{r}_i\equiv \frac{1}{\sqrt{r^2-{r_{2k+1}}^2-{r_{2k}}^2}}r_i ~~~(i=1,2,\cdots, 2k-1), ~~~~
\tan\theta \equiv \frac{1}{r_{2k}}\sqrt{r^2-{r_{2k+1}}^2-{r_{2k}}^2}.  
\ee 
$g(x)$ can also be regarded as a non-linear realization of $Spin(2k)$ associated with the symmetry breaking $SO(2k)~\rightarrow~SO(2k-1)$ with the broken generators $\gamma_i=2\sigma_{i,2k}~\in~Spin(2k)$.  
 $A$ and $A'$ are simply represented as 
\be
A=i\frac{1}{2r}(r-r_{2k+1})dg g^{\dagger},~~~~A'=-i\frac{1}{2r}(r+r_{2k+1})g^{\dagger}dg, 
\label{fieldconfigbyg}
\ee
where   
\be
-idg g^{\dagger} =\frac{2}{r^2-{r_{2k+1}}^2}{\sigma}_{mn}r_ndr_m,~~~~-ig^{\dagger}dg =-\frac{2}{r^2-{r_{2k+1}}^2}\bar{\sigma}_{mn}r_ndr_m.
\ee
The $k$th Chern number (\ref{defkthcsno}) can be expressed 
as  \cite{Hasebe-2014-1}
\be
c_k=\frac{(-i)^{k-1}}{(2k-1)!2^{k-1}{A}(S^{2k-1}_{\text{}})}\int_{S^{2k-1}_{\text{}}}\text{tr}(-ig^{\dagger}dg)^{2k-1}=(-i)^{k-1}\frac{1}{(2\pi)^k}\frac{(k-1)!}{(2k-1)!} \int_{S^{2k-1}_{\text{}}}\text{tr}(-ig^{\dagger}dg)^{2k-1} \label{windingnumbers2k-1}
\ee
where $A(S^{2k-1})$ signifies   the area of $(2k-1)$-sphere:   
\be
{A}(S^{2k-1}) =\frac{2\pi^k}{(k-1)!}. 
\ee
The associated topology is   indicated by 
\be
\pi_{2k-1}(SO(2k))~\simeq~\mathbb{Z}. 
\ee
Substituting  (\ref{so2kgroupele}) into (\ref{windingnumbers2k-1}), we have   
\be
N_{2k-1}=\frac{1}{{A}(S^{2k-1})} \int_{S^{2k-1}} \frac{1}{(2k-1)!}\epsilon_{a_1 a_2 \cdots a_{2k}} r_{a_{2k}} dr_{a_1} dr_{a_2} \cdots dr_{a_{2k-1}} =1, \label{s2k-1winding}
\ee
which denotes unit winding number from $S_{\text{phys.}}^{2k-1}$ to $S_{\text{field}}^{2k-1}$, and yields the same result as (\ref{kthchernunit}), as it should be.     
The equivalence between (\ref{kthchernunit}) and (\ref{s2k-1winding}) holds for other higher dimensional representations of gauge group matrix generators \cite{Hasebe-2017}. 
 We thus find that there are  two equivalent but superficially different representations of the $k$th Chern number for the monopole field configuration: 
\begin{enumerate} 
 \item Winding number  associated with $\pi_{2k}(S^{2k})\simeq \mathbb{Z}$.  
 \item Winding number  associated with  $\pi_{2k-1}(S^{2k-1})\simeq \mathbb{Z}$.  
\end{enumerate} 
We will utilize the first observation in the construction of the $O(2k+1)$ S-NLS models, and the second one in the construction  of the $O(2k)$ S-NLS models. This will also be important in the discussions of topological field configurations (Sec.\ref{subsec:topfield}). 

\subsection{Tensor gauge fields and extended objects}

The $k$th Chern number (\ref{defkthcsno}) can be expressed as 
\be
c_k =\frac{1}{k!(2\pi)^k }\int G_{2k}, 
\ee
where $G_{2k}$ denotes  a $2k$ rank tensor field strength  
\be
G_{2k} = \tr (F^k)=  \frac{1}{(2k)!}G_{a_1 a_2 \cdots a_{2k}} dr_{a_1} dr_{a_2} \cdots dr_{a_{2k}}  
\ee
or  
\be
G_{a_1 a_2\cdots a_{2k}} =\frac{1}{2^k} ~\tr(F_{[a_1 a_2}F_{a_3a_4}\cdots F_{a_{2k-1} a_{2k}]}  ) =\frac{1}{2^k} ~\tr(F_{[a_1a_2\cdots a_{2l-1} a_{2l}}F_{a_{2l+1} a_{2l+2} \cdots a_{2k-1} a_{2k} ]} ).    \label{kprodfab}
\ee
Here, we introduced the antisymmetric tensor field strength \cite{Tchrakian-1980}   
\be
F_{a_1a_2\cdots a_{2l}} \equiv \frac{1}{(2l)!}~F_{[\mu_1\mu_2} F_{a_3a_4}\cdots F_{a_{2l-1}a_{2l}]}. \label{astfs}
\ee
There are  $[k/2]$ independent ways for the decomposition (\ref{kprodfab}) in correspondence with  $l=1,2,\cdots , [k/2]$. $[k/2]$ signifies the maximum integer that does not exceed $k/2$. Apparently, there exists a local degree of freedom in the decomposition \cite{Ferreira-Shnir-2017}: 
\be
F_{a_1a_2\cdots a_{2l}} \cdot F_{a_{21+1}a_{2l+2}\cdots a_{2k}} =\lambda(x) ~F_{a_1a_2\cdots a_{2l}}\cdot  \frac{1}{\lambda(x)}  ~F_{a_{21+1}a_{2l+2}\cdots a_{2k}}.   
\label{doflambda}
\ee
For the non-Abelian monopole gauge field (\ref{fieldconfigstrength}), we can evaluate (\ref{kprodfab}) as   \cite{Hasebe-2014-1} 
\be
G_{2k} =\frac{1}{2^{k+1}r^{2k+1}}\epsilon_{a_1 a_2 \cdots a_{2k} a_{2k+1}} r_{a_{2k+1}}dr_{a_1} dr_{a_2}\cdots dr_{a_{2k}}, \label{tensormonodiffform}
\ee
or 
\be
G_{a_1 a_2 \cdots a_{2k}} =\frac{(2k)!}{2^{k+1} r^{2k+1} } \epsilon_{a_1 a_2 \cdots a_{2k+1}} r_{a_{2k+1}}, \label{tensormonofieldst2kcom}
\ee
which signifies the $2k$-rank tensor monopole field strength in its own right \cite{Nepomechie1985, Teitelboim1986}, and the $(2k-1)$-rank tensor gauge field  $(dC_{2k-1}=G_{2k})$ \cite{KalbRamond1974}  is coupled to $(2k-2)$-dimensionally extended objects, $i.e.$, $(2k-2)$-branes.   In the higher dimensional quantum Hall effect,   the size of the  gauge space  is comparable with the size of  the  base-manifold $S^{2k}$ \cite{Hasebe-2014-1}, and the whole system is regarded as a $(4k-1)$D space-time. The $(2k-2)$-brane current in $(4k-1)$D space-time is simply given by    
\be
J_{\mu_1\mu_2 \cdots \mu_{2k-1}} =\frac{1}{(2k)!}\epsilon_{\mu_1 \mu_2 \cdots \mu_{4k-1}}\epsilon_{a_1 a_2 \cdots a_{2k+1}}n_{a_1}\partial_{\mu_{2k}}n_{a_2} \partial_{\mu_{2k+1}}n_{a_3}\cdots \partial_{\mu_{4k-1}}n_{a_{2k+1}}, \label{topcurreosig}
\ee
where $n_a$ denote the internal field coordinates of the $(2k-2)$-brane (the blue sphere of left of Fig.\ref{duality.fig}). A simple subtraction, $(4k-1)-(2k-2)=2k+1$, implies that the dimension of the internal space of the $(2k-2)$-brane is   $2k$D  and is naturally described by the $S^{2k}$ field-manifold of   $O(2k+1)$ NLS models.  Indeed,   (\ref{topcurreosig}) is identical to the  topological current of the $O(2k+1)$ NLS model soliton   in  $(4k-1)$D space-time  
  with  coordinates $n_a$ subject to  $\sum_{a=1}^{2k+1} n_an_a=1$.  Notice that the obtained field-manifold is same as the original base-manifold $S^{2k}$. 
  Furthermore, the $(2k-2)$-brane current is coupled to the $(2k-1)$-rank tensor Chern-Simons field to realize a field theoretical description of  anyonic excitations in higher dimension. In this way, the  $O(2k+1)$ NLS model solitons necessarily appear in the context of the  higher dimensional quantum Hall effect.  

\section{1D promotion and the $O(5)$ S-NLS model}\label{sec:skyrmeO4andO5NLS}

 In Sec.\ref{sec:diffgeolandau}, we first considered two monopole gauge field configurations on $S^{2k}$ and later introduced their transition function  on the $S^{2k-1}$-equator of $S^{2k}$.   
In this section, we apply 
the $\it{reveres}$ process to derive the $O(5)$ S-NLS model from  the Skyrme's $S^3$ field-manifold.

\subsection{Translation to the field-manifold and 1D Promotion} 

While the base-manifold of the $SO(5)$ Landau model is $S^4$ and its equator is  $S^3$,  we reinterpret $S^4$ and $S^3$ as field-manifolds in  the NLS model side.  

\subsubsection{Skyrme's Field-manifold $S^3$}

The Skyrme's field  $n_m$ $(m=1,2,3,4)$ takes its values on $S^3_{\text{field}}$: 
\be
\sum_{m=1}^4{n_m}n_m=1. 
\ee
Instead of using  $n_m$  directly, we will represent the field in the form of  $SU(2)$ group element\footnote{(\ref{nmqm}) is known as the principal chiral field of mesons in  hadron physics.} 
\be
g=\sum_{m=1}^4 n_m \bar{q}_m,  \label{nmqm}
\ee
where  $\bar{q}_m\equiv \{-q_{i=1,2,3}, 1\}$ are (conjugate) quaternions that satisfy  
\be
{q_i}^2=-1, ~~~~q_i q_j =-q_jq_i=q_k~~(i\neq j).
\ee 
In a matrix representation, $q_i$ can be  represented as  
\be
q_i=-i\sigma_i. \label{2by2matquatern}
\ee
The associated gauge field  is simply a pure gauge on $S^{3}_{\text{field}}$: 
\be
\mathcal{A} = -ig^{\dagger} d g=-\bar{\eta}_{mn}^i\sigma_i n_n d n_m,~~~~
\mathcal{F}=d\mathcal{A} +i\mathcal{A}^2=\bar{\eta}_{mn}^i \sigma_i dn_m \wedge dn_n~(1-1)=0,  \label{puregaugeaandfield}
\ee
where $\bar{\eta}_{mn}^i\equiv \epsilon_{mn i4}-\delta_{mi}\delta_{n4} + \delta_{m4}\delta_{ni}$.  
Suppose that $n_m$ signify a field on $x_\alpha ~\in~\mathbb{R}^3$, and  the Skyrme's higher derivative term is  expressed as 
\be
(\partial_{\alpha}n_m)^2 (\partial_{\beta}n_n)^2 -( \partial_{\alpha}n_m \cdot \partial_{\beta}n_m)^2=-\frac{1}{8}\tr([\mathcal{A}_{\alpha}, \mathcal{A}_{\beta}]^2)=\frac{1}{8}\tr((\partial_{\alpha}\mathcal{A}_{\beta}-\partial_{\beta} \mathcal{A}_{\alpha})^2)   . \label{skyrmetermas}
\ee

\subsubsection{1D promotion}

\begin{figure}[tbph]
\center
\includegraphics*[width=160mm]{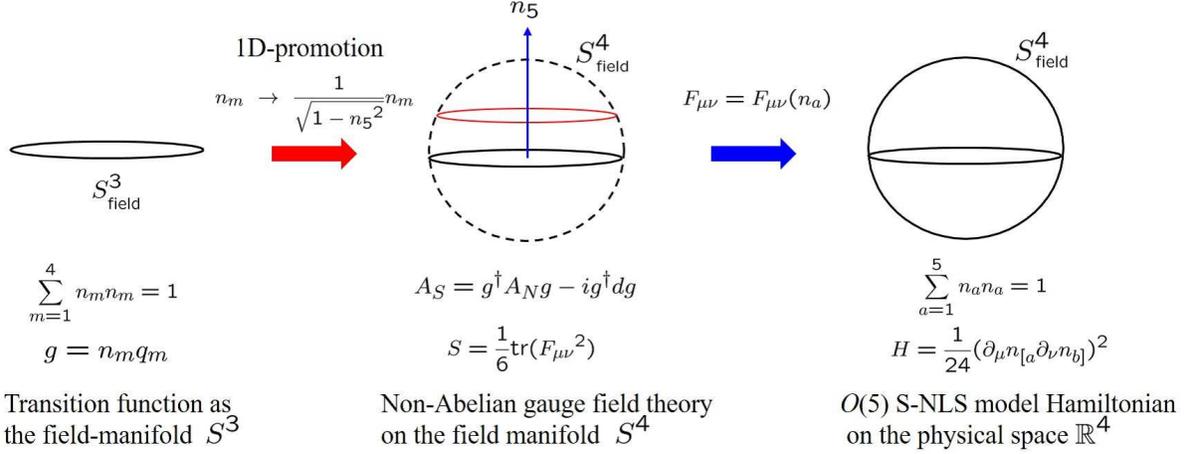}
\caption{ We first promote $S^3_{\text{field}}$ to $S^4_{\text{field}}$ (red arrow). Next, we  construct a gauge field theory on the field-manifold $S^4$ (middle). Expressing the gauge field by the NLS field  (blue arrow), we lastly derive   $O(5)$ S-NLS model Hamiltonian.   }
\label{3dto4d.fig}
\end{figure}

Stacking  $S^3_{\text{field}}$s along a virtual 5th direction, we form  a virtual $S^4_{\text{field}}$ (see the middle of Fig.\ref{3dto4d.fig}), in which the   radii of $S_3^{\text{field}}$s are  continuously tuned as   
\be
n_m ~\rightarrow~ \frac{1}{\sqrt{1-{n_5}^2}} ~n_m,     \label{changeeqlat}
\ee
so that $n_{a=1,2,3,4,5}$ realize the coordinates of $S^4_{\text{field}}$: 
\be
\sum_{a=1}^5 n_an_a=1.   
\ee
This process demonstrates   1D promotion from 3D to 4D and manifest the idea  of  dimensional hierarchy  \cite{Hasebe-2017, Hasebe-2014-2}.   
The $SU(2)$ group element (\ref{nmqm}) now turns to  
\be
g=\frac{1}{\sqrt{1-{n_5}^2}}\sum_{m=1}^4 n_m \bar{q}_m.  \label{defelementgroupg}
\ee
We regard $g$  as a transition function connecting  two gauge fields on the $S^3_{\text{field}}$-equator  of the virtual field-manifold $S^4_{\text{field}}$:  
\be
A' =g^{\dagger}A g -ig^{\dagger}dg. 
\ee
The corresponding gauge fields are   (\ref{fieldconfigbyg}):  
\be
A=i\frac{1}{2}(1-n_{5})dg g^{\dagger} =-\frac{1}{2(1+n_5)} 
\eta_{mn}^i 
n_n   \sigma_i dn_m ,    
~~~~A' = -i\frac{1}{2}(1+n_{5}) g^{\dagger}dg =-\frac{1}{2(1-n_5)} 
\bar{\eta}_{mn}^i 
n_n  \sigma_i dn_m. \label{aadashonna}
\ee
Let us assume that  $n_a$ denote a field representing a map from $x_{\mu}\in \mathbb{R}_{\text{phys.}}^4$ to $n_a \in S_{\text{field}}^4$, and then (\ref{aadashonna}) becomes  
\be
A= -\frac{1}{2(1+n_5)} 
\eta_{mn}^i 
n_n  \partial_{\mu}n_m \sigma_i dx_{\mu},   
~~~~A'
=-\frac{1}{2(1-n_5)} 
\bar{\eta}_{mn}^i  
n_n  \sigma_i \partial_{\mu}n_m dx_{\mu} .  
 \label{gaugeformexp}
\ee
Notice that (\ref{gaugeformexp}) represents field configurations on $\mathbb{R}^4_{\text{phys.}}$: 
\be
A_{\mu}(n_a(x)) =  -\frac{1}{2(1+n_5)} 
\eta_{mn}^i 
n_n  \partial_{\mu}n_m \sigma_i , ~~~~A'_{\mu}(n_a(x))
=-\frac{1}{2(1-n_5)} 
\bar{\eta}_{mn}^i 
n_n  \partial_{\mu}n_m \sigma_i.~~~~\label{gaugebpstcon}
\ee
The field strengths on $\mathbb{R}^4_{\text{phys.}}$ are   derived as 
\begin{align}
F_{\mu\nu}(n_a(x))&=\partial_{\mu}A_{\nu}-\partial_{\nu}A_{\mu}+i[A_{\mu}, A_{\nu}]\nn\\
&=\frac{1}{2} \eta_{mn}^i\partial_{\mu}n_m \partial_{\nu}n_n \sigma_i-\frac{1}{2(1+n_5)} \eta_{mn}^i
n_n (\partial_{\mu}n_m\partial_{\nu}n_5-\partial_{\nu}n_m\partial_{\mu}n_5)\sigma_i,  \nn\\
F'_{\mu\nu}(n_a(x))&=\partial_{\mu}A'_{\nu}-\partial_{\nu}A'_{\mu}+i[A'_{\mu}, A'_{\nu}]\nn\\
&=\frac{1}{2} \bar{\eta}_{mn}^i\partial_{\mu}n_m \partial_{\nu}n_n \sigma_i-\frac{1}{2(1-n_5)} \bar{\eta}_{mn}^i
n_n (\partial_{\mu}n_m\partial_{\nu}n_5-\partial_{\nu}n_m\partial_{\mu}n_5)\sigma_i.  \label{su2fieldstrength}
\end{align}
When  $n_a$ are given by the inverse stereographic coordinates on ${S}^4_{\text{phys.}}$ from $\mathbb{R}^4_{\text{phys.}}$:  
\be
r_a=\{r_{\mu}, r_5\}~\equiv ~\{\frac{2}{1+{x}^2}x_\mu,~ \frac{1-{x}^2}{1+{x}^2}\}, \label{soords4inv}
\ee
(\ref{gaugebpstcon}) and  (\ref{su2fieldstrength}) realize the BPST instanton configuration \cite{BPST-1975}:  
\be
A_{\mu}|_{n_a=r_a} =   -\frac{1}{{x}^2+1}\eta_{\mu\nu}^ix_{\nu} \sigma_i ,~~~~~~F_{\mu\nu}|_{n_a=r_a}=2\frac{1}{({x}^2+1)^2}\eta_{\mu\nu}^i\sigma_i,  \label{instantonconf}
\ee
which carries  unit 2nd Chern number. 
(\ref{instantonconf}) simply corresponds to the stereographic projection of the Yang's $SU(2)$ monopole gauge field (\ref{yangssu2mono}) on $S^4$ \cite{Jackiw-Rebbi-1976} (see Appendix \ref{append:stereo} for details).

\subsection{From the non-Abelian gauge theory to $O(5)$ S-NLS model}

The next step is to adopt a gauge theory action appropriate for the  construction of   NLS model Hamiltonian. 
A natural  choice may be the pure Yang-Mills action 
\be
S=\frac{1}{6}\int_{\mathbb{R}^4} d^4x ~\tr({F_{\mu\nu}}^2).  
\label{so5nlsffaction}
\ee
The previous studies \cite{Lukierski-1979, Kafiev-1980, Kafiev-1981, Gursey-Tze-1980, Jafarizadeh-Snyder-Tze-1980,Felsager-Leinas-1980} already showed that substitution of $F_{\mu\nu}$   (\ref{su2fieldstrength}) into (\ref{so5nlsffaction})   
 yields the $O(5)$ S-NLS model Hamiltonian 
\be 
H= \frac{1}{12}\int_{\mathbb{R}^4} d^4x~ \biggl( (\partial_{\mu}n_a)^2 (\partial_{\nu}n_b)^2 -(\partial_{\mu}n_a \partial_{\nu}n_a)^2   \biggr). \label{o5nlshamexp}
\ee 
One may notice that (\ref{o5nlshamexp}) is  a straightforward  4D generalization of the Skyrme term (\ref{skyrmetermas}). We revisit the  construction of the Hamiltonian from the view of the BPS equality.

\subsubsection{BPS inequality and Yang-Mills action}

Refs.\cite{Tchrakian-1980} and  \cite{Adam-Ferreira-Hora-Wereszczynski-Zakrzewski-2013,  Ferreira-Zakrzewski-2013, Ferreira-2017, Ferreira-Shnir-2017,  Amari-Ferreira-2018-1, Amari-Ferreira-2018-2} indicate a procedure to construct an action  from a given BPS inequality.\footnote{The author is indebted to Dr. Amari for the information.}  
Usually to describe a system   we set up an  action  at first, and  the BPS inequality is later derived, 
but here we take the reverse process: BPS inequality is firstly given, and  an appropriate action is later introduced so that the action can satisfy the given BPS inequality. As a preliminary, we demonstrate how this works in  the  4D Yang-Mills gauge theory. 
We first consider the BPS inequality:   
\be
\tr((F_{\mu\nu} -\tilde{F}_{\mu\nu})^2) 
~\ge~ 0 \label{4dcondfmunu}
\ee
or 
\be
\tr({F_{\mu\nu}}^2) +\tr({\tilde{F}_{\mu\nu}}^2)\ge 2\tr(F_{\mu\nu}\tilde{F}_{\mu\nu}), \label{bpsequine}
\ee
where $\tilde{F}_{\mu\nu}$ are defined as 
\be
\tilde{F}_{\mu\nu}\equiv \frac{1}{2}\epsilon_{\mu\nu\rho\sigma}F_{\rho\sigma}.   
\label{4dtildefmunu}
\ee
The integral of the right-hand side signifies  the second Chern number: 
\be
c_2 
=\frac{1}{16\pi^2}\int_{\mathbb{R}^4} d^4x ~\tr(F_{\mu\nu}\tilde{F}_{\mu\nu}) , 
\label{chern2ndff}
\ee
and from (\ref{bpsequine}) the action is constructed as  
\be
S_{4,2} \equiv  \frac{1}{12}\int_{\mathbb{R}^4} \biggl( \tr({F_{\mu\nu}}^2)+ \tr({\tilde{F}_{\mu\nu}}^2)\biggr)        ~\ge ~{A}(S^4)\cdot c_2, \label{bpsso5}
\ee
where ${A}(S^4)=\frac{8}{3}\pi^2$. 
From the special property in 4D,   
\be
{\tilde{F}_{\mu\nu}}^2={F_{\mu\nu}}^2, 
\ee
 $S_{4,2}$ (\ref{bpsso5})  ``accidentally'' coincides with the pure Yang-Mills action (\ref{so5nlsffaction}): 
\be
S_{4,2} = \frac{1}{6}\int_{\mathbb{R}^4} d^4x~\tr({F_{\mu\nu}}^2) .  
\label{energyso5fftil}
\ee
 In even higher dimensions, the corresponding actions are no longer Yang-Mills type but  higher  
tensor-field type   as we shall see in Sec.\ref{sec:evenhigherdim}.

\subsubsection{Construction of the $O(5)$ S-NLS model}\label{subsubsec:conso5}

We next substitute  (\ref{su2fieldstrength}) into the parent gauge theory action (\ref{energyso5fftil}) to obtain\footnote{
If one  adopted $F_{\mu\nu}'(n_a)$ (\ref{su2fieldstrength}) 
instead of $F_{\mu\nu}(n_a)$, the obtained Hamiltonian would be  the same due to the gauge invariace of the parent action (\ref{energyso5fftil}).   
} 
\be
 S_{4,2}~~\overset{F_{\mu\nu}=F_{\mu\nu}(n_a)}{\longrightarrow}~~
H_{4,2} = \frac{1}{12}\int_{\mathbb{R}^4} d^4x~  \partial_{\mu}n_a  \partial_{\nu}n_b \cdot \partial_{\mu}n_{[a}  \partial_{\nu}n_{b]}    = \frac{1}{24}\int_{\mathbb{R}^4} d^4x~ ( \partial_{\mu}n_{[a}  \partial_{\nu}n_{b]})^2,  \label{nnnnenergy}
\ee 
which is nothing but (\ref{o5nlshamexp}). 
Hereafter, $[\cdots]$ denotes the totally antisymmetric combination only about  the $\it{Latin}$ indices. 
For instance, 
\begin{align}
\partial_{\mu}n_{[a}\partial_{\nu}n_{b]} &\equiv \partial_{\mu}n_{a}\partial_{\nu}n_{b}-\partial_{\mu}n{_b}\partial_{\nu}n_{a}, \nn\\
\partial_{\mu}n_{[a}\partial_{\nu}n_{b}\partial_{\rho}n_{c]} &
\equiv \partial_{\mu}n_{a}\partial_{\nu}n_{b}\partial_{\rho}n_{c}-\partial_{\mu}n_{a}\partial_{\nu}n_{c}\partial_{\rho}n_{b}+\partial_{\mu}n_{b}\partial_{\nu}n_{c}\partial_{\rho}n_{a}-\partial_{\mu}n_{b}\partial_{\nu}n_{a}\partial_{\rho}n_{c}\nn\\
&~+\partial_{\mu}n_{c}\partial_{\nu}n_{a}\partial_{\rho}n_{b}-\partial_{\mu}n_{c}\partial_{\nu}n_{b}\partial_{\rho}n_{a}. 
\end{align}
Note that  the antisymmetricity of the Latin indices inherits the antisymmetricity of the Greek indices of the parent tensor field strengths.  
Similarly, the 2nd Chern number  (\ref{chern2ndff})  turns to the winding number: 
\be
c_2=\frac{1}{16\pi^2}\int_{\mathbb{R}^4}d^4x~\tr(F_{\mu\nu}\tilde{F}_{\mu\nu})~~\overset{F_{\mu\nu}=F_{\mu\nu}(n_a)}{\longrightarrow}~~N_4=\frac{1}{{A}(S^4)}\int_{\mathbb{R}^4} d^4x ~\epsilon_{\mu\nu\rho\sigma}~\frac{1}{4!}\epsilon_{abcde}n_e \partial_{\mu}n_a\partial_{\nu}n_b\partial_{\rho}n_c\partial_{\sigma}n_d, \label{windingnumbero5nls}
\ee
which indicates the homotopy 
\be
\pi_4(S^4)~\simeq~\mathbb{Z}. 
\ee
Since  we started from the BPS inequality of the gauge field (\ref{bpsequine}), the obtained $O(5)$ S-NLS model Hamiltonian necessarily satisfies the BPS inequality: 
\be
H_{4,2} ~\ge~ A(S^4) \cdot N_4.
\ee

Some technical comments are added here. 
It is a rather laborious task to derive (\ref{nnnnenergy}) by directly substituting (\ref{su2fieldstrength}) into (\ref{energyso5fftil}), but fortunately there exists a much  easier way. 
 First, we temporally neglect the clumsy parts associated with $n_5$ in (\ref{su2fieldstrength}); $F_{\mu\nu} ~\sim ~\frac{1}{2} \eta_{mn}^i\sigma_i\partial_{\mu}n_m \partial_{\nu}n_n$. With such  simplified $F_{\mu\nu}$, we next evaluate the Yang-Mills action $\tr(F_{\mu\nu}^2)$ to have $\frac{1}{2}( \partial_{\mu}n_m    \partial_{\nu}n_n \cdot  \partial_{\mu}n_{[m}    \partial_{\nu}n_{n]} )$. Lastly, we just recover  $n_5$-component in such a way that  $\frac{1}{2}( \partial_{\mu}n_m    \partial_{\nu}n_n \cdot  \partial_{\mu}n_{[m}    \partial_{\nu}n_{n]} )$ should  respect  the $SO(5)$ symmetry,  which is $\frac{1}{2}( \partial_{\mu}n_a    \partial_{\nu}n_b \cdot  \partial_{\mu}n_{[a}    \partial_{\nu}n_{b]} )$. 
 This short-cut method will be  useful  in deriving    S-NLS model Hamiltonians in even  higher dimensions.

From (\ref{nnnnenergy}), the equations of motion for the $O(5)$ NLS field are derived as 
\be
\partial_{\mu}(\partial_{\nu}n_b \partial_{\mu}n_{[a}\partial_{\nu}n_{b]})-\frac{\lambda}{2}n_a=0. \label{eomo5snlsm}
\ee
Here, $\lambda$ denotes the Lagrange multiplier and  is given by 
\be
\lambda= 2n_a \partial_{\mu}(\partial_{\nu}n_b \partial_{\mu}n_{[a}\partial_{\nu}n_{b]}). 
\label{lambdao5}
\ee
Eq.(\ref{eomo5snlsm}) is  highly non-linear, but a solution  is simply given by  $n_a=r_a$ 
with $r_a$ being the coordinates on $S^4_{\text{phys.}}$ (\ref{soords4inv}).  The solution also   carries the winding number 
 $N_4=1$ as expected from the discussions around (\ref{instantonconf}).

\section{$O(2k+1)$ S-NLS Models}\label{sec:evenhigherdim}

\begin{table}
\center 
   \begin{tabular}{|c|c|c|c|}\hline
     NLS model   &  $O(5)$   &   $O(7)$    & $O(2k+1)$ \\ \hline
    Base-manifold           &  $\mathbb{R}^4$     & $\mathbb{R}^6$    &  $\mathbb{R}^{2k}$  \\ \hline  
     Target manifold           &  $S^4$     & ${S}^6$    &  $S^{2k}$  \\ \hline  
Global symmetry     &  $SO(5)$      &   $SO(7)$     &       $SO(2k+1)$      \\ \hline 
Local symmetry      &  $SO(4)\simeq SU(2)(\otimes SU(2))$       &  $SO(6)\simeq SU(4)$   & $SO(2k)$ \\ \hline  
 Winding number    &  $\pi_4(S^4)\simeq \mathbb{Z}$   &  $\pi_6(S^6)\simeq \mathbb{Z}$       & $\pi_{2k}(S^{2k})\simeq \mathbb{Z}$ \\ \hline   
    \end{tabular}       
\caption{Geometric features of the $O(5)$ NLS model are naturally generalized in even higher dimensions.  
}
\label{table:correspnlsmodelon}
\end{table}

In this section,  we  present a general procedure to construct S-NLS models in arbitrary even dimensions  
and demonstrate the procedure to derive $O(7)$ S-NLS and $O(2k+1)$ S-NLS model Hamiltonians, respectively (Table \ref{table:correspnlsmodelon}).

\subsection{General Procedure }\label{subsec:generaproc}

The basic steps for the construction of higher dimensional S-NLS models are  as follows. 
\begin{enumerate}
\item Promote  $S^{2k-1}_{\text{field}}$-coordinates $n_m$ to $S^{2k}_{\text{field}}$-coordinates  $n_a$. 

First prepare a normalized field,  $n_{m=1,2,\cdots,2k}$, that represents a manifold $S^{2k-1}_{\text{field}}$. We assume that   $S^{2k-1}_{\text{field}}$ is realized as a latitude of a virtual 
$S^{2k}_{\text{field}}$:  
\be
n_m ~\rightarrow~ \frac{1}{\sqrt{1-{n_{2k+1}}^2}} ~n_m, \label{1dprgene}
\ee
where $n_m$ and  $n_{2k+1}$ on the right-hand side denote the coordinates on $S^{2k}_{\text{field}}$: 
\be
\sum_{a=1}^{2k+1} n_an_a=1. 
\ee
We also suppose that  NLS field $n_a(x)$ represents a map from $x_{\mu}~\in~\mathbb{R}^{2k}_{\text{phys.}}$ to $n_a~\in~S^{2k}_{\text{field}}$. 
Note that the dimension of the physical space is  same as  the dimension of the field space.

\item Derive  $SO(2k)$ gauge fields on the field-manifold $S^{2k}_{\text{field}}$  from the transition function.   

The $Spin(2k)$ group element is  expressed as 
\be
g=\sum_{m=1}^{2k} n_m \bar{g}_m , \label{noneqspin2k}
\ee
where $\bar{g}_m$ denote higher dimensional counterpart of the quaternions:  
\be
{g}_m=\{-i\gamma_i, ~1\}, ~~~~  
\bar{g}_m =\{i\gamma_i, ~1\}.   \label{matrrealiso2k}
\ee
Here, $\gamma_i$ $(i=1,2,\cdots, 2k-1)$ denote  the $SO(2k-1)$ gamma matrices. 
The basic algebras of the $g$ matrices are given by  [see Appendix \ref{sec:extintduality1} also]
\begin{align}
&g_{m}\bar{g}_{n}+g_{n}\bar{g}_{m}=\bar{g}_{m}{g}_{n}+\bar{g}_{n}{g}_{m}=2\delta_{mn}, \nn\\
&g_{m}\bar{g}_{n}-g_{n}\bar{g}_{m}=4i \bar{\sigma}_{mn}, ~~\bar{g}_{m}{g}_{n}-\bar{g}_{n}{g}_{m}=4i {\sigma}_{mn},  
\end{align}
where either of $\sigma_{mn}$ and $\bar{\sigma}_{mn}$ denote  $Spin(2k)$ matrix generators.   
By the 1D promotion (\ref{1dprgene}), (\ref{noneqspin2k}) becomes  
\be
g=\frac{1}{\sqrt{1-{n_{2k+1}}^2}}\sum_{m=1}^{2k} n_m \bar{g}_m , \label{noneqspin2kprom}
\ee
which acts as a transition function that connects the $SO(2k)$ monopole gauge fields defined on the field-manifold $S^{2k}_{\text{field}}$: 
\be
A' =g^{\dagger}A g -ig^{\dagger}dg.  \label{so4gaugetransaa}
\ee
The gauge field is  expressed as 
\be
A_{\mu}(n_a(x))=i\frac{1}{2}(1-n_{2k+1})\partial_{\mu}g ~g^{\dagger} 
= -\frac{1}{1+n_{2k+1}} 
\sigma_{mn} 
n_n  \partial_{\mu}n_m,  \label{afieldstrengthnls2k+1}
\ee
and the  field strength $F_{\mu\nu} =\partial_{\mu}A_{\nu}-\partial_{\nu}A_{\mu}+i[A_{\mu}, A_{\nu}]$ is  
\be
F_{\mu\nu}(n_a(x)) 
=\sigma_{mn}\partial_{\mu}n_m \partial_{\nu}n_n -\frac{1}{1+n_{2k+1}} \sigma_{mn}
n_n (\partial_{\mu}n_m\partial_{\nu}n_{2k+1}-\partial_{\nu}n_m\partial_{\mu}n_{2k+1}). \label{fieldstrengthnls2k+1}
\ee 

\item Make use of the BPS inequality to construct  tensor field theory actions.   

With the totally antisymmetric tensor field strength  
\be
F_{\mu_1\mu_2\cdots\mu_{2l}} \equiv \frac{1}{(2l)!}F_{[\mu_1\mu_2}F_{\mu_3\mu_4}\cdots F_{\mu_{2l-1}\mu_{2l}]}, \label{totalanfietchr}
\ee
and its dual tensor field strength\footnote{(\ref{defdualf}) satisfies 
\be 
\tilde{\tilde{F}}_{\mu_1\mu_2\cdots \mu_{2l}} ={F}_{\mu_1\mu_2\cdots \mu_{2l}} . 
\ee
} 
\be
\tilde{F}_{\mu_1\mu_2\cdots \mu_{2l}} \equiv \frac{1}{(2k-2l)!}\epsilon_{\mu_1\mu_2\cdots \mu_{2k}}{F}_{\mu_{2l+1}\mu_{2l+2}\cdots \mu_{2k}}, \label{defdualf}
\ee
the $k$th Chern number can be expressed as 
\begin{align}
c_k&=\frac{1}{k!(4\pi)^k}\int d^{2k}x~\epsilon_{\mu_1\mu_{2}\cdots \mu_{2k}}~\tr(F_{\mu_1\mu_2}F_{\mu_3\mu_4}\cdots F_{\mu_{2k-1}\mu_{2k}}) \nn\\
&=\frac{(2k-2l)!}{k!(4\pi)^k}\int d^{2k}x ~\tr(F_{\mu_1\mu_2\cdots\mu_{2l}} \tilde{F}_{\mu_1\mu_2\cdots\mu_{2l}}), 
\end{align}
where 
\be
l=1,2,\cdots, [k/2]. \label{rageoflfromto}
\ee
Following to the idea of \cite{Tchrakian-1980} and \cite{Adam-Ferreira-Hora-Wereszczynski-Zakrzewski-2013}, 
we construct  tensor gauge theory action so that the action can satisfy the BPS inequality: 
\be
S_{2k,2l}~ \ge~ {A}(S^{2k}_{\text{phys.}})\cdot c_k,   \label{newYM2k+1hamgauge}
\ee
which is\footnote{Here, we added the coefficients in front of   $F^2$ and $\tilde{F}^2$ for the later convenience. Recall that there exists the local degree of freedom indicated by $\lambda(x)$  in (\ref{doflambda}).}  
\begin{align}
S_{2k,2l}& =\frac{(2k-2l)!}{(2k)!}\int_{\mathbb{R}^{2k}} d^{2k}x ~\tr\biggl({\frac{1}{2^{k-2l}}~F_{\mu_1\mu_2\cdots \mu_{2l}}^2}+2^{k-2l}~{\tilde{F}_{\mu_1\mu_2\cdots \mu_{2l}}^2}\biggr)\nn\\
&= \frac{1}{(2k)!}\int_{\mathbb{R}^{2k}} d^{2k}x ~\tr\biggl((2k-2l)!~{\frac{1}{2^{k-2l}}~F_{\mu_1\mu_2\cdots \mu_{2l}}^2}+(2l)!~ 2^{k-2l}~{F}_{\mu_{2l+1}\mu_{2l+2}\cdots \mu_{2k}}^2\biggr), \label{actiontensorgaugefieldaction}
\end{align}
where we used  
\be
\frac{1}{(2l)!}F_{\mu_1 \mu_2 \cdots \mu_{2l}}^2 = \frac{1}{(2k-2l)!}{\tilde{F}_{\mu_1 \mu_2 \cdots \mu_{2k-2l}}}^2.  \label{squareoffieldstr}
\ee
According to the distinct decompositions of the $k$th Chern number (\ref{rageoflfromto}), there exist $[k/2]$ different tensor gauge theory actions.\footnote{See Appendix \ref{appendix:tensorfieldtheory} for details about the tensor gauge field theory.}  
(\ref{actiontensorgaugefieldaction}) has the symmetry 
\be
S_{2k,2l} =S_{2k, 2k-2l},  
\ee
and hence there are $[k/2]$  independent actions  $S_{2k,2l}$ 
in accordance with  (\ref{rageoflfromto}).

\item Express the tensor  gauge theory action by the NLS field. 

Substitute  (\ref{fieldstrengthnls2k+1}) into (\ref{actiontensorgaugefieldaction}) to express $S_{2k,2l}$ with  the NLS field:  
\be
S_{2k,2l}~~~
{\rightarrow}~~~H_{2k,2l}=\frac{(2k-2l)!}{(2k)!}\int_{\mathbb{R}^{2k}} d^{2k}x ~\tr\biggl({\frac{1}{2^{k-2l}}~F_{\mu_1\mu_2\cdots \mu_{2l}}^2}+2^{k-2l}~{\tilde{F}_{\mu_1\mu_2\cdots \mu_{2l}}^2}\biggr)\biggr|_{F_{\mu\nu}=F_{\mu\nu}(n_a)}. \label{fromstoh2k+1}
\ee
(\ref{fromstoh2k+1}) realizes our $O(2k+1)$ S-NLS model Hamiltonian.  
Similarly,  $k$th Chern number turns to   
\begin{align}
c_k~~&\overset{F_{\mu\nu}=F_{\mu\nu}(n_a)}{\longrightarrow}~~N_{2k}= \frac{1}{{A}(S^{2k})}\int_{\mathbb{R}^{2k}_{\text{phys.}}}~ d^{2k}x~\frac{1}{(2k)!} \epsilon_{a_1 a_2 \cdots a_{2k+1}}n_{a_{2k+1}} \partial_{1}n_{a_1} \partial_{2}n_{a_2} \cdots \partial_{{2k}}n_{a_{2k}}, 
\end{align}
which stands for the $O(2k+1)$ NLS model winding number associated with $\pi_{2k}(S^{2k}) ~\simeq~\mathbb{Z}$ \cite{Patani-Schlindwein-Shafi-1976}. 
The BPS inequality (\ref{newYM2k+1hamgauge}) is  rephrased as 
\be
H_{2k,2l}\ge {A}(S^{2k}_{\text{phys.}})\cdot N_{2k}. \label{newYM2k+1hamgaugenls}
\ee

\end{enumerate} 

Two important features of the tensor field gauge theory are inherited to the obtained S-NLS models. One is  the local symmetry and the other is the BPS inequality.  As the tensor field strength action (\ref{actiontensorgaugefieldaction}) enjoys the $SO(2k)$ gauge symmetry,  the  S-NLS model Hamiltonian $\it{necessarily}$ possesses the local $SO(2k)$  symmetry.  Similarly, as the tensor gauge field action is constructed so as to satisfy the BPS inequality,  the  S-NLS model  Hamiltonian  $\it{automatically}$ satisfies the BPS inequality. 


One should not confuse the present  local symmetry with the hidden local symmetry of \cite{Bando-Kugo-Yamawaki-1988,Bando-Kugo-Uehara-Yamawaki-Yanagida-1985, Ma-Harada-2016} (see Appendix \ref{append:hls}). The present $SO(2k)$ local symmetry stems from  the  gauge symmetry of the particular form of the parent tensor field action, while the  hidden $SO(2k)$ local symmetry  exists in $\it{any}$ NLS models whose field-manifold is $S^{2k}$.   

\subsection{$O(7)$ S-NLS model} 

From the general procedure, we explicitly  construct the $O(7)$ S-NLS  model Hamiltonian.   
The steps  1 and 2  are obvious.  From (\ref{fieldstrengthnls2k+1}),  
the $SO(6)$ gauge field strength is given by 
\be
F_{\mu\nu} 
=\sigma_{mn}\partial_{\mu}n_m\partial_{\nu}n_n-\frac{1}{1+n_7}\sigma_{mn}n_n(\partial_{\mu}n_m\partial_{\nu}n_7 - \partial_{\nu}n_m \partial_{\mu}n_7),  \label{fso7nn}
\ee 
where $\sigma_{mn}$ denote  the $Spin(6)$ generators, 
%
and (\ref{totalanfietchr}) yields the totally antisymmetric four-rank tensor 
\be
F_{\mu\nu\rho\sigma} \equiv \frac{1}{4!}F_{[\mu\nu}F_{\rho\sigma]} 
=\frac{1}{6}(\{F_{\mu\nu},F_{\rho\sigma}\} +\{F_{\mu\rho},F_{\sigma\nu}\} +\{F_{\mu\sigma}, F_{\nu\rho}\}) , 
\ee
and 
its dual 
\be
\tilde{F}_{\mu\nu} 
=\frac{1}{4!}\epsilon_{\mu\nu\rho\sigma\kappa\tau}F_{\rho\sigma}F_{\kappa\tau} =\frac{1}{4!}\epsilon_{\mu\nu\rho\sigma\kappa\tau}F_{\rho\sigma\kappa\tau}.
\ee
The BPS inequality, 
\be
S_{6,2} ~\ge~ {A}(S^6) \cdot c_3 ,  
\ee
introduces   the tensor gauge field action as  
\begin{align}
S_{6,2} &\equiv \frac{1}{60} \int_{\mathbb{R}^6} d^6x ~ \tr({F_{\mu\nu}}^2 +4{\tilde{F}_{\mu\nu}}^2) =\frac{1}{60} \int_{\mathbb{R}^6} d^6x ~ \tr({F_{\mu\nu}}^2 +\frac{1}{3}{{F}_{\mu\nu\rho\sigma}}^2)  \nn\\
&=\frac{1}{60} \int_{\mathbb{R}^6} d^6x ~ \tr( {F_{\mu\nu}}^2 + \frac{1}{18}( {F_{\mu\nu}}^2)^2 -\frac{2}{9}F_{\mu\nu}F_{\rho\sigma}F_{\mu\rho}F_{\nu\sigma} +\frac{1}{18}(F_{\mu\nu}F_{\rho\sigma})^2 ) . \label{so7ffffhaml} 
\end{align}
Here, we used ${A}(S^6) =\frac{16}{15}\pi^3$  and 
\be
c_3 =\frac{1}{3!(4\pi)^3}\int d^6x~\epsilon_{\mu\nu\rho\sigma\kappa\tau}~\tr(F_{\mu\nu}F_{\rho\sigma}F_{\kappa\tau})  
=\frac{1}{2 (2\pi)^3}~\int d^6x~\tr(F_{\mu\nu}\tilde{F}_{\mu\nu}). \label{thirdchern} 
\ee
(\ref{so7ffffhaml}) is essentially  the 6D action constructed by Tchrakian \cite{Tchrakian-1980}.\footnote{ 
Another 6D action   
of  a  triple form of the field strengths, $\frac{1}{6} f^{abc}F_{\mu\nu}^a F_{\nu\rho}^b F_{\rho\mu}^c$, is constructed in \cite{Saclioglu1986}, but it  is not positive definite in general. 
 Meanwhile,  $S_{6,2}$ (\ref{so7ffffhaml}) only with even powers of the field strengths  does not have such a  problem. }

With (\ref{fso7nn}) and the properties of the $Spin(6)$ generators 
\be
\tr(\sigma_{mn}\sigma_{pq}) =\delta_{mp}\delta_{nq} -\delta_{mq}\delta_{np},  ~~~
\sigma_{[mn}\sigma_{pq]} =3~ \epsilon_{mnpqst}\sigma_{st},   
\ee
we can express  the two terms of $S_{6,2}$ as 
\begin{align}
&\tr({F_{\mu\nu}}^2)|_{F_{\mu\nu}=F_{\mu\nu}(n_a)} = (\partial_{\mu}n_a)^2  (\partial_{\nu}n_b)^2 -(\partial_{\mu}n_a \partial_{\nu}n_a)^2 
= \partial_{\mu}n_a \partial_{\nu}n_b \cdot \partial_{\mu}n_{[a} \partial_{\nu}n_{b]} = \frac{1}{2}( \partial_{\mu}n_{[a} \partial_{\nu}n_{b]} )^2, \nn\\
&\tr({\tilde{F}_{\mu\nu}}^2) |_{F_{\mu\nu}=F_{\mu\nu}(n_a)}
= \frac{1}{ 2\cdot 4!}  \partial_{\mu}n_a \partial_{\nu}n_b\partial_{\rho}n_c  \partial_{\sigma}n_d \cdot \partial_{\mu}n_{[a} \partial_{\nu}n_b\partial_{\rho}n_c  \partial_{\sigma}n_{d]} =\frac{1}{ 2\cdot (4!)^2} ( \partial_{\mu}n_{[a} \partial_{\nu}n_b\partial_{\rho}n_c  \partial_{\sigma}n_{d]})^2,   \label{calftilftilmunu}
\end{align}
and then  
\be
H_{6,2}=\frac{1}{60}\int d^6x ~\biggl(  \partial_{\mu}n_a    \partial_{\nu}n_b \cdot   \partial_{\mu}n_{[a}    \partial_{\nu}n_{b]}  
 + \frac{1}{12} \cdot \partial_{\mu}n_a \partial_{\nu}n_b\partial_{\rho}n_c  \partial_{\sigma}n_d \cdot \partial_{\mu}n_{[a} \partial_{\nu}n_b\partial_{\rho}n_c  \partial_{\sigma}n_{d]}\biggr).\label{o7nlsmodelhader}
\ee 
The first quartic derivative term of $H_{6,2}$ acts to shrink a  soliton configuration, while    
the second octic derivative term  acts to expand the configuration just like  the original Skyrme 
term 
  and is expanded as 
\begin{align}
&\partial_{\mu}n_a \partial_{\nu}n_b \partial_{\rho}n_c \partial_{\sigma}n_d \cdot \partial_{\mu}n_{[a} \partial_{\nu}n_{b}\partial_{\rho}n_{c} \partial_{\sigma}n_{d]} \nn\\
&=((\partial_{\mu}n_a)^2 )^4 + 3 ((\partial_{\mu}n_a\partial_{\nu}n_a)^2)^2- 6 ((\partial_{\mu}n_a)^2 )^2 (\partial_{\nu}n_b\partial_{\rho}n_b)^2  \nn\\
&-6 (\partial_{\mu}n_a\partial_{\nu}n_a)(\partial_{\nu}n_b\partial_{\rho}n_b)(\partial_{\rho}n_c\partial_{\sigma}n_c)(\partial_{\sigma}n_d\partial_{\mu}n_d)  + 8(\partial_{\mu}n_a)^2 (\partial_{\nu}n_b \partial_{\rho}n_b)(\partial_{\rho}n_c \partial_{\sigma}n_c)(\partial_{\sigma}n_d \partial_{\nu}n_d).  \label{delnquadra}
\end{align}
The third Chern number $c_3$ turns to the $O(6)$ NLS model winding number of $\pi_6(S^6)\simeq \mathbb{Z}$: 
\be 
N_6 = \frac{1}{{A}(S^6_{\text{phys.}})}\int_{\mathbb{R}^6_{\text{phys.}}}~ d^6x~\frac{1}{6!} \epsilon_{\mu\nu\rho\sigma\kappa\tau}\epsilon_{abcdefg}n_g \partial_{\mu}n_a \partial_{\nu}n_b \partial_{\rho}n_c\partial_{\sigma}n_d \partial_{\kappa}n_e \partial_{\tau}n_f . 
\ee


\subsection{$O(2k+1)$ S-NLS models}\label{sec:nlsardim}
 
In low dimensions, the  numbers of the S-NLS model Hamiltonians are  counted as  %
\be
O(5)~:~1,~~O(7)~:~1,~~O(9)~:~2,~~O(11)~:~2. 
\ee
For the previous  $O(5)$ and $O(7)$ cases, we have single S-NLS model Hamiltonian, but for $O(2k+1)$, we have $[k/2]$ Hamiltonians.   
 In the following, we construct $O(2k+1)$ NLS model Hamiltonians for two typical cases,  $2+(2k-2)$  and  $k+k$.

\subsubsection{$2+(2k-2)$ decomposition}

In $2+(2k-2)$ decomposition, 
the tensor gauge theory action is given by  
\begin{align}
S_{2k,2}&= \frac{1}{2k(2k-1) } \int d^{2k}x ~ \tr\biggl(\frac{1}{2^{k-2}}~{F_{\mu\nu}}^2 +2^{k-2}~{\tilde{F}_{\mu\nu}}^2\biggr)\nn\\
&=\frac{1}{(2k)! } \int d^{2k}x ~ \tr\biggl(\frac{1}{2^{k-2}}~(2k-2)!~{F_{\mu\nu}}^2 +2^{k-2}~2!~{{F}_{\mu_1\mu_2\mu_3 \cdots \mu_{2k-2}}}^2\biggr).  \label{so2k+1ffffhaml}
\end{align}
From the properties of the $Spin(2k)$ generators 
\begin{align} 
&\tr(\sigma_{mn}\sigma_{pq}) =2^{k-3} (\delta_{mp}\delta_{nq} -\delta_{mq}\delta_{np}),  \nn\\
&\sigma_{[m_1m_2}\sigma_{m_3m_4}\cdots\sigma_{m_{2k-3}, m_{2k-2}]} =\frac{(2k-2)!}{2^{k-1}}\epsilon_{m_1m_2m_3\cdots m_{2k}} \sigma_{m_{2k-1}, m_{2k}}, 
\end{align} 
the two terms of $S_{2k,2}$  (\ref{so2k+1ffffhaml}) can be represented as 
\begin{align}
&\tr({F_{\mu\nu}}^2)|_{F_{\mu\nu}=F_{\mu\nu}(n_a)} 
=2^{k-3} \partial_{\mu}n_a \partial_{\nu}n_b \cdot \partial_{\mu}n_{[a} \partial_{\nu}n_{b]}, \nn\\
&\tr({\tilde{F}_{\mu\nu}}^2)|_{F_{\mu\nu}=F_{\mu\nu}(n_a)} =\frac{1}{2^{k-2}~ (2k-2)!}~\partial_{\mu_1}n_{a_1} \partial_{\mu_2}n_{a_2}\cdots \partial_{\mu_{2k-2}}n_{a_{2k-2}}\cdot \partial_{\mu_1}n_{[a_1} \partial_{\mu_2}n_{a_2}\cdots \partial_{\mu_{2k-2}}n_{a_{2k-2]}},   
\end{align}
and so  
\begin{align}
&H_{2k,2}=\frac{1}{4k(2k-1)}\int_{\mathbb{R}^{2k}} d^{2k}x ~\times \nn\\
&\biggl(  \partial_{\mu_1}n_{a_1}   \partial_{\mu_2}n_{a_2} \cdot \partial_{\mu_1}n_{[a_1}   \partial_{\mu_2}n_{a_2]} +\frac{2}{ (2k-2)!}~
\partial_{\mu_1}n_{a_1} \partial_{\mu_2}n_{a_2}\cdots \partial_{\mu_{2k-2}}n_{a_{2k-2}}\cdot \partial_{\mu_1}n_{[a_1} \partial_{\mu_2}n_{a_2}\cdots \partial_{\mu_{2k-2}}n_{a_{2k-2]}}\biggr). \label{hamgene1sto2k+1}
\end{align}
Notice that the first term is a quartic derivative term while the second term is a $4(k-1)$th derivative term. Their competing scaling effect determines the size of  soliton configurations (except for the scale invariant case $k=2$). 
For $k=2$ and $3$, (\ref{hamgene1sto2k+1}) indeed reproduces the previous  $O(5)$ (\ref{nnnnenergy})\footnote{For $O(5)$ $(k=2)$, the first and second terms on the right-hand side of (\ref{hamgene1sto2k+1}) coincide,  
and so (\ref{hamgene1sto2k+1}) is reduced to (\ref{nnnnenergy}). 
} and $O(7)$ (\ref{o7nlsmodelhader}) NLS model Hamiltonians, respectively.

\subsubsection{$k+k$ decomposition for even $k$}\label{subsec:k+kdecomp}

In the special case $(d,2l)=(2k, k)$: 
\be
(d,k) =(4,2), ~(8,4),~(12, 6), ~(16,4),~\cdots,  
\ee
${F_{\mu_1\mu_2\cdots\mu_{k}}}^2={\tilde{F}_{\mu_1\mu_2\cdots\mu_{k}}}^2$ holds, and so 
 (\ref{fromstoh2k+1}) is reduced to   a scale invariant action: 
\be
S_{2k,k}=
2\frac{k!}{(2k)!}\int_{\mathbb{R}^{2k}} d^{2k}x ~\tr ({F_{\mu_1\mu_2\cdots\mu_{k}}}^2   ). 
\ee
The equations of motion are derived as 
\be
D_{\mu_1}F_{\mu_1\mu_2\cdots \mu_{k}}\equiv \partial_{\mu_1}F_{\mu_1\mu_2\cdots \mu_{k}}+i[A_{\mu_1}, F_{\mu_1\mu_2\cdots \mu_{k}}]=0. \label{eomk+k}
\ee
The tensor gauge field strength ${F}_{\mu_1\mu_2\cdots\mu_{k}}=\frac{1}{k!} F_{[\mu_1\mu_2}F_{\mu_3\mu_4}\cdots F_{\mu_{k-1}\mu_k]}$ made of the $SO(2k)$ ``instanton'' configuration\footnote{The $SO(2k)$ instanton configuration (\ref{instskonr2k}) is  a  stereographic projection of the $SO(2k)$ monopole field configuration on $S^{2k}$ (\ref{fieldconfigstrength}) (Appendix \ref{append:stereo}). 
} 
\be
F_{\mu\nu}|_{n_a=r_a} =\frac{4}{(x^2+1)^2}\sigma_{\mu\nu},  \label{instskonr2k}
\ee
is given by 
\be
F_{\mu_1\mu_2\cdots\mu_k} =\frac{1}{k!}\biggl(\frac{2}{{x}^2+1}\biggr)^k \sigma_{[\mu_1\mu_2}\sigma_{\mu_3\mu_4}\cdots \sigma_{\mu_{k-1}\mu_{k}]}, \label{tensorscalesoltens}
\ee
which carries  unit $k$th Chern number.  (\ref{tensorscalesoltens})  
satisfies the self-dual equation \cite{Grossmanetal1984, Tchrakian-1985, OSe-Tchrakian-1987, Takesue-2017}
\be
\tilde{F}_{\mu_1\mu_2\cdots\mu_{k}} ={F}_{\mu_1\mu_2\cdots\mu_{k}}, \label{tenssenfdual}
\ee
due to the property of  the $Spin(2k)$ matrix generators:\footnote{ 
Generally, the $Spin(2k)$ generators  satisfy  
\be
\frac{1}{(2l)!} \sigma_{[\mu_1\mu_2} \sigma_{\mu_3\mu_4}\cdots  \sigma_{\mu_{2l-1}\mu_{2l}]} = 2^{k-2l}\frac{1}{((2k-2l)!)^2}  ~ \epsilon_{\mu_1\mu_2\mu_3\cdots \mu_{2k}}  \sigma_{[\mu_{2l+1} \mu_{2l+2}}\cdots \sigma_{\mu_{2k-1}\mu_{2k}]},   \label{dualarb}
\ee
which is reduced to (\ref{selfdusigma}) in the special case $k=2l$. 
The tensor instanton configuration (\ref{tensorscalesoltens}) also satisfies 
\be
F_{\mu_1\mu_2\cdots\mu_{2l}} = \biggl(\frac{(x^2+1)^2}{2}\biggr)^{k-2l}~\tilde{F}_{\mu_1\mu_2\cdots\mu_{2l}},   \label{tenssenfdualgene} 
\ee
which  reproduces  (\ref{tenssenfdual}) when $k=2l$.  
} 
\be
 \sigma_{[\mu_1\mu_2} \sigma_{\mu_3\mu_4}\cdots  \sigma_{\mu_{k-1}\mu_{k]}} = \frac{1}{k!}  ~ \epsilon_{\mu_1\mu_2\mu_3\cdots \mu_{2k}}  \sigma_{[\mu_{k+1} \mu_{k+2}}\cdots \sigma_{\mu_{2k-1}\mu_{2k}]}. \label{selfdusigma}
\ee
Because of the Bianchi identity for  tensor fields, the self-dual tensor field (\ref{tensorscalesoltens}) is a solution  of the equations of motion (\ref{eomk+k}) (see Appendix \ref{appendix:tensorfieldtheory} for details).\footnote{Note that while  (\ref{tensorscalesoltens}) realizes a solution of (\ref{eomk+k}), (\ref{instskonr2k}) is $\it{not}$ a solution of the pure Yang-Mills field equation except for $k=2$ (see Appendix \ref{append:nonabemono}).} 
 In low dimensions, one may directly confirm that (\ref{tensorscalesoltens})  satisfies  (\ref{eomk+k}) with  
\be
A_{\mu} =-\frac{2}{x^2+1}\sigma_{\mu\nu}x_{\nu}.  \label{gaugescaleinvconfig}
\ee

 To express  the tensor gauge theory action in terms of $O(2k+1)$ NLS field, we  utilize the short-cut method  mentioned in Sec.\ref{subsubsec:conso5}. We truncate the field strength  $F_{\mu\nu}~\rightarrow ~\sigma_{mn}\partial_{\mu}n_m\partial_{\nu}n_n$ to have 
\begin{align}
&\tr(F_{\mu_1\mu_2\cdots\mu_{k}}^2) ~\rightarrow~   \nn\\
&\biggl(\frac{1}{k!}\biggr)^2 ~\tr(\sigma_{m_1 m_2}\cdots \sigma_{m_{k-1}m_k}\sigma_{m_1'm_2'}\cdots \sigma_{m_{k-1}'m_k'} )~\partial_{\mu_1}n_{[m_1} \partial_{\mu_2}n_{m_2}\cdots \partial_{\mu_k}n_{m_k]} ~\partial_{\mu_1}n_{[m_1'} \partial_{\mu_2}n_{m_2'} \cdots\partial_{\mu_k}n_{m_k']}.  
\label{ffourthsq}
\end{align}
$\partial_{\mu_1}n_{[m_1}\partial_{\mu_2}n_{m_2}\cdots\partial_{\mu_k}n_{m_k]}$ consists of $k!$ terms of  totally antisymmetric combination  about the Latin indices, $m_1,m_2,\cdots,m_k$. 
The $Spin(2k)$ matrix part of (\ref{ffourthsq}) can be expressed as 
\begin{align}
&\tr(\sigma_{m_1 m_2}\sigma_{m_3 m_4}\cdots  \sigma_{m_{2k-1} m_{2k}}) \nn\\
&= \frac{1}{2}\biggl(-i\frac{1}{4}\biggr)^k ~\tr(\gamma_{m_1}\gamma_{m_2}\gamma_{m_3}\cdots \gamma_{m_{2k}})\cdot (1-P_{m_1 m_2})(1-P_{m_3 m_4})\cdots (1-P_{m_{2k-1} m_{2k}})+\frac{1}{2}\epsilon_{ m_1 m_2 m_3\cdots m_{2k}}. 
\label{trasigmas2k}
\end{align}
Here, $P_{mn}$ signifies an operation that interchanges  $m$ and $n$, $i.e.$ $P_{mn}(\gamma_m\gamma_n)=\gamma_n\gamma_m$, and  
in the present case, due to  the antisymmetricity of $m$s, we can just replace $(1-P_{mn})$ with $2$. Besides the epsilon tensor part of (\ref{trasigmas2k}) obviously has no effect in  (\ref{ffourthsq}), and thereby  
\be
\tr(\sigma_{m_1 m_2}\sigma_{m_3 m_4}\cdots  \sigma_{m_{2k-1} m_{2k}})~~\rightarrow~~ \frac{1}{2}\biggl(-i\frac{1}{2}\biggr)^k ~\tr(\gamma_{m_1}\gamma_{m_2}\gamma_{m_3}\cdots \gamma_{m_{2k}})~~\rightarrow~~\frac{1}{2} ~k!~\delta_{m_1 m_2}\delta_{m_3 m_4}\cdots \delta_{m_{2k-1} m_{2k}}. 
\ee
In the last arrow we assumed that $k$ is even. 
Eventually,  we obtain  
\be
\tr({F_{\mu_1\mu_2\cdots\mu_{k}}}^2) = \tr({\tilde{F}_{\mu_1\mu_2\cdots\mu_{k}}}^2) 
= \frac{1}{2} (\partial_{\mu_1}n_{a_1}\partial_{\mu_2}n_{a_2} \cdots \partial_{\mu_k}n_{a_k})\cdot (\partial_{\mu_1}n_{[a_1}\partial_{\mu_2}n_{a_2} \cdots \partial_{\mu_k}n_{a_k]}), 
\ee 
which implies 
\begin{align}
H_{2k,k}&=\frac{k!}{(2k)!} ~\int_{\mathbb{R}^{2k}} d^{2k}x~ (\partial_{\mu_1}n_{a_1}\partial_{\mu_2}n_{a_2} \cdots \partial_{\mu_k}n_{a_k})\cdot (\partial_{\mu_1}n_{[a_1}\partial_{\mu_2}n_{a_2} \cdots \partial_{\mu_k}n_{a_k]})\nn\\
&= \frac{1}{(2k)!} ~\int_{\mathbb{R}^{2k}} d^{2k}x~ (\partial_{\mu_1}n_{[a_1}\partial_{\mu_2}n_{a_2} \cdots \partial_{\mu_k}n_{a_k]})^2.  
\label{k+ko2k+1NLsham}
\end{align}
$H_{2k,k}$  accommodates scale invariant soliton solutions as we shall discuss in Sec.\ref{subsec:eofscaling}.     
For $k=2$, (\ref{k+ko2k+1NLsham}) is reduced to the $O(5)$ S-NLS model Hamiltonian (\ref{nnnnenergy}).

\section{$O(2k)$ S-NLS Models}\label{sec:o2knls}

\begin{figure}[tbph]
\center
\includegraphics*[width=160mm]{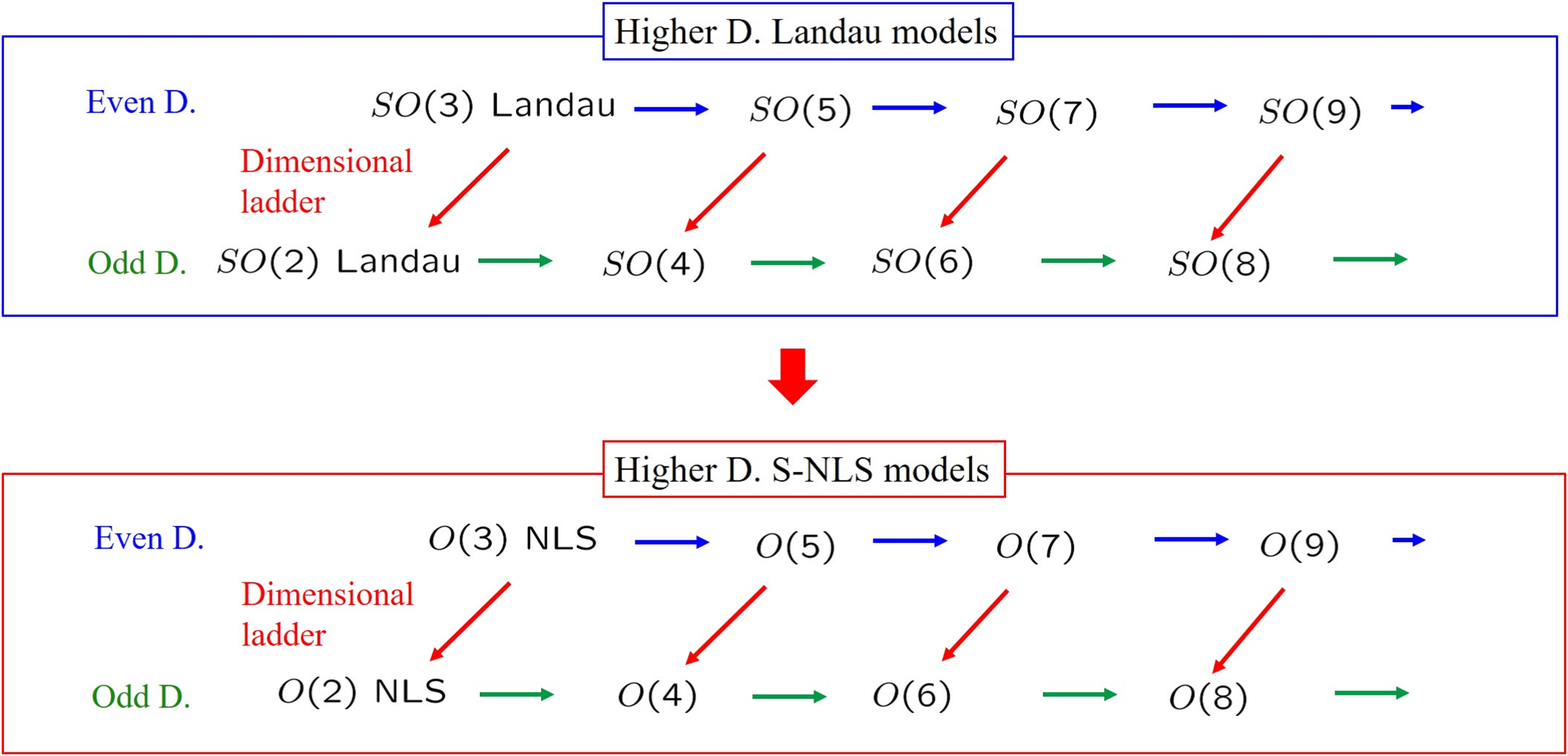}
\caption{The  dimensional ladder of the higher dimensional Landau models and that of the higher dimensional S-NLS models. }
\label{ExtendDim.fig}
\end{figure}

In this section, based on the Chern-Simons term expression of the $k$th Chern number,  we construct $O(2k)$  S-NLS model Hamiltonians in $(2k-1)$D. 
The dimensional hierarchy of the  Landau models  \cite{Hasebe-2017, Hasebe-2020-1} suggests that the dimensional reduction  of the $O(2k+1)$ NLS model may yield  the $O(2k)$ NLS model (Fig.\ref{ExtendDim.fig}).   
More specifically, the 1D reduction of 
$H_{2k,2l}$ gives rise to  two  $O(2k)$ Hamiltonians, $H_{2k-1, 2l-1}$ and $H_{2k-1, 2l}$. 
By removing  duplications  from the symmetry $H_{2k-1, 2l}=H_{2k-1, 2k-1-2l}$, we have $(k-1)$ distinct  $O(2k)$ Hamiltonians that exhaust all possible S-NLS model Hamiltonians in $(2k-1)$D. 
For instance,\footnote{The soliton configuration of $O(2)$  NLS model is given by the  Nielson-Olsen vortex \cite{Nielsen-Olsen-1973}.}   
\begin{align}
&k=2:~~O(5)~\text{S-NLS~model}:~H_{4,2} ~~~~~~~~~~~~\rightarrow~~~~~~~O(4)~\text{S-NLS~model}:~H_{3,1} ,\nn \\
&k=3:~~O(7)~\text{S-NLS~model}:~H_{6,2} ~~~~~~~~~~~~\rightarrow~~~~~~~O(6)~\text{S-NLS~model}:~H_{5,1},~~H_{5,2},\nn \\
&k=4:~~O(9)~\text{S-NLS~model}:~H_{8,2},~H_{8,4} ~~~~\rightarrow~~~~~~~O(8)~\text{S-NLS~model}:~H_{7,1},~~H_{7,2},~~H_{7,3}. \label{dimhieosnls}
\end{align}
The solitons described by the $O(2k)$ S-NLS model naturally appear as anyonic objects in the BF effective field theory of the odd dimensional quantum Hall effect \cite{Hasebe-2017}. 

\subsection{The Chern-Simons term and  the action of pure gauge fields}

As is well known, the Chern number (density) can be expressed by 
\be
 \tr(F^k) =d L_{\text{CS}}^{(2k-1)}[A]
\ee
where $L_{\text{CS}}^{(2k-1)}[A]$ signifies the  $(2k-1)$D Chern-Simons term 
\be
L_{\text{CS}}^{(2k-1)}[A] =k\int_0^1 dt ~\tr(A(tdA+it^2A^2)^{k-1}). 
\label{compcsterm}
\ee
In low dimensions, (\ref{compcsterm}) reads as 
\be
L_{\text{CS}}^{(1)}[A] =\tr A,~~~~L_{\text{CS}}^{(3)}[A] =\tr (AF-\frac{1}{3}iA^3), ~~~~L_{\text{CS}}^{(5)}[A] =\tr (AF^2-\frac{1}{2}iA^3 F -\frac{1}{10}A^5). 
\ee
We make use of the Chern-Simons field description of the Chern number to construct $O(2k)$ S-NLS model Hamiltonians.   
Recall that the transition function (\ref{noneqspin2k}) represents  
 $S^{2k-1}_{\text{field}}$, and the associated gauge field is  given by a pure gauge\footnote{$\mathcal{A}$ (\ref{puregauge2k-1}) naturally appears  in the context of the hidden local symmetry also  (see Appendix \ref{append:o2khls}).}  
\be
\mathcal{A}=-ig^{\dagger}dg, ~~~~\mathcal{F}=d\mathcal{A}+i\mathcal{A}^2=0.  
\label{puregauge2k-1}
\ee
For the pure gauge (\ref{puregauge2k-1}), the  Chern-Simons term (\ref{compcsterm}) is reduced to 
\begin{align}
L_{\text{CS}}^{(2k-1)}[\mathcal{A}] &=(-i)^{k-1}\frac{k!(k-1)!}{(2k-1)!} ~\tr(\mathcal{A}^{2k-1}) \nn\\
&=
(-i)^{k-1}\frac{k!(k-1)!}{(2k-1)!} ~d^{2k-1}x~\epsilon_{\alpha_1\alpha_2\cdots \alpha_{2k-1}}\tr(\mathcal{A}_{\alpha_1}\mathcal{A}_{\alpha_2}\cdots \mathcal{A}_{\alpha_{2k-1}}), 
\end{align}
where   we used 
$\int_0^1 dt ~(t-t^2)^{k-1}=\frac{((k-1)!)^2}{(2k-1)!}$ and 
assumed that $\mathcal{A}$ is  one-form on $x_{\alpha} ~\in ~\mathbb{R}^{2k-1}_{\text{phys.}}$: 
\be
\mathcal{A}=\sum_{\alpha=1}^{2k-1}\mathcal{A}_{\alpha}dx_{\alpha}.  \label{aexpnfield}
\ee
We introduce $p$-rank  tensor field associated with the pure gauge as 
\be
\mathcal{A}_{\alpha_1\alpha_2\cdots\alpha_p} \equiv (-i)^{\frac{1}{2}p(p-1)}\frac{1}{p!}\mathcal{A}_{[\alpha_1} \mathcal{A}_{\alpha_2} \cdots \mathcal{A}_{\alpha_p]}, 
\label{deftensorgaugefield}
\ee
and its dual 
\begin{align}
\tilde{\mathcal{A}}_{\alpha_1 \alpha_2\cdots \alpha_p} &\equiv \frac{1}{(d-p)!} \epsilon_{\alpha_1\alpha_2\cdots \alpha_d} \mathcal{A}_{\alpha_{p+1} \alpha_{p+2}\cdots \alpha_d}\nn\\
& = (-i)^{\frac{1}{2}(d-p)(d-p-1)}\frac{1}{(d-p)!}~ \epsilon_{\alpha_1\alpha_2\cdots \alpha_d} \mathcal{A}_{\alpha_{p+1}} \mathcal{A}_{\alpha_{p+2}} \cdots \mathcal{A}_{\alpha_d},  \label{dualtensfamus}
\end{align}
which satisfies 
\be
\frac{1}{p!}{\tilde{\mathcal{A}}_{\alpha_1\alpha_2\cdots\alpha_p}}^2=\frac{1}{(2k-1-p)!}{\mathcal{A}_{\alpha_1\alpha_2\cdots\alpha_{2k-1-p}}}^2. 
\ee
In (\ref{deftensorgaugefield}), $(-i)^{\frac{1}{2}p(p-1)}$ is added so that $\mathcal{A}_{\alpha_1\alpha_2\cdots\alpha_p}$ may be Hermitian. 
For instance, 
\begin{align}
&\mathcal{A}_{\alpha\beta} =-i\frac{1}{2}[\mathcal{A}_{\alpha}, \mathcal{A}_{\beta}]=\frac{1}{2}\partial_{[\alpha}\mathcal{A}_{\beta]}, \nn\\
&\mathcal{A}_{\alpha\beta\gamma} =i\frac{1}{3!}\mathcal{A}_{[\alpha} \mathcal{A}_{\beta} \mathcal{A}_{\gamma]}=-\frac{1}{3}(\mathcal{A}_{\alpha}\mathcal{A}_{\beta\gamma} +   \mathcal{A}_{\beta}\mathcal{A}_{\gamma\alpha}+\mathcal{A}_{\gamma}\mathcal{A}_{\alpha\beta}  ), \nn\\
&\mathcal{A}_{\alpha\beta\gamma\delta} =-\frac{1}{4!}\mathcal{A}_{[\alpha} \mathcal{A}_{\beta} \mathcal{A}_{\gamma} \mathcal{A}_{\delta ]}=\frac{1}{6} ( \{ \mathcal{A}_{\alpha\beta}, \mathcal{A}_{\gamma\delta} \}  - \{ \mathcal{A}_{\alpha\gamma}, \mathcal{A}_{\beta\delta} \} + \{ \mathcal{A}_{\alpha\delta}, \mathcal{A}_{\beta\gamma} \}   ). 
\end{align}
In a similar manner to Sec.\ref{subsec:generaproc},  
  we  represent  the Chern-Simons action of the pure gauge as\footnote{With $g$ (\ref{puregauge2k-1}) being a non-linear sigma field, (\ref{puregaugecsterm}) becomes the Wess-Zumino action \cite{Wess-Zumino-1971} in $(2k-1)$D: 
\be
\Gamma_{\text{WZ}}^{(2k-1)}[g] =\mathcal{S}_{\text{CS}}^{(2k-1)}[\mathcal{A}]\biggr|_{\mathcal{A}=-ig^{\dagger}dg} =-\frac{1}{(2\pi)^{k}} i^{k}\frac{(k-1)!}{(2k-1)!} \int ~\tr((g^{\dagger}dg)^{2k-1}).  
\ee
 } 
\begin{align}
\mathcal{S}_{\text{CS}}^{(2k-1)}[\mathcal{A}] &\equiv \frac{1}{k!(2\pi)^k}\int L_{\text{CS}}^{(2k-1)}[\mathcal{A}] \nn\\
&=
\frac{1}{(2\pi)^k} \frac{(k-1)! (2k-1-p)!}{(2k-1)!} \int_{\mathbb{R}^{2k-1}_{\text{phys.}}} d^{2k-1}x~\tr(\mathcal{A}_{\alpha_1\alpha_2\cdots \alpha_{p}} \tilde{\mathcal{A}}_{\alpha_1\alpha_2\cdots \alpha_{p}} )  , \label{puregaugecsterm}
\end{align}
where 
\be
p=1,2,\cdots, k-1.
\ee
In low dimensions, (\ref{puregaugecsterm}) provides  
\begin{align}
\mathcal{S}_{\text{CS}}^{(3)}[\mathcal{A}]&
=\frac{1}{12\pi^2}\int d^3x~\tr(\mathcal{A}_{\alpha}\tilde{\mathcal{A}}_{\alpha}),\nn \\
\mathcal{S}_{\text{CS}}^{(5)}[\mathcal{A}]&
=\frac{1}{20\pi^3}\int d^5x~\tr(\mathcal{A}_{\alpha}\tilde{\mathcal{A}}_{\alpha})
=\frac{1}{80\pi^3}\int d^5x~ \tr(\mathcal{A}_{\alpha\beta}\tilde{\mathcal{A}}_{\alpha\beta}).
\end{align}
From the BPS inequality 
\be
S_{2k-1, p}[\mathcal{A}] ~\ge~ {A}(S^{2k-1}) \cdot \mathcal{S}_{\text{CS}}^{(2k-1)}[\mathcal{A}], \label{bpsineqpuregauge}
\ee
we construct an action made of the pure gauge tensor field:\footnote{As explained around (\ref{doflambda}), there  exists a local degree of freedom in decomposing    $\mathcal{A}_{\alpha_1\alpha_2\cdots \alpha_{p}}\times \tilde{\mathcal{A}}_{\alpha_1\alpha_2\cdots \alpha_{p}}$. 
 }     
\begin{align}
{S}_{2k-1, p}[\mathcal{A}]&= \frac{1}{2^k} \frac{(2k-1-p)!}{(2k-1)!}~\int_{\mathbb{R}^{2k-1}}  d^{2k-1}x~\biggl({\tr(\mathcal{A}_{\alpha_1\alpha_2\cdots \alpha_{p}}}^2)+\tr({\tilde{\mathcal{A}}_{\alpha_1\alpha_2\cdots \alpha_{p}}}^2) \biggr) \nn\\
&=\frac{1}{2^k (2k-1)!}~\int_{\mathbb{R}^{2k-1}}d^{2k-1}x ~\biggl((2k-1-p)!~\tr({\mathcal{A}_{\alpha_1\alpha_2\cdots \alpha_{p}}}^2)+p!~\tr({\mathcal{A}_{\alpha_{p+1}\alpha_{p+2}\cdots\alpha_{2k-1}}}^2) \biggr). 
\label{defhsk-1p}
\end{align}
Notice that we can also obtain (\ref{defhsk-1p}) by  the  following formal replacement  in  the $2k$D tensor gauge field action $S_{2k,2l}$ (\ref{actiontensorgaugefieldaction}) with the dimensional reduction ($2k ~\rightarrow ~2k-1$): 
\begin{subequations}
\label{replacementftoamathcal}
\begin{align}
F_{\mu_1\mu_2\cdots\mu_{2l}}, ~~~F_{\mu_{2l+1}\mu_{2l+2}\cdots\mu_{2k}}~~~&\longrightarrow~~~\mathcal{A}_{\alpha_1\alpha_2\cdots\alpha_{p=2l-1}}, ~~~\mathcal{A}_{\alpha_{p+1=2l}\alpha_{p+2}\cdots\alpha_{2k-1}}, \\
&~~\text{or} \nn\\
F_{\mu_1\mu_2\cdots\mu_{2l}}, ~~~F_{\mu_{2l+1}\mu_{2l+2}\cdots\mu_{2k}}~~~&\longrightarrow~~~\mathcal{A}_{\alpha_1\alpha_2\cdots\alpha_{p=2l}}, ~~~\mathcal{A}_{\alpha_{p+1=2l+1}\alpha_{p+2}\cdots\alpha_{2k-1}}. 
\end{align}
\end{subequations}
Unlike the $2k$D action  (\ref{actiontensorgaugefieldaction}), (\ref{defhsk-1p})  consists  of the ``bare'' tensor gauge fields  (not  the field strengths), and  
so  $S_{2k-1,p}$ does not  have gauge symmetry. 
Viewing the above process inversely, we may say there always exists one-dimension higher tensor gauge field theory behind every  odd D Skyrme model.

\subsection{Explicit constructions}

With (\ref{noneqspin2k}),    the pure gauge field 
(\ref{puregauge2k-1})  can be represented as 
\be
\mathcal{A}_{\alpha}(n_m)=-ig^{\dagger}\partial_{\alpha}g=-2\bar{\sigma}_{mn}n_n \partial_{\alpha} 
n_m,  \label{compgaugepure}
\ee
where $\bar{\sigma}_{mn}$ denote the $Spin(2k)$ matrix generators. 
Substituting (\ref{compgaugepure}) into (\ref{deftensorgaugefield}),  we can derive the NLS field expression of  $\mathcal{A}_{\alpha_1\alpha_2\cdots\alpha_p}$.  For instance 
\be
\mathcal{A}_{\alpha\beta}\biggl|_{\mathcal{A}_{\alpha}=\mathcal{A}_{\alpha}(n_m)}=-2i\bar{\sigma}_{mp}\bar{\sigma}_{nq}n_p n_q \partial_{\alpha}n_{[m} \partial_{\beta}n_{n]}=-\bar{\sigma}_{mn}\partial_{\alpha} n_{[n} \partial_{\beta}n_{m]}.   
\ee
Just as in the tensor gauge field strength in Sec.\ref{sec:evenhigherdim}, the antisymmetricity of the $\it{Greek}$ indices of the parent tensor gauge field  is inherited to that of the $\it{Latin}$ indices of the NLS field.
With such substitutions, the $O(2k)$ S-NLS model Hamiltonian is obtained from $S_{2k-1, p}$:   
\begin{align}
&S_{2k-1, p}  ~~{\rightarrow} ~~\nn\\
& H_{2k-1, p} =\frac{1}{2^k (2k-1)!}~\int_{\mathbb{R}^{2k-1}}d^{2k-1}x ~\biggl((2k-1-p)!~\tr({\mathcal{A}_{\alpha_1\alpha_2\cdots \alpha_{p}}}^2)+p!~\tr({\mathcal{A}_{\alpha_1\alpha_2\cdots\alpha_{2k-1-p}}}^2) \biggr) \biggl|_{{\mathcal{A}_{\alpha}=\mathcal{A}_{\alpha}(n_m)}}.  
\label{2k-1nlsmfiehamil}
\end{align}
Similarly, the Chern-Simons term (\ref{puregaugecsterm}) turns to  the winding number of  
$\pi_{2k-1}(S^{2k-1}) \simeq \mathbb{Z}$:    
\begin{align}
\mathcal{S}_{\text{CS}}^{(2k-1)}  ~~{\rightarrow} ~~ N_{2k-1} 
 = \frac{1}{{A}(S^{2k-1})}\int_{\mathbb{R}^{2k-1}_{\text{phys.}}} d^{2k-1}x~\epsilon_{m_1m_2\cdots m_{2k}} n_{m_{2k}} \partial_{1}n_{m_1}\partial_{2}n_{m_2}\cdots \partial_{{2k-1}}n_{m_{2k-1}}.   \label{expwindingsk-1}
\end{align}
As in the previous $O(2k+1)$ S-NLS models, the parent BPS inequality (\ref{bpsineqpuregauge}) guarantees 
the BPS inequality of the $O(2k)$ S-NLS models: 
\be
H_{2k-1, p} ~\ge~ {A}(S^{2k-1}) \cdot N_{2k-1}. 
\ee
Since the parent pure actions (\ref{defhsk-1p}) do not have gauge symmetries, the corresponding $O(2k)$ S-NLS models do not either. This ``explains'' the  non-existence of the gauge symmetry
of the Skyrme models in odd dimensions.  
In the following, we demonstrate the above procedure to derive the $O(2k)$ S-NLS model Hamiltonians  for $d=3$ and $d=5$. 

\subsubsection{The Skyrme model: $O(4)$ S-NLS model}\label{subsec:skyrmemodel}
 
For $d=3$,  the pure gauge field action is given by 
\be
S_{3,1}= \frac{1}{12} \int_{\mathbb{R}_{\text{phys.}}^3} d^3x~\tr ({\mathcal{A}_{\alpha}}^2+{\tilde{\mathcal{A}}_{\alpha}}^2) 
=\frac{1}{12}\int_{\mathbb{R}_{\text{phys.}}^3} d^3x~ (\tr({\mathcal{A}_{\alpha}}^2) + \frac{1}{2}\tr({\mathcal{A}_{\alpha\beta}}^2)) =S_{3,2}, 
\ee
where 
$\mathcal{A}_{\alpha}$  
and its dual field $\tilde{\mathcal{A}}_{\alpha}$ are represented as 
\be
\mathcal{A}_{\alpha}= 2\bar{\sigma}_{mn}n_m\partial_{\alpha}n_n,~~~~
\tilde{\mathcal{A}}_{\alpha} =\frac{1}{2}\epsilon_{\alpha\beta\gamma}A_{\beta\gamma} =\epsilon_{\alpha\beta\gamma}\bar{\sigma}_{mn}\partial_{\beta}n_m\partial_{\gamma}n_n,   
\ee
with   $Spin(4)$ matrix generators:  
\be
\bar{\sigma}_{mn} =\frac{1}{2}\bar{\eta}_{mn}^{i}\sigma_i. \label{so4weylsigma}
\ee
From the following formula
\footnote{The $U(2)$ generators (the Pauli matrices and the unit matrix) span the $2\times 2$ matrix space, and so the product of  two $SU(2)$ Pauli matrices or  $Spin(4)$ matrix generators can be represented as a linear combination of  
the $U(2)$ generators.  }    
\be
\bar{\sigma}_{mn}\bar{\sigma}_{pq}  =\frac{1}{4}(\delta_{mp}\delta_{nq}-\delta_{mq}\delta_{np} -\epsilon_{mnpq})1_2 +i\frac{1}{2}(\delta_{mp}\bar{\sigma}_{nq}-\delta_{mq}\bar{\sigma}_{np}+\delta_{nq}\bar{\sigma}_{mp}-\delta_{np}\bar{\sigma}_{mq}),   \label{sigsigprd}
\ee
we can readily  show   
\be
\tr({\mathcal{A}_{\alpha}}^2)|_{\mathcal{A}=\mathcal{A}(n_m)}
=2 (\partial_{\alpha}n_m)^2, ~~~~~~\tr({\tilde{\mathcal{A}}_{\alpha}}^2)|_{\mathcal{A}=\mathcal{A}(n_m)}
=\frac{1}{2}(\partial_{\alpha}n_{[m}\partial_{\beta}n_{n]})^2  
\ee 
to have  
\be
H_{3,1}=
\frac{1}{6}\int_{\mathbb{R}_{\text{phys.}}^3} d^3x~     \biggl(   (\partial_{\alpha}n_m)^2  +\frac{1}{4}(\partial_{\alpha}n_{[m}\partial_{\beta}n_{n]})^2 \biggr). \label{o4skyrhamex31}
\ee
Thus, the  $O(4)$ S-NLS model  Hamiltonian is nothing but the original Skyrme Hamiltonian. 
As mentioned before, the anti-symmetricity of the indices of  $\mathcal{A}_{\alpha\beta}$ is inherited to the anti-symmetricity of the Latin indices of $O(4)$ NLS field of the Skyrme term. 

\subsubsection{$O(6)$ S-NLS models}

Next we consider the case  $d=5$. There exist  two distinct actions: 
\begin{subequations}
\begin{align}
S_{5,1}&=  \frac{1}{40}\int_{\mathbb{R}^5_{\text{phys.}}} d^5x~ \tr ({\mathcal{A}_{\alpha}}^2+{\tilde{\mathcal{A}}_{\alpha}}^2)  =\frac{1}{40} \int_{\mathbb{R}^5_{\text{phys.}}} d^5x~\tr ({\mathcal{A}_{\alpha}}^2+\frac{1}{4!}{{\mathcal{A}}_{\alpha\beta\gamma\delta}}^2),  \\
S_{5,2}&=  \frac{1}{160}\int_{\mathbb{R}^5_{\text{phys.}}} d^5x~\tr ({\mathcal{A}_{\alpha\beta}}^2+{\tilde{\mathcal{A}}_{\alpha\beta}}^2)   =\frac{1}{160} \int_{\mathbb{R}^5_{\text{phys.}}} d^5x~ \tr ({\mathcal{A}_{\alpha\beta}}^2 +\frac{1}{3} {\mathcal{A}_{\alpha\beta\gamma}}^2), 
\end{align}\label{s5152}
\end{subequations}
$\mathcal{A}_{\alpha}$ is given by  (\ref{compgaugepure}) with $Spin(6)$ matrix generators $\bar{\sigma}_{mn}$. 
From the isomorphism $Spin(6) \simeq SU(4)$, we can express the $Spin(6)$ matrices $\bar{\sigma}_{mn}$ as a linear combination  of the $SU(4)$ Gell-Mann matrices  $\lambda_{A=1,2,\cdots, 15}$ \cite{Greiner-Muller-book-1981}:    
\be
\bar{\sigma}_{mn} =\frac{1}{2}\sum_{A=1}^{15} \bar{\eta}_{mn}^A \lambda_A. 
\ee
Here we introduced  an $SU(4)$-generalized 't Hooft symbol, $\bar{\eta}_{mn}^A =\text{tr}(\lambda_A\bar{\sigma}_{mn})$ (see Appendix \ref{appendix:genethooft} for detail properties). 
The product of two $Spin(6)$ generators is explicitly given by\footnote{The $SU(4)$  Gell-Mann matrices  \cite{Greiner-Muller-book-1981} are ortho-normalized as $\text{tr}(\lambda_A\lambda_B)=2\delta_{AB}$, and   with the $4\times 4$ unit matrix they constitute the $U(4)$ matrix generators that  span  the whole  $4\times 4$ matrix space. }   
\be
\bar{\sigma}_{mn}\bar{\sigma}_{pq} =\frac{1}{4}(\delta_{mp}\delta_{nq}-\delta_{mq}\delta_{nq})1_4+i\frac{1}{2}(\delta_{mp}\bar{\sigma}_{nq}-\delta_{mq}\bar{\sigma}_{np}+\delta_{nq}\bar{\sigma}_{mp}-\delta_{np}\bar{\sigma}_{mq}) -\frac{1}{4}\epsilon_{mnpqrs}\bar{\sigma}_{rs}. \label{so6matcomprel}
\ee
From this formula,  the pure tensor gauge fields  can be expressed as   
\begin{align}
&\mathcal{A}_{\alpha\beta} = -\bar{\sigma}_{mn}\partial_{\alpha}n_{[n}\partial_{\beta}n_{m]}   ,\nn\\
&\mathcal{A}_{\alpha\beta\gamma} 
=-\frac{1}{3}(\mathcal{A}_{\alpha\beta}\mathcal{A}_{\gamma}+\mathcal{A}_{\beta\gamma}\mathcal{A}_{\alpha} +\mathcal{A}_{\rho\mu}A_{\nu}) 
=\epsilon_{mnpqrs}\partial_{\alpha}n_m\partial_{\beta}n_n\partial_{\gamma}n_p n_q \bar{\sigma}_{rs}, \nn\\
&\mathcal{A}_{\alpha\beta\gamma\delta}  
=\frac{1}{3!}(\{\mathcal{A}_{\alpha\beta}, \mathcal{A}_{\gamma\delta}\} - \{\mathcal{A}_{\alpha\gamma}, \mathcal{A}_{\beta\delta}\}+\{\mathcal{A}_{\alpha\delta}, \mathcal{A}_{\beta\gamma}\}) 
=-\epsilon_{mnpqrs}\bar{\sigma}_{rs}\partial_{\alpha}n_m\partial_{\beta}n_n\partial_{\gamma}n_p\partial_{\delta}n_q, \label{tenspugans}
\end{align}
where we used 
\begin{align}
&\mathcal{A}_{\alpha\beta}\mathcal{A}_{\gamma} =2in_p \bar{\sigma}_{mp}\partial_{\gamma}n_n(\partial_{\alpha}n_m\partial_{\beta}n_n-\partial_{\beta}n_m\partial_{\alpha}n_n) -\epsilon_{mnpqrs}\partial_{\alpha}n_m\partial_{\beta}n_n\partial_{\gamma}n_q n_p\bar{\sigma}_{rs}, \nn\\
&\{\mathcal{A}_{\alpha\beta}, \mathcal{A}_{\gamma\delta}\} = 2(\partial_{\alpha}n_m\partial_{\gamma}n_m\cdot \partial_{\beta}n_n\partial_{\delta}n_n -\partial_{\alpha}n_m\partial_{\delta}n_m\cdot \partial_{\beta}n_n\partial_{\gamma}n_n)1_4-2\epsilon_{mnpqrs}\partial_{\alpha}n_m \partial_{\beta}n_n \partial_{\rho}n_p \partial_{\sigma}n_r \bar{\sigma}_{rs}. 
\end{align}
Substituting (\ref{tenspugans}) into (\ref{s5152}), we obtain the  $O(6)$ S-NLS model Hamiltonians: 
\begin{subequations}
\begin{align}
H_{5,1}& 
=\frac{1}{10} \int_{\mathbb{R}^5_{\text{phys.}}} d^5x~ \biggl((\partial_{\alpha}n_m)^2+\frac{1}{ (4!)^2} (\partial_{\alpha}n_{[m}\partial_{\beta}n_n\partial_{\gamma}n_p\partial_{\delta}n_{q]}  )^2\biggr) ,  \label{h51o6ham} \\ 
H_{5, 2}& 
=\frac{1}{80} \int_{\mathbb{R}^5_{\text{phys.}}} d^5x~ \biggl((\partial_{\alpha}n_{[m} \partial_{\beta}n_{n]})^2 +\frac{1}{9}~ (\partial_{\alpha}n_{[m}\partial_{\beta}n_n \partial_{\gamma}n_{p]})^2\biggr) . \label{h52o6ham}
\end{align}\label{h52o6hamtot}
\end{subequations}
The octic derivative term of $H_{5,1}$  is similarly given by (\ref{delnquadra})  
and  the sextic derivative term of $H_{5,2}$ is 
\be
(\partial_{\alpha}n_{[m} \partial_{\beta}n_{n}\partial_{\gamma}n_{p]})^2 = 6((\partial_{\alpha}n_m)^2)^3    -18 (\partial_{\alpha}n_m)^2 (\partial_{\beta}n_n\partial_{\gamma}n_p)^2  +12 (\partial_{\alpha}n_m\partial_{\beta}n_m)(\partial_{\beta}n_n\partial_{\gamma}n_n)(\partial_{\gamma}n_p\partial_{\alpha}n_p) . 
\ee
$H_{5,1}$ and $H_{5,2}$ respectively correspond to the Type I and Type II Skyrme Hamiltonians on $S^5$  \cite{Amari-Ferreira-2018-2}.  

The mathematical structure of the $O(6)$ S-NLS model Hamiltonians is quite similar to that of the Skyrme's $O(4)$ Hamiltonian (\ref{o4skyrhamex31}).  Each partial derivative acts to every component of the NLS field and  all of the Latin indices of the components  are totally antisymmetrized to build the constituent terms of the  Hamiltonian. Recall that the $O(2k+1)$ S-NLS model Hamiltonians exhibited the  similar  structures. 
Such common structures between $O(2k+1)$ and $O(2k)$ S-NLS models 
suggest an  existence of a unified formulation that covers all of the S-NLS models. We shall explore the formulation in Sec.\ref{sec:od+1nls}. 

As a final comment of this section, we mention about relationship to  the formerly derived 7(+1)D  Skyrmion  model 
 by the Atiyah-Manton construction \cite{Nakamula-Sasaki-Takesue-2017}. For $k=4$ and $p=3$, (\ref{2k-1nlsmfiehamil}) yields an $O(8)$ S-NLS Hamiltonian: 
\be
H_{7,3} =\frac{1}{3360}\int_{\mathbb{R}^7}d^7x ~\text{tr}(\mathcal{A}_{\alpha\beta\gamma}^2 +\tilde{\mathcal{A}}_{\alpha\beta\gamma}^2)
=\frac{1}{3360}\int_{\mathbb{R}^7}d^7x ~\text{tr}(\mathcal{A}_{\alpha\beta\gamma}^2 +\frac{1}{4}{\mathcal{A}}_{\alpha\beta\gamma\delta}^2). 
\label{h73hamsky}
\ee
Interestingly, (\ref{h73hamsky}) takes the same form as the 7D Skyrme Hamiltonian obtained in \cite{Nakamula-Sasaki-Takesue-2017}. Although   detail relations between  
the  present  and  Atiyah-Manton constructions need to be excavated, both of them are based on the hierarchical construction from instantons and practically apply the replacement (\ref{replacementftoamathcal}) to the gauge theory actions to  yield  same Skyrme Hamiltonians.  

\section{$O(d+1)$ S-NLS Models}\label{sec:od+1nls}

We discuss a general construction of the S-NLS models from the expression of higher winding number. 
This construction actually reproduces all of the  S-NLS model Hamiltonians previously derived  
and also supplements other S-NLS model Hamiltonians of the type $H_{2k, \text{odd}}$ in even D that eluded the previous discussions based on the tensor gauge theories.

\subsection{$O(d+1)$ S-NLS models and their basic properties}  

\subsubsection{General $O(d+1)$ S-NLS model Hamiltonians}

The winding number of the $O(d+1)$ NLS model associated with 
\be
\pi_d(S^d) \simeq \mathbb{Z} 
\ee
is given by \cite{Patani-Schlindwein-Shafi-1976}
\begin{align}
N_d &=\frac{1}{{A}(S^d_{\text{phys.}})}
~\frac{1}{d!}\int_{\mathbb{R}^d_{\text{phys.}}} d^d x~\epsilon_{a_1 a_2 \cdots a_{d+1}}  \epsilon_{\mu_1 \mu_2 \cdots \mu_{d}} n_{a_{d+1}}\partial_{\mu_1}n_{a_1} \partial_{\mu_2}n_{a_2}  \cdots \partial_{\mu_d}n_{a_d} \nn\\
&=\frac{1}{{A}(S^d_{\text{phys.}})}~\int_{\mathbb{R}^d_{\text{phys.}}} d^d x~\epsilon_{a_1 a_2 \cdots a_{d+1}}  n_{a_{d+1}}\partial_{1}n_{a_1} \partial_{2}n_{a_2}  \cdots \partial_{d}n_{a_d}, \label{ndnnns}
\end{align}
where $n_a(x)$ denote the $O(d+1)$ NLS model field on $x_{\mu} \in \mathbb{R}^{d}$ subject to  
\be
\sum_{a=1}^{d+1}n_an_a =1~:~S^{d}. 
\ee
As in the previous cases, we first decompose the winding number (\ref{ndnnns}) as   
\be
N_d = \frac{1}{\mathcal{A}({S^d})}\frac{p!(d-p)!}{d!}\int_{\mathbb{R}^d}d^dx~N^{a_1 a_2\cdots a_{p}}_{\mu_1\mu_2\cdots\mu_p}~\tilde{N}^{a_1 a_2\cdots a_{p}}_{\mu_1\mu_2\cdots\mu_p}
\ee
where 
\begin{subequations}
\begin{align}
N^{a_1 a_2\cdots a_{p}}_{\mu_1\mu_2\cdots\mu_p}&\equiv \frac{1}{p!} \partial_{\mu_1}n_{[a_1} \partial_{\mu_2}n_{a_2}\cdots \partial_{\mu_p}n_{a_p]}, \label{nass}\\
\tilde{N}^{a_1 a_2\cdots a_{p}}_{\mu_1\mu_2\cdots\mu_p}&\equiv \frac{1}{p!(d-p)!}\epsilon_{\mu_1\mu_2\cdots\mu_d}\epsilon_{a_1 a_2\cdots a_{d+1}} n_{a_{d+1}} N_{\mu_{p+1}\mu_{p+2}\cdots\mu_{d}}^{a_{p+1}a_{p+2}\cdots a_d}\nn\\
&=\frac{1}{p!(d-p)!}\epsilon_{\mu_1\mu_2\cdots\mu_d}\epsilon_{a_1 a_2\cdots a_{d+1}} n_{a_{d+1}} \partial_{\mu_{p+1}}n_{a_{p+1}} \partial_{\mu_{p+2}}n_{a_{p+2}}\cdots \partial_{\mu_{d}}n_{a_{d}}.   \label{tilnass}
\end{align}
\end{subequations}
The BPS inequality,  
$(N^{a_1 a_2\cdots a_{p}}_{\mu_1\mu_2\cdots\mu_p}-\tilde{N}^{a_1 a_2\cdots a_{p}}_{\mu_1\mu_2\cdots\mu_p})^2 \ge 0$,  
or 
\be
H_{d,p} \ge {A}(S^d)\cdot N_d,    
\ee
yields  the $O(d+1)$ S-NLS model Hamiltonian:  
\be
H_{d,p} =H_{d,p}^{(1)} + H_{d,p}^{(2)} \label{hdpgeneralham}
\ee
with 
\begin{subequations}
\begin{align}
&H_{d,p}^{(1)}= \frac{(d-p)!}{2 ~d! p!}\int_{\mathbb{R}_{\text{phys.}}^d}d^dx~ ~(\partial_{\mu_1}n_{[a_1}  \partial_{\mu_2}n_{a_2}  \cdots \partial_{\mu_p}n_{a_p]}  )^2 , \label{h1dp}\\
&H_{d,p}^{(2)}=\frac{p!}{2~d!(d-p)!}\int_{\mathbb{R}_{\text{phys.}}^d}d^dx ~(\partial_{\mu_{1}}n_{[a_{1}}  \partial_{\mu_{2}}n_{a_{2}}  \cdots \partial_{\mu_{d-p}}n_{a_{d-p}]}  )^2.  
\end{align} \label{hdpstwo}
\end{subequations}
The BPS equation,  
${N_{\mu_1\mu_2\cdots\mu_{p}}^{a_1 a_2\cdots a_p}}={\tilde{N}_{\mu_1\mu_2\cdots\mu_{p}}^{a_1 a_2\cdots a_p}}$, is rephrased as  
\be
 \partial_{\mu_1}n_{[a_1} \partial_{\mu_2}n_{a_2}\cdots \partial_{\mu_p}n_{a_p]}=\frac{1}{(d-p)!}\epsilon_{\mu_1\mu_2\cdots\mu_d}\epsilon_{a_1 a_2\cdots a_{d+1}} n_{a_{d+1}} \partial_{\mu_{p+1}}n_{a_{p+1}} \partial_{\mu_{p+2}}n_{a_{p+2}}\cdots \partial_{\mu_{d}}n_{a_{d}}. \label{genebpsforns}
\ee
Notice that the $O(d+1)$ Hamiltonian is invariant under the interchange $p~\leftrightarrow~d-p$: 
\be
H_{d,p} =H_{d, d-p}. 
\ee
Therefore, there are $[d/2]$ distinct Hamiltonians in correspondence with $p=1,2,\cdots [d/2]$. 
One may readily check that (\ref{hdpgeneralham}) reproduces  the $O(2k+1)$ S-NLS model Hamiltonians, (\ref{hamgene1sto2k+1}) and (\ref{k+ko2k+1NLsham}), 
and also the $O(2k)$ S-NLS model Hamiltonians,  (\ref{o4skyrhamex31}) and (\ref{h52o6hamtot}). Not only do $H_{d,p}$ cover all of the previously derived S-NLS model Hamiltonians, but $H_{d,p}$ also provide other S-NLS model Hamiltonians that eluded the previous derivations.   
In low dimensions, from (\ref{hdpgeneralham})  such S-NLS model Hamiltonians 
are obtained as  
\begin{align}
&H_{2,1}=\frac{1}{2}\int_{\mathbb{R}^2_{\text{phys.}}} d^2x ~(\partial_{\mu}n_a)^2, \nn\\
&H_{4,1} = \frac{1}{8}  \int_{\mathbb{R}^4_{\text{phys.}}}d^4 x~ \biggl(   (\partial_{\mu}n_a)^2 +\frac{1}{36}  (\partial_{\mu}n_{[b}  \partial_{\nu}n_{c}  \partial_{\rho}n_{d]}  )^2  \biggr), \nn\\
&H_{6,1} 
=\frac{1}{12}  \int_{\mathbb{R}^6_{\text{phys.}}}d^6 x~ \biggl(  (\partial_{\mu}n_a)^2 +\frac{1}{14400}  (\partial_{\mu_1}n_{[a_1}  \partial_{\mu_2}n_{a_2}  \cdots \partial_{\mu_5}n_{a_5]}  )^2  \biggr), \nn\\
&H_{6,3} 
=\frac{1}{720} \int_{\mathbb{R}^6_{\text{phys.}}}d^6 x~ (\partial_{\mu_1}n_{[a_1}  \partial_{\mu_2}n_{a_2}  \cdots \partial_{\mu_6}n_{a_6]}  )^2. 
\end{align}
Note that $H_{2,1}$ represents the well known $O(3)$ NLS model  Hamiltonian.   

It is not difficult also to incorporate 
non-derivative term ($p=0$) in the present formalism.  With 
\be
H_{d,p=0}^{(1)} =\frac{1}{2}\int_{\mathbb{R}_{\text{phys.}}^d}d^dx ~U(n) , ~~~~~~H_{d,p}^{(2)} =\frac{1}{2~{d!}^2}\int_{\mathbb{R}_{\text{phys.}}^d}d^dx ~(\partial_{\mu_{1}}n_{[a_{1}}  \partial_{\mu_{2}}n_{a_{2}}  \cdots \partial_{\mu_{d}}n_{a_{d}]}  )^2,  
\ee 
we have 
\be
H_{d,p=0}= \frac{1}{2}\int_{\mathbb{R}_{\text{phys.}}^d}d^dx \biggl( \frac{1}{{d!}^2} ~(\partial_{\mu_{1}}n_{[a_{1}}  \partial_{\mu_{2}}n_{a_{2}}  \cdots \partial_{\mu_{d}}n_{a_{d}]}  )^2 +U(n)\biggr),    \label{hamp0}
\ee
which is exactly equal to the Hamiltonian  introduced in \cite{Chakrabarti-Piette-Tchrakian-Zakrzewski-1992}.  
The BPS inequality is given by 
\be
H_{d,p=0} ~\ge~\frac{1}{d!}\int_{\mathbb{R}_{\text{phys.}}^d}d^dx ~\epsilon_{\mu_1\mu_2\cdots\mu_d} \epsilon^{a_1 a_2 \cdots a_{d+1}}   \sqrt{U(n)}~n_{a_{d+1}}\partial_{\mu_1}n_{a_1}\partial_{\mu_2}n_{a_2}\cdots \partial_{\mu_d}n_{a_d}.  \label{bpsp0} 
\ee
For $U=1$, the right-hand side of (\ref{bpsp0}) is reduced to the usual topological number 
$A(S^d) \cdot N_d$.   
(\ref{hamp0}) realizes the restricted baby Skyrme model for $d=2$ \cite{Gisiger-Paranjape-1996,Gisiger-Paranjape-1997, Adam-Romanczukiewicz-SanchezGuillen-Wereszczynski-2010} and the BPS Skyrme model for $d=3$ \cite{Adam-SanchezGuillen-Wereszczynski-2010-1, Adam-SanchezGuillen-Wereszczynski-2010-2}.\footnote{Recently, BPS Skyrme models attracted much attention for the reason that  they can  lower the binding energy compared to the original Skyrme model \cite{ Adam-Naya-SanchezGuillen-Wereszczynski-2013,Harland-2014,Gillard-Harland-Speight-2015,Gudnason-Zhang-Ma-2016}.} 

\subsubsection{Equations of motion and the scaling arguments}

From (\ref{hdpgeneralham}), it is not difficult to derive the equations of motion: 
\begin{align}
&\partial_{\mu_1}\biggl( (d-p-1)! ~ \partial_{\mu_2}n_{a_2}\partial_{\mu_3}n_{a_3}\cdots \partial_{\mu_p}n_{a_p} \cdot \partial_{\mu_1}n_{[a_1}\partial_{\mu_2}n_{a_2}\partial_{\mu_3}n_{a_3}\cdots \partial_{\mu_p}n_{a_p]}  \nn\\
&~~~+ (p-1)!~\partial_{\mu_2}n_{a_2}\partial_{\mu_3}n_{a_3}\cdots \partial_{\mu_{d-p}}n_{a_{d-p}} \cdot \partial_{\mu_1}n_{[a_1}\partial_{\mu_2}n_{a_2}\partial_{\mu_3}n_{a_3}\cdots \partial_{\mu_{d-p}}n_{a_{d-p}]} \biggr)-\lambda n_{a_1}=0, \label{geneeomd+1}
\end{align}
where $\lambda$ denotes the Lagrange multiplier  
\begin{align}
\lambda=n_{a_1}\partial_{\mu_1}&\biggl( (d-p-1)! ~ \partial_{\mu_2}n_{a_2}\partial_{\mu_3}n_{a_3}\cdots \partial_{\mu_p}n_{a_p} \cdot \partial_{\mu_1}n_{[a_1}\partial_{\mu_2}n_{a_2}\partial_{\mu_3}n_{a_3}\cdots \partial_{\mu_p}n_{a_p]}  \nn\\
&+ (p-1)!~\partial_{\mu_2}n_{a_2}\partial_{\mu_3}n_{a_3}\cdots \partial_{\mu_{d-p}}n_{a_{d-p}} \cdot \partial_{\mu_1}n_{[a_1}\partial_{\mu_2}n_{a_2}\partial_{\mu_3}n_{a_3}\cdots \partial_{\mu_{d-p}}n_{a_{d-p}]} \biggr).  
\end{align}
For $(d,p)=(2k,2l)$, (\ref{geneeomd+1}) signifies the equations of motion of the $O(2k+1)$ S-NLS model Hamiltonian (\ref{fromstoh2k+1}). 
In particular for $(d, p)=(2k, 2)$, (\ref{geneeomd+1}) becomes   
\be
\partial_{\mu_1} \biggl(\partial_{\mu_2} n_{b}   \cdot \partial_{\mu_1}{n}_{[a}  \partial_{\mu_2}{n}_{b]}  
+\frac{1}{(2k-3)!}~
\partial_{\mu_2}{n}_{a_2}\cdots \partial_{\mu_{2k-2}}{n}_{a_{2k-2}} ~\cdot ~\partial_{\mu_1}{n}_{[a} \partial_{\mu_2}n_{a_2}\cdots \partial_{\mu_{2k-2}}{n}_{{a_{2k-2}}]}   \biggr)-\frac{1}{(2k-3)!}\lambda n_a=0, \label{eomso2k+1}
\ee 
which represents the equations of motion of (\ref{hamgene1sto2k+1}). 
In low dimensions, (\ref{geneeomd+1}) gives 
\begin{align}
&(d,p)=(2,1)~:~2{\partial_{\mu}}^2n_a-\lambda n_a=0, \nn\\
&(d,p)=(3,1)~:~1!~{\partial_{\mu}}^2n_a +\partial_{\mu}(\partial_{\nu}n_b \partial_{\mu}n_{[a}\partial_{\nu}n_{b]})-\lambda n_a=0, \nn\\
&(d,p)=(4,1)~:~2!~{\partial_{\mu}}^2n_a +\partial_{\mu}(\partial_{\nu}n_n \partial_{\rho}n_p\partial_{\mu}n_{[a}\partial_{\nu}n_{b}\partial_{\rho}n_{c]})-\lambda n_a=0, \nn\\
&(d,p)=(4,2)~:~2\partial_{\mu}(\partial_{\nu}n_b\partial_{\mu}n_{[a}\partial_{\nu}n_{b]})-\lambda n_{a}=0, \nn\\ 
&(d,p)=(5,1)~:~3!~{\partial_{\mu}}^2n_a +\partial_{\mu}(\partial_{\nu}n_b \partial_{\rho}n_c\partial_{\sigma}n_d\partial_{\mu}n_{[a}\partial_{\nu}n_{b}\partial_{\rho}n_{c}\partial_{\sigma}n_{d]})-\lambda n_a=0, \nn\\
&(d,p)=(5,2)~:~2\partial_{\mu}(\partial_{\nu}n_b \partial_{\mu}n_{[a}\partial_{\nu}n_{b]})+\partial_{\mu}(\partial_{\nu}n_b\partial_{\rho}n_c \partial_{\mu}n_{[a} \partial_{\nu}n_{b}\partial_{\rho}n_{c]} ) -\lambda n_a=0. 
\label{slippedexamples}
\end{align}
The equations of motion of the $O(3)$ NLS model and the $O(5)$ S-NLS model are realized for $(d,p)=(2,1)$ and $(4,2)$ in (\ref{slippedexamples}), respectively.  For the $O(3)$ NLS model,  soliton solutions with arbitrary winding number are derived in \cite{Belavin-Polyakov-1975, Faddeev-1974}, but for other S-NLS models,  to solve the equations of motion (\ref{geneeomd+1}) is rather formidable in general.  

 Instead of solving the equations of motion, we prepare one(-scale)-parameter family of  field configurations and  evaluate the size of the  configuration based on  the scaling argument of Derrick \cite{Derrick-1964}.    
The mass dimensions of the quantities inside the integrals of  $H_{d,p}^{(1)}$ and $H_{d,p}^{(2)}$ (\ref{hdpstwo}) are  $2p-d$ and $d-2p$, respectively.\footnote{Both $H_{d,p}^{(1)}$ and $H_{d,p}^{(2)}$ should have mass dimension one, and so, to be precise,  some dimensionful parameters are necessary in front of them to adjust the dimension counting.}  Suppose that the energy of a given field  configuration $n_a(x)$ is given by  $E_{d,p}=E_{d,p}^{(1)}+E_{d,p}^{(2)}$. Under the scale transformation  
\be
n_a(x)~\rightarrow~n^{(R)}_a(x)\equiv n_a(x/{R} ), 
\ee
$E_{d,p}^{(1)}$ and $E_{d,p}^{(2)}$ are  transformed as 
\be
E_{d,p}=E_{d,p}^{(1)}+E_{d,p}^{(2)} ~\rightarrow~E_{d,p}(R) =E_{d,p}^{(1)}(R)+E_{d,p}^{(2)}(R),   
\label{scalinglaw}
\ee
where 
\be
E_{d,p}^{(1)}(R) = {R^{d-2p}}E_{d,p}^{(1)}, ~~~E_{d,p}^{(2)}(R)=\frac{1}{R^{d-2p}}E_{d,p}^{(2)}. 
\ee
The scale parameter $R$ can be considered as a variational parameter that represents the size of the field configuration. 
For $p > [d/2]$, as $R$ increases, $E_{d,p}^{(1)}(R)$ monotonically increases while  $E_{d,p}^{(2)}(R)$ monotonically decreases. 
This implies that $E_{d,p}^{(1)}(R)$ term energetically favors a smaller  size field configuration while   $E_{d,p}^{(2)}(R)$ favors a larger size configuration.  
These two competing effects determine an optimal size of the field configuration of soliton. 
More specifically,  we  
take the derivative of $E_{d,p}(R)$ (\ref{scalinglaw}) with respect to $R$ to obtain a local energy minimum and have 
\be
\bar{R}_{d,p}=\biggl(\frac{E^{(2)}_{d,p}}{E_{d,p}^{(1)}}\biggr)^{\frac{1}{2(d-2p)}}. \label{optimalradius}
\ee
 The present S-NLS models  thus realize soliton configurations with  the finite size  (\ref{optimalradius}) (except for the scale invariant case $P=[d/2]$), and   two competing energies are exactly balanced at the point: 
\be
E_{d,p}^{(1)}(\bar{R}) =(E_{d,p}^{(1)}E_{d,p}^{(2)})^{\frac{1}{2(d-2p)}} = E_{d,p}^{(2)}(\bar{R}), 
\ee
which signifies the virial relation in higher dimensions.  

\subsubsection{Scale invariant solutions}\label{subsec:eofscaling}

Next let us consider $(d, p)=(2k, k)$,  in which the two competing Hamiltonians coincide,  $H_{2k,k}^{(1)}=H_{2k,k}^{(2)}$,  to realize scale invariant field solutions.\footnote{Other types of scale invariant solitons associated with the Hopf map are proposed in \cite{Deser-Duff-Isham-1976,Nicole-1978} and \cite{Aratyn-Ferreira-Zimerman-1999}.}  The S-NLS model Hamiltonian (\ref{hdpgeneralham}) becomes 
\be
H_{2k,k}= \frac{1}{(2k)!}\int_{\mathbb{R}^d}d^dx~ (\partial_{\mu_1}n_{[a_1}  \partial_{\mu_2}n_{a_2}  \cdots \partial_{\mu_p}n_{a_k]}  )^2. \label{scaleinvhamaorbitk} 
\ee
When $k$ is even,  (\ref{scaleinvhamaorbitk}) is exactly equal to the former scale invariant Hamiltonian  (\ref{k+ko2k+1NLsham}). 
The equations of motion 
 (\ref{geneeomd+1}) and the BPS equation (\ref{genebpsforns}) are reduced to     
\begin{subequations}
\begin{align}
&\partial_{\mu_1}\biggl((\partial_{\mu_2}n_{a_2} \cdots \partial_{\mu_p}n_{a_k})\cdot (\partial_{\mu_1}n_{[a_1}\partial_{\mu_2}n_{a_2} \cdots \partial_{\mu_k}n_{a_k]})\biggr)-\frac{1}{2(k-1)!}\lambda n_{a_1}=0, \label{eomselfdual} \\
&\partial_{\mu_1}n_{[a_1} \partial_{\mu_2}n_{a_2}\cdots \partial_{\mu_k}n_{a_{k}]}-\frac{1}{k!}\epsilon_{\mu_1\mu_2\cdots\mu_{2k}}\epsilon_{a_1 a_2\cdots a_{2k+1}} n_{a_{2k+1}} \partial_{\mu_{k+1}}n_{a_{k+1}} \partial_{\mu_{k+2}}n_{a_{k+2}}\cdots \partial_{\mu_{2k}}n_{a_{2k}}=0. \label{genebpsforns2k}
\end{align}
\end{subequations}
Especially for $d=4$, 
 (\ref{eomselfdual}) reproduces the  $(d,p)=(4,2)$ equation of (\ref{slippedexamples}). 
The equations of motion (\ref{eomselfdual}) are highly non-linear equations, but the  inverse stereographic coordinate configuration
\be
n_a(x)=r_a\equiv \{\frac{2}{1+x^2}x_{\mu}, \frac{1-x^2}{1+x^2}\},    \label{solscaleinv}
\ee
realizes a simple solution of (\ref{eomselfdual}) and  satisfies the BPS equation 
(\ref{genebpsforns2k}) also.\footnote{Also recall the results of  Sec.\ref{subsec:k+kdecomp} where  the tensor instanton configuration  satisfies  the BPS equation and the equations of motion. }    
From the one-to-one correspondence between the points on  $\mathbb{R}^{2k}$ and those on  $S^{2k}$,   
it may be obvious that (\ref{solscaleinv}) also  represents a field configuration of the winding number 1. One can explicitly confirm this as  
\be
N_d|_{n_a=r_a}=\frac{1}{A(S^d)}\int_{\mathbb{R}_{\text{phys.}}^d} d^dx~\epsilon_{a_1a_2\cdots a_{d+1}}r_{a_{d+1}}\partial_1 r_{a_1}\partial_2 r_{a_2}\cdots \partial_{d}r_{a_d} 
=\frac{A(S^{d-1})}{A(S^d)}\int_0^{\infty} dx~ x^{d-1}  \frac{2^d}{(1+x^2)^d}=1. 
\ee
The energy density for (\ref{solscaleinv}) is also  evaluated as  
\be
\frac{1}{d!}  (\partial_{\mu_1}n_{[a_1}  \partial_{\mu_2}n_{a_2}  \cdots \partial_{\mu_p}n_{a_{d/2}]}  )^2\biggr|_{n_a=r_a} =\frac{2^d}{(1+x^2)^d}, 
\ee
which implies that (\ref{solscaleinv}) signifies a solitonic field configuration localized around the origin.

\subsection{Topological field configurations }\label{subsec:topfield}

Recall that the $k$th Chern number has two equivalent expressions, $N_{2k-1}$ and $N_{2k}$ (Sec.\ref{subsec:difftoplandau2k+1}). This equivalence may imply intimate relations between topological field configurations of $O(2k)$ and $O(2k+1)$ S-NLS models with  same winding number. In this Section, we utilize the idea of  dimensional hierarchy to construct  topological field configurations with higher winding numbers. 

\subsubsection{Topological field configurations in odd D }\label{sec:winding2k-1}

The transition function $g$ (\ref{so2kgroupele})\footnote{(\ref{nonlinearsigma}) is a non-linear realization of $SO(2k)$ matrix with broken generators $\sigma_{i,2k} =2\gamma_i$ for the coset $S^{2k-1}\simeq SO(2k)/SO(2k-1)$  \cite{Salam-Strathdee-1982}. } 
\be
g=e^{i\theta \sum_{i=1}^{2k-1}\gamma_{i} \hat{r}_{i}} =\sum_{\mu=1}^{2k}r_{\mu}\bar{g}_{\mu}
~~~~~~(\sum_{i=1}^{2k-1} \hat{r}_{i} \hat{r}_i=\sum_{\mu=1}^{2k} r_{\mu} {r}_{\mu}=1)
\label{nonlinearsigma}
\ee
represents  $N_{2k-1}=1$  associated with the homotopy $\pi_{2k-1}(S^{2k-1})\simeq \mathbb{Z}$. 
Using (\ref{nonlinearsigma}), we can construct a map from $r_{\mu} ~\in~ {S}_{\text{phys.}}^{2k-1}$ to $n_{\mu}~\in~S^{2k-1}_{\text{field}}$ with arbitrary winding number $N$:   
\be
g^N=e^{i(N\theta) \sum_{i=1}^{2k-1}\gamma_i \hat{r}_{i}}=\sum_{\mu=1}^{2k}n_{\mu}\bar{g}_{\mu}.  
\ee
Here, $n_{\mu}$  is given by 
\be
n_{\mu}=\{n_i, n_{2k}\}\equiv \{\sin(N\theta)~{r}_i,~ \cos(N\theta)\}.  \label{xdashntheta}
\ee
The argument of the trigonometric function in (\ref{xdashntheta}) is  $N\cdot\theta$, meaning that   
 when the azimuthal angle $\theta$ sweeps $S^{2k-1}_{\text{phys.}}$ once,  (\ref{xdashntheta}) sweeps $S^{2k-1}_{\text{field}}$ $N$ times. 
For small $N$, (\ref{xdashntheta}) is  given by 
\begin{align}
&N=1~:~n_{\mu}=\{n_i, n_{2k}\}=\{\sin(\theta)~\hat{r}_i,~ \cos(\theta)\}=r_{\mu}, \nn\\ 
&N=2~:~n_{\mu}=\{n_i, n_{2k}\}=\{\sin(2\theta)~\hat{r}_i,~ \cos(2\theta)\}=\{ 2r_{2k} r_i, -{r_i}^2+{r_{2k}}^2  \}, \nn\\
&N=3~:~n_{\mu}=\{n_i, n_{2k}\}=\{\sin(3\theta)~\hat{r}_i,~ \cos(3\theta)\}=\{ -({r_j}^2-3{r_{2k}}^2) r_i, -(3{r_j}^2-{r_{2k}}^2) r_{2k}  \}.  \label{confign2k-1theta} 
\end{align}
One may notice  that the map associated with the winding number $N$ is given  by the $N$th polynomials of ${r}$s.  
For  (\ref{xdashntheta}), $N_{2k-1}$ (\ref{ndnnns})  is actually evaluated  as 
\be 
N_{2k-1}
=\frac{1}{A(S^{2k-1}_{\text{phys.}})}\int_{S^{2k-1}_{\text{phys.}}} N  ~\sin^{2k-2}(N \theta) ~ d\theta~ d\Omega_{2k-2} =N\frac{1}{A(S^{2k-1})}\int_{S^{2k-1}_{\text{phys.}}}d\Omega_{2k-1} =N, 
\ee
where we used 
\be
\int_{0}^{\pi}d\theta \sin^{{2k}}(N\theta) =\pi\frac{(2k-1)!!}{(2k)!!}=\int_{0}^{\pi}d\theta \sin^{{2k}}(\theta). 
\ee
Regarding $n_{\mu}$ as the $O(2k)$ NLS field, we treat (\ref{xdashntheta}) as  topological field configuration  on $S^{2k-1}_{\text{phys.}}$ with  winding number $N$.  
To construct   topological field configurations on $\mathbb{R}^{2k-1}_{\text{phys.}}$, we apply the stereographic projection in the {physical} space:  
\be
r_{\mu}~\in~S^{2k-1}_{\text{phys.}}~~\longrightarrow~~x_i=\frac{R}{R+{r}_{2k}}{r}_i ~\in~\mathbb{R}^{2k-1}_{\text{phys.}}~~~(i=1,2,\cdots, 2k-1)
\ee
or 
\be
{r}_i =\frac{2R^2}{R^2+x^2}~x_i, ~~~{r}_{2k} =\frac{R^2-{x}^2}{R^2+{x}^2}R. \label{rfromRrx}
\ee
Here, we took the radius of $S_{\text{phys.}}^{2k-1}$ as $R$. Substituting (\ref{rfromRrx}) into the expressions of $n_{\mu}$ such as (\ref{confign2k-1theta}),  
we obtain   one-parameter family of the $O(2k)$ NLS field configurations on $\mathbb{R}^{2k-1}_{\text{phys.}}$: 
\be
n_{\mu}^{(R)} (x_i) =n_{\mu}(x_i/R). 
\ee
For instance,  
\begin{align}
&N=1~:~n^{(R)}_i(x)=  \frac{2R}{x^2+R^2}x_i, ~~ n^{(R)}_{2k}(x)=-\frac{x^2-R^2}{x^2+R^2}   , \nn\\
&N=2~:~n^{(R)}_i(x)= -\frac{4R}{(x^2+R^2)^2}(x^2-R^2)x_i,~~n^{(R)}_{2k}(x)= \frac{1}{(x^2+R^2)^2} (-4R^2x^2+(x^2-R^2)^2), \nn\\
&N=3~:~n^{(R)}_i(x)=  -\frac{2R}{(x^2+R^2)^3} (4R^2 x^2-3(x^2-R^2)^2)x_i, \nn\\
&~~~~~~~~~~~~~ n^{(R)}_{2k}(x)=\frac{1}{(x^2+R^2)^3}(12 R^2x^2 -(x^2-R^2)^2)(x^2-R^2).  \label{confign2k-1y} 
\end{align}
Substituting (\ref{confign2k-1y})  into  (\ref{ndnnns}), one may explicitly confirm that (\ref{confign2k-1y}) represents the topological field configurations of  $N_{2k-1}=1,2,3$. While $R$ originally denotes the radius of sphere,   
$R$ in (\ref{confign2k-1y}) signifies  the size of the soliton configuration. This is intuitively explained as follows. Since the soliton configuration on $\mathbb{R}^{2k-1}$ is related to the  field configuration on $S^{2k-1}$ through the stereographic projection,   as the size of the sphere becomes larger, the ``concentration'' of the soliton field  around the origin   will be thinner, and consequently the size of the soliton becomes larger.  
Treating  $R$ as a variational parameter of $n_{\mu}^{(R)}(x)$, we consider  minimal energy configuration in each topological sector. 
 The previous scaling argument (\ref{optimalradius}) indicates 
\be
R_{2k-1,p}(N)=\biggl(\frac{E_{2k-1,p}^{(2)}(N)}{E_{2k-1,p}^{(1)}(N)}\biggr)^{\frac{1}{2(2k-2p-1)}}, 
\ee
which is the optimal size of the $O(2k)$ NLS field configuration with a given topological number $N$.

\subsubsection{Topological field configurations in even D}
 
\begin{figure}[tbph]
\center
\includegraphics*[width=160mm]{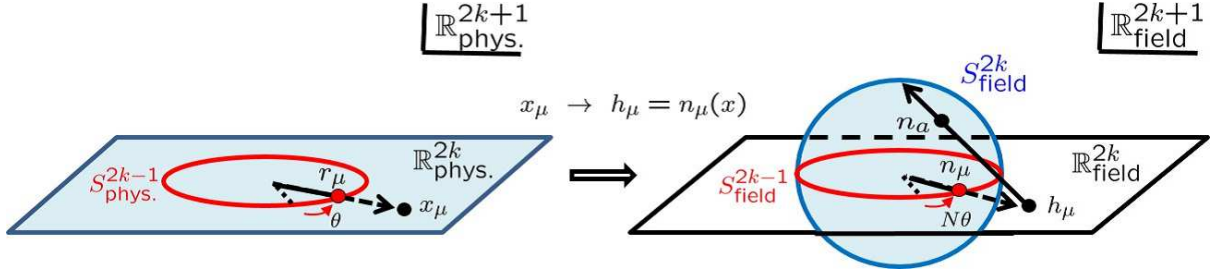}
\caption{The $O(2k+1)$ NLS field with the winding number $\pi_{2k}(S^{2k})\simeq \mathbb{Z}$ is  constructed by the $O(2k)$ NLS field with the   winding number $\pi_{2k-1}(S^{2k-1})\simeq \mathbb{Z}$.  
}
\label{windingno.fig}
\end{figure}

Using the set-up of $(2k-1)$D, we  construct  $O(2k+1)$ topological field configuration  on $\mathbb{R}^{2k}$ for  
\be
\pi_{2k}(S^{2k})~\simeq~\mathbb{Z}. \label{hopfo2k+1sec}
\ee
We add  radial direction to $S^{2k-1}_{\text{phys.}}$ and consider 1D higher space, 
$\mathbb{R}^{2k}_{\text{phys.}}$ (left of Fig.\ref{windingno.fig}). 
The original map from $r_{\mu} ~\in~S^{2k-1}_{\text{phys.}}$ to $n_{\mu} ~\in~S^{2k-1}_{\text{field}}$ is now transformed to (Fig.\ref{windingno.fig})
\be
x_{\mu} ~\in~\mathbb{R}^{2k}_{\text{phys.}} ~~\rightarrow~~h_{\mu}\equiv n_{\mu}(x) ~\in~\mathbb{R}^{2k}_{\text{field}}. \label{inheritwinding}
\ee
The radial direction has no effect about the winding in (\ref{hopfo2k+1sec}), and the winding number associated with the map  (\ref{inheritwinding}) 
can be  accounted for by the winding  from $S^{2k-1}_{\text{phys.}}$ on $\mathbb{R}^{2k}_{\text{phys.}}$ to the $S^{2k-1}_{\text{field}}$ on $\mathbb{R}^{2k}_{\text{field}}$ (Fig.\ref{windingno.fig}),  which is nothing but the previous $(2k-1)$D winding,  $\pi_{2k-1}(S^{2k-1})\simeq \mathbb{Z}$.  
In correspondence with (\ref{confign2k-1theta}),    we have 
\begin{align}
&N=1~:~h_{\mu}=\frac{1}{R}x_{\mu}, \nn\\ 
&N=2~:~h_{\mu}=\{h_i, h_{2k}\}=\frac{1}{R^2}\{ 2x_{2k} x_i, -{x_i}^2+{x_{2k}}^2  \} , \nn\\ 
&N=3~:~h_{\mu}=\{h_i, h_{2k}\}=\frac{1}{R^3}\{ -({x_j}^2-3{x_{2k}}^2) x_i, -(3{x_j}^2-{x_{2k}}^2) x_{2k}  \}. \label{confign2k} 
\end{align}

To realize topological field configurations with field-manifold  $S^{2k}_{\text{field}}$,  we apply the inverse stereographic projection in the field space (right  of Fig.\ref{windingno.fig}): 
\be
h_{\mu}~\in~\mathbb{R}_{\text{field}}^{2k}~~\longrightarrow~~{n}_{\mu} =\frac{2}{1+{h_{\nu}}^2}~h_{\mu}, ~~~~~{n}_{2k+1} =\frac{1-{h_{\nu}}^2}{1+{h_{\nu}}^2}~\in~{S}_{\text{field}}^{2k}.  \label{defhatna}
\ee
Substituting (\ref{confign2k}) into (\ref{defhatna}), we obtain the $O(2k+1)$ topological field configurations on $\mathbb{R}^{2k}_{\text{phys.}}$:  
\begin{align}
&N=1:~n_{\mu}^{(R)}(x)=\frac{2R}{{x_{\nu}}^2+R^2}x_{\mu},~~n_{2k+1}^{(R)}(x)=-\frac{{x_{\nu}}^2-R^2}{{x_{\nu}}^2+R^2}, \nn\\ 
&N=2:~n_{i}^{(R)}(x)=\frac{4R^2}{({x_{\nu}}^2)^2+R^4}x_{2k} x_i,~~n_{2k}^{(R)}(x) =\frac{2R^2}{({x_{\nu}}^2)^2+R^4}(-{x_i}^2+{x_{2k}}^2),~~n^{(R)}_{2k+1}(x) =-\frac{({x_{\nu}}^2)^2-R^4}{({x_{\nu}}^2)^2+R^4},  \nn\\ 
&N=3:~n_{i}^{(R)}(x)= -\frac{2R^3}{({x_{\nu}}^2)^3+R^6}({x_j}^2-3{x_{2k}}^2)x_i,~~~n^{(R)}_{2k} (x)=-\frac{2R^3}{({x_{\nu}}^2)^3+R^6}(3{x_j}^2-{x_{2k}}^2)x_{2k}, \nn\\
&~~~~~~~~~~~~~n^{(R)}_{2k+1}(x)=-\frac{({x_{\nu}}^2)^3-R^6}{({x_{\nu}}^2)^3+R^6}. \label{confign2kxx} 
\end{align}
One can explicitly check  that (\ref{confign2kxx}) describes topological  field configurations of  $N_{2k}=1,2,3$ with  (\ref{ndnnns}).    
The scaling argument  (\ref{optimalradius})
 determines the parameter $R$ as 
\be
R_{2k,p}(N)=\biggl(\frac{E_{2k,p}^{(2)}(N)}{E_{2k,p}^{(1)}(N)}\biggr)^{\frac{1}{4(k-p)}}. 
\ee

Here, we add some comments about the scale invariant case. 
For the $O(3)$ NLS model being scale invariant,  soliton solutions with arbitrary topological numbers are given by the holomorphic functions  on $\mathbb{C}\simeq \mathbb{R}^2$ \cite{Belavin-Polyakov-1975, Faddeev-1974}, and the power of the complex coordinates  indicates  the winding  number \cite{Duff-Isham-1977, Belavin-Polyakov-1975}. Meanwhile for the scale invariant $O(5)$ S-NLS model ($H_{4,2}$), though the topological field  configuration is simply obtained by the multiple of quaternionic analytic function \cite{Gursey-Jafarizadeh-Tze-1979,Jafarizadeh-Snyder-Tze-1980},  soliton solutions are not easily derived except for $N=1$. Similarly as demonstrated in Sec.\ref{subsec:eofscaling}, the $O(2k+1)$ topological field configuration (\ref{confign2kxx}) with $N=1$ realizes a scale invariant solution of the equations of motion (\ref{eomselfdual}),   but other configurations of higher winding number ((\ref{confign2kxx}) with $N \ge 2$) do not realize scale invariant solutions.

\section{Summary}\label{sec:summdisc}

We performed a systematic construction of S-NLS models in arbitrary dimensions based on  the Landau/NLS model correspondence.  
Exploiting the differential geometry of the Landau models, we introduced  the  $[k/2]$ distinct parent tensor gauge theories on the field-manifold $S^{2k}$ and subsequently derived  the $[k/2]$   $O(2k+1)$ S-NLS models on  $\mathbb{R}_{\text{phys.}}^{2k}$.  
 The $SO(2k)$  gauge symmetry and the BPS inequality of the parent tensor gauge theories are necessarily inherited to the  obtained $O(2k+1)$ S-NLS models. 
As a dimensional reduction from $2k$D to $(2k-1)$D, 
 we adopted the Chern-Simons term description of the Chern number. Representing the  transition function by $O(2k)$ NLS field, we  derived the $O(2k)$ S-NLS model Hamiltonians from pure tensor gauge fields, which indeed realize the original 3D Skyrme model, and formerly derived 5D and 7D Skyrme models   as  the special cases.  Since the parent field theories do not have gauge symmetries,   the obtained $O(2k)$ S-NLS models do not possess  gauge symmetries, either. 
Further, the dimensional reduction implies that there always exists one-dimension higher tensor gauge field theory behind every  odd D Skyrme model.  
 From  the NLS field expression of the higher winding number, we explored a unified $O(d+1)$ formulation of the S-NLS models.   Among the $O(d+1)$ S-NLS model Hamiltonians, $H_{d=2k,p=2l}$ $(l=1,2,\cdots, [k/2])$ are identical to  the $O(2k+1)$ S-NLS Hamiltonians derived from the tensor gauge actions and enjoy the hidden $O(2k)$ gauge symmetry. (As emphasized in the main text, this should not be confused with the  hidden local symmetry.)    
We derived  the equations of motion  and   constructed a scale invariant solution with unit  winding number. 
Topological field configurations  with arbitrary winding number are also constructed  by exploiting the idea of the dimensional hierarchy. 
The  topological field configurations depend on 
the variational scaling parameter, which is determined by the scaling arguments.  
A particular feature of the present model is that 
the decomposition of the topological number necessarily yields the Hamiltonian of two competing scale terms  and their competing   results in  a  finite size soliton configuration.
  
 Analytic derivation of  explicit solutions is not easy even for the original Skyrme model.\footnote{For $O(6)$ S-NLS models,   explicit solutions  were recently derived in toroidal coordinates \cite{Amari-Ferreira-2018-2}, and  an $O(8)$ S-NLS model was also numerically analyzed in \cite{Nakamula-Sasaki-Takesue-2017}.}  
 Similarly, though we obtained the equations of motion of the higher dimensional S-NLS models,  their explicit solutions  have not been generally derived.   
  One apparent  direction is to investigate the soliton solutions  by  numerical methods.    
 Another direction will be a generalization of the S-NLS models based on other symmetries.  
 While in this work we were focused on the $O(N)$ S-NLS models that are closely related to the $SO(N)$ Landau models,  
 many Landau models with different symmetries, including  supersymmetry \cite{Ivanov-Mezincescu-Townsend-2004, Hasebe-Kimura-2004}, have been constructed in the developments of the higher dimensional quantum Hall effect.   
The topological table also accommodates  various cousins of the Landau models with  different symmetries \cite{RyuSFL2009}. It is  tempting to derive other  NLS models that originate from such various Landau models.    
 It should also be emphasized that the Skyrmions    
have played crucial roles  in the non-perturbative analysis  of strongly correlated systems, such as QCD, 2D quantum Hall system.  
 As the S-NLS model solitons emerge as collective excitations in  the higher dimensional quantum Hall effect,  their roles  will be indispensable in  understanding   topological phases in higher dimensions. 

\section*{Acknowledgments}

I am glad to thank Yuki Amari for useful email correspondences and a lecture  about the recent developments of NLS models. I am also grateful to   Shin Sasaki and Atsushi Nakamula  for fruitful discussions and arranging a seminar at  Kitasato University.    This work was supported by JSPS KAKENHI Grant Numbers~16K05334 and 16K05138.

\appendix

\section{Stereographic projection and $SO(2k)$ instanton configurations}\label{append:stereo}

Here, we review the stereographic projection from $S^{2k}$ to $\mathbb{R}^{2k}$ and explore the relationship between the monopole gauge field on $S^{2k}$ and the instanton field on $\mathbb{R}^{2k}$  \cite{Banerjee-2004,   OSe-Tchrakian-1987, Jackiw-Rebbi-1976, Adler-1972}. 


First we introduce a general  map from $\mathbb{R}^{2k}$ to $S^{2k}$: 
\be
x_{\mu}~\in~\mathbb{R}^{2k}~~\rightarrow~~n_a(x)~\in~S^{2k}
\ee
where $n_a$ are subject to 
\be
\sum_{a=1}^{2k+1}n_an_a=1. 
\ee
We introduce gauge fields $A_\mu$ on $\mathbb{R}^{2k}$ and $A_{a}$ on $S^{2k}$:    
\be
A=A_{\mu}dx_{\mu}=A_a dn_a, ~~~~F=dA+iA^2=\frac{1}{2}F_{\mu\nu}dx_{\mu}dx_{\nu} = \frac{1}{2}F_{ab}dn_{a}dn_{b}. 
\ee
Since $dn_a =\frac{\partial n_a}{\partial x_{\mu}}dx_{\mu}$, they are related as 
\be
A_{\mu} =\frac{\partial n_a}{\partial x_{\mu}} A_a, ~~~~~F_{\mu\nu} =\frac{\partial n_a}{\partial x_{\mu}}\frac{\partial n_{b}}{\partial x_\nu} F_{ab}.  \label{transdiffmanif}
\ee
The $SO(2k)$ monopole gauge field on $S^{2k}$  is expressed as 
\be
A_m =-\frac{1}{1+n_{2k+1}}\sigma_{mn} n_n,~~A_{2k+1}=0, \label{aorffmonoons2k}  
\ee
and the  monopole field strength  $F_{ab}=\partial_aA_b-\partial_bA_a+i[A_a, A_b]$   is 
\be
F_{mn}=\sigma_{mn}-n_mA_n+n_nA_m, ~~~F_{m, 2k+1}=-F_{2k+1,m}=(1+n_{2k+1})A_m. \label{orffmonoons2k}  
\ee
(\ref{aorffmonoons2k}) and (\ref{orffmonoons2k}) 
are  related to (\ref{afieldstrengthnls2k+1}) and (\ref{fieldstrengthnls2k+1}) through (\ref{transdiffmanif}). 

\subsection{Stereographic projection and gauge  theory on a sphere}

We choose $n_a$ as the inverse stereographic coordinates on $S^d$: 
\be
r_{\mu=1,2,\cdots,d} =\frac{2}{1+x^2}x_{\mu}, ~~r_{d+1}=\frac{1-x^2}{1+x^2}. 
\label{stereograexp}
\ee
Through (\ref{transdiffmanif}),   
 the monopole  configuration  on $S^{2k}$ 
\begin{align}
&\hat{A}_{\mu} =-\frac{1}{1+r_{d+1}}\sigma_{\mu\nu}r_{\nu} ,~~~\hat{A}_{d+1}=0 \nn\\
&\hat{F}_{\mu\nu} =-r_{\mu}\hat{A}_{\nu}+r_{\nu}\hat{A}_{\mu}+\sigma_{\mu\nu}, ~~~\hat{F}_{\mu, d+1}=-\hat{F}_{d+1, \mu} =(1+r_{2k+1})\hat{A}_{\mu},  \label{monogau}
\end{align} 
is transformed to  the ``instanton'' configuration  on $\mathbb{R}^{2k}$, (\ref{gaugescaleinvconfig}) and (\ref{instskonr2k}), as  
\be 
A_{\mu}=-2\frac{1}{x^2+1}\sigma_{\mu\nu}x_\nu, ~~~~~F_{\mu\nu}=4\frac{1}{(x^2+1)^2}\sigma_{\mu\nu} .  \label{socallinstantonconfigf}
\ee 
For $k=2$, (\ref{socallinstantonconfigf}) represents the BPST instanton configuration.  In this paper we  call  (\ref{socallinstantonconfigf}) the ``instanton'' configuration even for arbitrary $k$,  although (\ref{socallinstantonconfigf})  is no longer  a solution of the pure Yang-Mills field equations except for 
$k=2$ (Appendix \ref{append:nonabemono}).   
 Notice that the moduli size-parameter of the instanton (\ref{socallinstantonconfigf}) is identified with the radius of  $S^{2k}$ on which the monopole gauge field lives. Indeed, under the scale transformation 
\be 
r_a ~\rightarrow~R ~r_a
\ee
 or 
$x~\rightarrow~\frac{1}{R}x$,  
(\ref{socallinstantonconfigf}) is transformed as 
\be
A~\rightarrow~ 
-\frac{2}{{x}^2+R^2}\sigma_{\mu\nu} x_{\nu}d x_{\mu}.
\ee
Since the instanton configuration can be obtained by the stereographic projection of the monopole configuration on the sphere, it may be obvious that the size of the  instanton corresponds to the size of the sphere.

From (\ref{monogau}), we can obtain the tensor monopole  field strength on $S^{2k}$ \cite{Hasebe-2014-1}: 
\be
\hat{G}_{a_1 a_2 \cdots a_{2k}} \equiv \frac{1}{2^k}\tr(\hat{F}_{[a_1 a_2} \hat{F}_{a_3a_4} \cdots \hat{F}_{a_{2k-1} a_{2k}]}) =\frac{(2k)!}{2^{k+1}}~
\epsilon_{a_1a_2\cdots a_{2k+1}} r_{a_{2k+1}},  \label{exptensromos2k}
\ee
and similarly the tensor instanton field strength on $\mathbb{R}^{2k}$:   
\be
{G}_{\mu_1 \mu_2 \cdots \mu_{2k}} \equiv \frac{1}{2^k}\tr({F}_{[\mu_1 \mu_2} {F}_{\mu_3\mu_4} \cdots {F}_{\mu_{2k-1} \mu_{2k}]})|_{n_a=r_a} =(2k)! {2^{k-1}} \biggl( \frac{1}{1+x^2} \biggr)^{2k}~
\epsilon_{\mu_1 \mu_2\cdots \mu_{2k}},    \label{exptensromor2k}
\ee
where we used 
\be
\tr(\sigma_{[\mu_1 \mu_2}\sigma_{\mu_3\mu_4}\cdots \sigma_{\mu_{2k-1}\mu_{2k}]}     ) = \frac{1}{2}(2k)! ~\epsilon_{\mu_1\mu_2\mu_3\cdots \mu_{2k}}. 
\ee
$\hat{C}_{a_1 a_2 \cdots a_{2k-1}}$  and  ${C}_{\mu_1 \mu_2 \cdots \mu_{2k-1}}$ that satisfy  
\begin{subequations}
\begin{align}
&\hat{G}_{a_1 a_2 \cdots a_{2k}}=\frac{1}{(2k-1)!}\hat{\partial}_{[a_1} \hat{C}_{a_2 a_3 \cdots a_{2k]}}, \\
&G_{\mu_1 \mu_2 \cdots \mu_{2k}}=\frac{1}{(2k-1)!}{\partial}_{[\mu_1} {C}_{\mu_2 \mu_3 \cdots \mu_{2k]}}, 
\end{align}
\end{subequations}
are obtained from the Chern-Simons term: 
\begin{subequations}
\begin{align}
&\frac{1}{(2k-1)!}\hat{C}_{a_1a_2\cdots a_{2k-1}}dr_{a_1}\wedge dr_{a_2}\cdots dr_{a_{1k-1}} =L_{\text{CS}}^{(2k-1)}[\hat{A}], \label{hatctens}\\
&\frac{1}{(2k-1)!}{C}_{\mu_1 \mu_2\cdots \mu_{2k-1}}dx_{\mu_1}\wedge dx_{\mu_2}\cdots dx_{\mu_{1k-1}} =L_{\text{CS}}^{(2k-1)}[{A}]. \label{exptensromor2kchern}
\end{align}
\end{subequations}
In low dimensions, 
 (\ref{exptensromor2kchern}) is expressed as 
\begin{align}
&k=1~:~C_\mu=\text{tr}A_\mu,
\nn\\
&k=2~:~C_{ \mu\nu\rho}=\tr(A_{[ \mu}\partial_{ \nu}A_{ \rho]}+\frac{2}{3}iA_{[ \mu}A_{ \nu}A_{ \rho]})=\frac{1}{2}\tr(A_{[ \mu}F_{ \nu\rho]}-\frac{2}{3}iA_{[ \mu}A_{ b\nu}A_{ c\rho ]}), 
\nn\\
&k=3~:~C_{\mu_1\mu_2\mu_3\mu_4\mu_5}=\frac{1}{4}\tr(A_{[a\mu_1}F_{\mu_2\mu_3}F_{\mu_4\mu_5]} 
-i A_{[\mu_1}A_{\mu_2}A_{\mu_3}F_{\mu_5]}-\frac{2}{5}A_{[\mu_1}A_{\mu_2}A_{\mu_3}A_{\mu_4}A_{\mu_5]}). 
\label{correpondcsandcfield}
\end{align}
For the instanton configuration (\ref{socallinstantonconfigf}), (\ref{correpondcsandcfield}) becomes\footnote{The explicit forms of $\hat{C}_{a_1 a_2 \cdots a_{2k-1}}$ (\ref{hatctens}) are derived in \cite{Hasebe-2014-1}. }    
\begin{align}
&
k=1~:~C_{\mu}=-\frac{1}{1+x^2}\epsilon_{\mu\nu}x_{\nu}, ~~\nn\\
&
k=2~:~C_{\mu\nu\rho}=-\biggl( \frac{2}{1+x^2}\biggr)^3 \biggl(1+\frac{1+x^2}{2}\biggr)\epsilon_{\mu\nu\rho\sigma}x_{\sigma},  \nn\\
&
k=3~:~C_{\mu_1\mu_2\cdots\mu_5} = -9\biggl( \frac{2}{1+x^2}\biggr)^5 \biggl(1+\frac{1+x^2}{2}+\frac{2}{3}\biggl(\frac{1+x^2}{2}\biggr)^2 \biggr)\epsilon_{\mu_1\mu_2\cdots \mu_{6}}x_{\mu_6}. 
&
\end{align}
(\ref{transdiffmanif}) implies  the following transformation between the monopole and instanton tensor fields:  
\begin{align}
G_{\mu_1\mu_2\cdots \mu_{2k}} &=\hat{G}_{a_1a_2\cdots a_{2k}}\frac{\partial r_{a_1}}{\partial x_{\mu_1}} \frac{\partial r_{a_2}}{\partial x_{\mu_2}}\cdots \frac{\partial r_{a_{2k}}}{\partial x_{\mu_{2k}}} =  \biggl( \frac{2}{1+x^2} \biggr)^{4k} K_{a_1}^{\mu_1}K_{a_2}^{\mu_2} \cdots K_{a_{2k}}^{\mu_{2k}} ~ \hat{G}_{a_1a_2\cdots a_{2k}},  \nn\\ 
C_{\mu_1\mu_2\cdots \mu_{2k-1}} &=\hat{C}_{a_1a_2\cdots a_{2k-1}}\frac{\partial r_{a_1}}{\partial x_{\mu_1}} \frac{\partial r_{a_2}}{\partial x_{\mu_2}}\cdots \frac{\partial r_{a_{2k-1}}}{\partial x_{\mu_{2k-1}}} =  \biggl( \frac{2}{1+x^2} \biggr)^{2(2k-1)} K_{a_1}^{\mu_1}K_{a_2}^{\mu_2} \cdots K_{a_{2k-1}}^{\mu_{2k-1}} ~ \hat{C}_{a_1a_2\cdots a_{2k-1}},  \label{tensfrel}
\end{align}
which can be explicitly confirmed with  the expressions of the fields. 
In (\ref{tensfrel}), we introduced an important quantity 
\be
K_a^{\mu} \equiv \biggl( \frac{1+x^2}{2} \biggr)^2~\frac{\partial r_a}{\partial x_{\mu}},  
\ee
or 
\be
K^{\mu}_{\nu}=\frac{1+x^2}{2}\delta^{\mu}_{\nu}-x_{\mu}x_{\nu}, 
~~~K^{\mu}_{2k+1}=-x_{\mu}. 
\label{sterekillsumkk}
\ee 
$K_a^{\mu}$ are known  as the conformal Killing vectors \cite{Banerjee-2004} that satisfy  the conformal Killing equations  
\be
\partial^{\mu}K^{\nu}+\partial^{\nu}K^{\mu} =\frac{2}{d}\partial^{\lambda}K^{\lambda}\delta_{\mu\nu},~~~~ ~~(\mu,\nu=x_1, x_2, \cdots, x_{d}) 
\ee
and the  transversality condition 
\be
r_a K^{\mu}_a=0. 
\ee
 The conformal Killing vectors  have the following properties:   
\begin{align}
&K_a^{\mu}K_a^{\nu}=\biggl(\frac{1+x^2}{2}\biggr)^2\delta^{\mu\nu}, ~~~~~K_a^{\mu}K_b^{\mu}=\biggl(\frac{1+x^2}{2}\biggr)^2(\delta_{ab}-r_ar_b), \nn\\ 
&\epsilon_{a_1a_2\cdots a_{d+1}}r_{a_{d+1}} K_{a_1}^{\mu_1} K_{a_2}^{\mu_2}  \cdots K_{a_d}^{\mu_d}  = \biggl(  \frac{1+x^2}{2} \biggr)^d  ~\epsilon_{\mu_1\mu_2\cdots \mu_{d}}.  \label{relrkkkkeps}
\end{align}
For more detail properties about $K_a^{\mu}$, see \cite{Banerjee-2004}.

We here discuss somewhat in detail about the formulation of the field theory on sphere  by adding some more information to  \cite{Banerjee-2004, Adler-1972}. 
 Apparently,  the gauge fields on $\mathbb{R}^{2k}$ and on $S^{2k}$  are generally related as 
\be 
A_{\mu} =\biggl( \frac{2}{1+x^2}\biggr)^2 K_a^{\mu}\hat{A}_a, ~~~~
F_{\mu\nu} =\biggl( \frac{2}{1+x^2}\biggr)^4 K_a^{\mu}K_b^{\nu}\hat{F}_{ab},\label{transffs}
\ee
or 
\be 
\hat{A}_{a} = K_a^{\mu}{A}_\mu,~~~~
\hat{F}_{ab} = K_a^{\mu}K_b^{\nu} {F}_{\mu\nu}. \label{hatfields2}
\ee
The derivative on $S^{2k}$ is constructed as 
\be
\hat{\partial}_a \equiv K_a^{\mu} \frac{\partial}{\partial x_{\mu}}~=   \frac{\partial}{\partial r_a} -r_a r_b\frac{\partial}{\partial r_b} = ir_b L_{ba}, \label{defderivs2k}
\ee
where 
\be
L_{ab} =-ir_a \frac{\partial}{\partial r_b} + i r_b \frac{\partial}{\partial r_a} =-ir_a\hat{\partial}_b +ir_b\hat{\partial}_a=-iK_a^{\mu}\frac{\partial K_b^{\nu}}{\partial x_{\mu}}\frac{\partial}{\partial x_{\nu} }  +iK_b^{\mu}\frac{\partial K_a^{\nu}}{\partial x_{\mu}}\frac{\partial}{\partial x_{\nu} } . 
\ee 
Although $r_a$ are the coordinates on $S^{d}$ subject to $\sum_{a=1}^{d+1}{r_a}r_a=1$,  we can treat $r_a$ as if they are $\it{independent}$ parameters in using (\ref{defderivs2k}). 
$\hat{\partial}_a$ are non-commutative operators that satisfy the $SO(d+1, 1)$ algebra  with $L_{ab}$:\footnote{
Under the identification $L_{a,d+2}=-i\hat{\partial}_a$ $(a=1,2,\cdots,d+1)$, $L_{AB}$ $(A,B=1,2,\cdots, d+2)$ realize the $SO(d+1,1)$ algebra:  
\be
[L_{AB} , L_{CD}] =i\eta_{AC}L_{BD} - i\eta_{AD}L_{BC} + i\eta_{BD}L_{AC} - i\eta_{BC}L_{AD} 
\ee
with $\eta_{AB} =\text{diag}(\overbrace{+,+,\cdots, +}^{d+1}, -)$.  
} 
\begin{align}
&[-i\hat{\partial}_a , -i\hat{\partial}_b] = -iL_{ab}, 
~~~~[L_{ab}, -i\hat{\partial}_c] = i\delta_{ac}(-i\hat{\partial}_b) -i\delta_{bc}(-i\hat{\partial}_a), \nn\\
&[L_{ab} , L_{cd}] =i\delta_{ac}L_{bd} - i\delta_{ad}L_{bc} + i\delta_{bd}L_{ac} - i\delta_{bc}L_{ad}. 
\end{align}
The field strength on $S^{2k}$ is given by\footnote{ (\ref{fatfieldstfromgau}) is simply related to the three-rank antisymmetric field strength  \cite{Adler-1972}
\be
\hat{F}_{abc} =i(L_{ab}\hat{A}_c +L_{bc}\hat{A}_a +L_{ca}\hat{A}_b ) +i(r_a[\hat{A}_b, \hat{A}_c] +r_b[\hat{A}_c, \hat{A}_a] +r_c[\hat{A}_a  \hat{A}_b]  )
\ee
as 
\be
\hat{F}_{ab} =r_c \hat{F}_{abc}. 
\ee
} 
\be
\hat{F}_{ab} = \hat{\partial}_a \hat{A}_b -\hat{\partial}_b \hat{A}_a +i[\hat{A}_a, \hat{A}_b] +ir_c L_{ab}\hat{A}_c.  
\label{fatfieldstfromgau}
\ee 
Note  the existence of the last term on the right-hand side of (\ref{fatfieldstfromgau}).  
Substituting (\ref{hatfields2}) and (\ref{defderivs2k})  into (\ref{fatfieldstfromgau}), we have  
\be
\hat{F}_{ab}= K_a^{\mu}K_{b}^{\nu}F_{\mu\nu} +K_a^{\mu} (\partial_{\mu}K_b^{\nu})A_{\nu} - K_b^{\mu}(\partial_{\mu}K_a^{\nu})A_{\nu} +ir_c L_{ab}\hat{A}_c.  \label{fatfabfromfmunu}
\ee
The validity of (\ref{fatfabfromfmunu}) can be easily confirmed for the monopole and instanton configurations. 
For the monopole field (\ref{monogau})  and  the instanton field  (\ref{socallinstantonconfigf}), we can show  
\be
K_a^{\mu} (\partial_{\mu}K_b^{\nu})A_{\nu} - K_b^{\mu}(\partial_{\mu}K_a^{\nu})A_{\nu}=r_a\hat{A}_b-r_b\hat{A}_a =-ir_c L_{ab}\hat{A}_c. 
\ee
Therefore, only the first term on the right-hand side of (\ref{fatfabfromfmunu}) survives to yield $\hat{F}_{ab}= K_a^{\mu}K_{b}^{\nu}F_{\mu\nu}$, which is (\ref{transffs}).

For tensor fields, (\ref{fatfieldstfromgau}) may be generalized as 
\be
\hat{G}_{a_1 a_2 \cdots a_{2k}} =\frac{1}{(2k-1)!}\hat{\partial}_{[a_1}\hat{C}_{a_2 \cdots a_{2k}]}+i\frac{1}{2 ~(2k-2)!} r_{a_{2k+1}}L_{[a_1 a_2} \hat{C}_{a_3 \cdots a_{2k}] a_{2k+1}}. 
\ee

\subsection{Yang-Mills action and Chern number}

With  the area element of $S^{d}$ 
\be
d\Omega_d =\biggl(\frac{2}{1+x^2} \biggr)^d~d^dx   \label{areaelesd}
\ee
 and 
\be
{\hat{F}_{ab}}^2 =\bigg(\frac{1+x^2}{2}\biggr)^4 ~{F_{\mu\nu}}^2, 
\ee
the Yang-Mills action is expressed as  
\be
\int_{S^{2k}}~d\Omega_{2k} ~\tr({\hat{F}_{ab}}^2) =\int_{\mathbb{R}^{2k}} d^{2k}x ~ \bigg(\frac{1+x^2}{2}\biggr)^{4-2k} \tr({F_{\mu\nu}}^2). \label{intff}
\ee
For the special case $2k=4$, the conformal factor on the right-hand side of (\ref{intff}) vanishes, and so 
 (\ref{intff}) becomes   
\be
\int_{S^{4}}~d\Omega_{4} ~\tr({\hat{F}_{ab}}^2) =\int_{\mathbb{R}^{4}} d^{4}x ~ \tr({F_{\mu\nu}}^2),  
\ee
which yields the equations of motion: 
\be
\hat{D}_a\hat{F}_{ab}|_{2k=4} = D_{\mu}F_{\mu\nu}|_{2k=4}=0.  \label{k2dafdmuf}
\ee
Meanwhile, the $k$th Chern number is expressed as  
\begin{align}
c_k &= \frac{1}{(2\pi)^k k!}\int_{S^{2k}} ~\tr(\hat{F}^k) \nn\\
&=\frac{1}{(4\pi)^k k!} \int_{S^{2k}} \tr\biggl(\hat{F}_{a_1a_2}\cdots \hat{F}_{a_{2k-1}a_{2k}} \biggr) \epsilon_{a_1 a_2 \cdots a_{2k+1}} r_{a_{2k+1}} d\Omega_{2k} \nn\\
&=\frac{1}{(4\pi)^k k!}\int_{\mathbb{R}^{2k}}\tr(F_{\mu_1\mu_2}F_{\mu_3\mu_4}\cdots F_{\mu_{2k-1}\mu_{2k}}) \epsilon_{\mu_1\mu_2\cdots\mu_{2k}}d^{2k}x \nn\\
&= \frac{1}{(2\pi)^k k!}\int_{\mathbb{R}^{2k}} ~\tr({F}^k).   \label{chernnumberfinstmono}
\end{align}
In the third equation,  we used  (\ref{relrkkkkeps}) and (\ref{hatfields2}).  
The Chern number of the instanton configuration on $\mathbb{R}^{2k}$ is exactly equal to that of the monopole configuration  
on $S^{2k}$.  Indeed for instance, (\ref{monogau}) and (\ref{socallinstantonconfigf}) yield the same result $c_k=1$ in (\ref{chernnumberfinstmono}).     

\subsection{Equations of motion for the monopole fields and the instanton fields}\label{append:nonabemono}

For the monopole gauge field $\hat{A}_a$ (\ref{monogau}), 
 the corresponding field strength  is obtained from (\ref{fatfieldstfromgau}):  
\be
\hat{F}_{\mu\nu}  =-r_{\mu}\hat{A}_{\nu} +r_{\nu}\hat{A}_{\mu} +\sigma_{\mu\nu}, ~~~\hat{F}_{\mu, d+1} =-\hat{F}_{d+1, \mu} =-\sigma_{\mu\nu}r_{\nu}=(1+r_{d+1})\hat{A}_{\mu}, \label{monogaust}
\ee
where we used 
\begin{align}
&\partial_{\mu}\hat{A}_{\nu} -\hat{\partial}_{\nu}\hat{A}_{\mu} +i[\hat{A}_{\mu}, \hat{A}_{\nu}] =\sigma_{\mu\nu},~~~ir_\rho L_{\mu\nu}\hat{A}_{\rho} =-r_{\mu}\hat{A}_{\nu}+r_{\nu}\hat{A}_{\mu}, \nn\\
&\partial_{\mu}\hat{A}_{d+1} -\hat{\partial}_{d+1}\hat{A}_{\mu} +i[\hat{A}_{\mu}, \hat{A}_{d+1}] =\hat{A}_{\mu},~~~ir_\rho L_{\mu, d+1}\hat{A}_{\rho} =r_{d+1}\hat{A}_{\mu}. 
\end{align}
(\ref{monogaust}) is identical to  (\ref{monogau}).  
We can check that the monopole gauge field satisfies the pure Yang-Mills equation on $S^{2k}$: 
\be
\hat{D}_a \hat{F}_{ab} \equiv \hat{\partial}_a \hat{F}_{ab} +i[\hat{A}_a, \hat{F}_{ab}] =0,  \label{monopoleymeq} 
\ee
where we used 
\be
\hat{\partial}_a\hat{F}_{ab} =(2-d)\hat{A}_{b} =-i[\hat{A}_a, \hat{F}_{ab}]. 
\ee
(\ref{monopoleymeq}) is  expected from the previous result (\ref{EOMso2kmono}). 

Meanwhile, the instanton configuration (\ref{socallinstantonconfigf}) satisfies 
\be
D_\mu F_{\mu\nu} +\biggl(\frac{2}{1+x^2}\biggr)^2(4-2k)A_{\nu}=0 , \label{generainstaneq}
\ee
where 
\be
D_{\mu}F_{\mu\nu}\equiv \frac{\partial}{\partial x_{\mu}}F_{\mu\nu} +i[A_{\mu}, F_{\mu\nu}]. 
\ee
Notice that in the special case $2k=4$, the second term on the left-hand side of (\ref{generainstaneq}) vanishes, and so  the instanton configuration realizes a solution of the pure Yang-Mills field equation: 
\be
D_{\mu}F_{\mu\nu}|_{2k=4}=0.   
\ee
For general $k$, the instanton configuration (\ref{socallinstantonconfigf}) does not satisfy the pure Yang-Mills equation.

Using  (\ref{monogau}) and (\ref{socallinstantonconfigf}), 
we can directly show  
\be
\hat{D}_a\hat{F}_{ab} =\bigg(\frac{1+x^2}{2}\biggr)^2 ~K_b^{\nu}~\biggl(D_{\mu}F_{\mu\nu}+\bigg(\frac{2}{1+x^2}\biggr)^2(4-2k)A_{\nu}\biggr)  \label{dafab}
\ee
or 
\be
D_{\mu}F_{\mu\nu}+\bigg(\frac{2}{1+x^2}\biggr)^2(4-2k)A_{\nu}  =\bigg(\frac{2}{1+x^2}\biggr)^4 ~K_b^{\nu}~\hat{D}_a\hat{F}_{ab}. \label{dmufmunu}
\ee
Here, we used 
\be
\hat{D}_a\hat{F}_{ab} =\bigg(\frac{1+x^2}{2}\biggr)^2\bigg(K_b^{\nu}D_{\mu}F_{\mu\nu}+R_b\biggr)
\ee
with 
\be
R_a\equiv \biggl(\frac{\partial}{\partial x_{\mu}}K_a^{\nu} +\frac{2(2-d)}{1+x^2}x_{\mu}K_a^{\nu}\biggr)F_{\mu\nu} =\biggl(\frac{2}{1+x^2}\biggr)^2 (4-d)K_a^{\mu}{A}_\mu. 
\ee
(\ref{dafab}) or (\ref{dmufmunu}) implies that 
\be
\hat{D}_a\hat{F}_{ab} =0 ~~\leftrightarrow~~ D_{\mu}F_{\mu\nu}+\bigg(\frac{2}{1+x^2}\biggr)^2(4-2k)A_{\nu}=0,  \label{necsuffdf}
\ee
which is consistent with  (\ref{monopoleymeq}) and (\ref{generainstaneq}).

\section{$g$ matrices and the $SU(4)$-generalized 't Hooft symbol}\label{sec:extintduality}

\subsection{Properties of $g$ matrices}\label{sec:extintduality1}

$g$ matrices are a higher dimensional analogue of the quaternions:\footnote{For $k=4$, $g$ matrices yield $e_{M=1,2,\cdots, 8}$ in \cite{Nakamula-Sasaki-Takesue-2017}.}    
\be
g_{m}\equiv \{-i\gamma_i,~ 1_{2^{k-1}}\}, ~~~(m=1,2,\cdots, 2k) \label{defgm}
\ee
and 
\be
\bar{g}_{m}\equiv \{i\gamma_i,~ 1_{2^{k-1}}\} ={g_m}^{\dagger},   \label{defgbarm}
\ee
where $\gamma_i$ $(i=1,2,\cdots, 2k-1)$ are the $SO(2k-1)$ gamma matrices: 
\be
\{\gamma_i, \gamma_j\}=2\delta_{ij}. 
\ee
 The $SO(2k+1)$ gamma matrices, $\Gamma_a$, and $SO(2k+1)$ matrix generators, $\Sigma_{ab} =-i\frac{1}{4}[\Gamma_a, \Gamma_b]$, are constructed as  
\begin{align}
&\Gamma_{m} =\begin{pmatrix}
0 & \bar{g}_{m} \\
g_{m} & 0 
\end{pmatrix}, ~~~\Gamma_{2k+1} =\begin{pmatrix}
1_{2^{k-1}} & 0 \\
0 & -1_{2^{k-1}}
\end{pmatrix},  \nn\\
&\Sigma_{mn} 
=\begin{pmatrix}
\sigma_{mn} & 0 \\
0 & \bar{\sigma}_{mn}
\end{pmatrix},  ~~~\Sigma_{m , 2k+1}=-\Sigma_{2k+1, m} 
=i\frac{1}{2}\begin{pmatrix}
0 & \bar{g}_{m} \\
-g_{m} & 0 
\end{pmatrix}, 
\end{align}
where  $Spin(2k)$ generators are given by   
\be
\sigma_{mn} = -i\frac{1}{4}(\bar{g}_{m}g_{n}-\bar{g}_{n}g_{m}), ~~~~
\bar{\sigma}_{mn}= -i\frac{1}{4}({g}_{m}\bar{g}_{n}-{g}_{n}\bar{g}_{m}). 
\ee 
With $\sigma_{mn}$ and $\bar{\sigma}_{mn}$,  $g$ matrices  satisfy  
\begin{subequations}
\begin{align}
&g_{m}\bar{g}_{n}+g_{n}\bar{g}_{m}=\bar{g}_{m}{g}_{n}+\bar{g}_{n}{g}_{m}=2\delta_{mn}, \\
&g_{m}\sigma_{np}-\bar{\sigma}_{np}g_{m}=-i(\delta_{mn}g_{p}-\delta_{mp}g_{n}), 
~~~\bar{g}_{m}\bar{\sigma}_{np}-{\sigma}_{np}\bar{g}_{m}=-i(\delta_{mn}\bar{g}_{p}-\delta_{mp}\bar{g}_{n}). \label{sigmagammarelations}
\end{align}
\end{subequations}

\subsection{Generalized 't Hooft symbol}\label{appendix:genethooft}

\subsubsection{The original 't Hooft symbol}

The $SO(4)$ gamma matrices and 
matrix generators are expressed as\footnote{
The components of $\sigma_{mn}$ and $\bar{\sigma}_{mn}$ are 
\be
\sigma_{ij}=\bar{\sigma}_{ij}=\frac{1}{2}\epsilon_{ijk}\sigma_k, ~~~~~\sigma_{i 4}=-\bar{\sigma}_{i4}=\frac{1}{2}\sigma_i. ~~~~(i,j=1,2,3)
\ee
}  
\begin{align}
&\gamma_m =\begin{pmatrix}
0 & \bar{q}_m \\
q_m & 0 
\end{pmatrix}, ~~~(q_i=-i\sigma_i, ~q_4=1_2)\label{so4gamma} \nn\\
&\Sigma_{mn} = \begin{pmatrix}
\sigma_{mn} & 0 \\
0 &  \bar{\sigma}_{mn}
\end{pmatrix} \equiv 
\frac{1}{2}\begin{pmatrix} 
\eta_{mn}^i\sigma_i & 0 \\
0 & \bar{\eta}_{mn}^i \sigma_i 
\end{pmatrix},  
\end{align}
where ${\eta}_{mn}^{i}$ and $\bar{\eta}_{mn}^{i}$ are the 't Hooft symbols \cite{tHooft-1976}: 
\be
{\eta}_{mn}^{i}=\epsilon_{mni4} + \delta_{mi}\delta_{n4} -  \delta_{m4}\delta_{ni}, ~~~~~\bar{\eta}_{mn}^{i}=\epsilon_{mni4} -\delta_{mi}\delta_{n4} + \delta_{m4}\delta_{ni}. 
\ee
The Pauli matrices are inversely represented as 
\be
\sigma_i =\frac{1}{4}\eta_{mn}^i\sigma_{mn} = \frac{1}{4}\bar{\eta}_{mn}^i\bar{\sigma}_{mn}.  
\ee
The $Spin(4)$ matrix generators satisfy  the self-dual and the anti-self-dual equations,  
\be
\sigma_{mn} =\frac{1}{2}\epsilon_{mnpq}\sigma_{pq}, ~~~\bar{\sigma}_{mn} =-\frac{1}{2}\epsilon_{mnpq}\bar{\sigma}_{pq}, 
\ee
and 
\begin{align}
&\sigma_{mn} \sigma_{pq} = i\frac{1}{2}(\delta_{mp}\sigma_{nq} - \delta_{mq}\sigma_{np} + \delta_{nq}\sigma_{mp} - \delta_{np}\sigma_{mq} ) + \frac{1}{4} (\delta_{mp}\delta_{nq}-\delta_{mq}\delta_{np}) 1_2 + \frac{1}{4}\epsilon_{mnpq}1_2,   \nn\\
&\bar{\sigma}_{mn} \bar{\sigma}_{pq} = i\frac{1}{2}(\delta_{mp}\bar{\sigma}_{nq} - \delta_{mq}\bar{\sigma}_{np} + \delta_{nq}\bar{\sigma}_{mp} - \delta_{np}\bar{\sigma}_{mq} ) + \frac{1}{4} (\delta_{mp}\delta_{nq}-\delta_{mq}\delta_{np}) 1_2 - \frac{1}{4}\epsilon_{mnpq}1_2,    \nn\\
&\sigma_{mn}\bar{\sigma}_{mn} = \bar{\sigma}_{mn} {\sigma}_{mn} =2\frac{1}{4}(3-3)1_2={0}_2.  \label{sigsigbarsu2relmat}
\end{align}
The above relations can be rephrased as the properties of the   't Hooft symbol:  
\begin{subequations}
\begin{align}
&\eta_{mn}^i=\frac{1}{2}\epsilon_{mnpq}\eta_{pq}^i, \\
&{\eta}_{mn}^{i}~{\eta}_{pq}^{i} =\delta_{mp}\delta_{nq} -\delta_{mq}\delta_{np} +  \epsilon_{mnpq}, \label{epconcetasu2}\\
&\epsilon_{ijk}{\eta}_{mn}^{i}{\eta}_{pq}^{j} = \delta_{mp} {\eta}_{nq}^{k} - \delta_{mq}{\eta}_{np}^{k} + \delta_{nq}{\eta}_{mp}^{k} - \delta_{np}{\eta}_{mq}^{k} ,  \label{selfdualthooft} 
\end{align}
\end{subequations}
and 
\be
{\eta}_{mn}^{i}{\eta}_{mn}^{j}=4\delta^{ij}, ~~~~{\eta}_{mn}^{i}{\eta}_{np}^{j}{\eta}_{pm}^{k}=4\epsilon^{ijk}. 
\ee
Note  that $\epsilon_{ijk}=-i\frac{1}{2}\text{tr}(\sigma_i\sigma_j\sigma_k)$ are  the structure constants  of 
the $SU(2)$.   
Except for (\ref{selfdualthooft}) and  (\ref{epconcetasu2}), all relations also hold for 
$\bar{\eta}_{mn}^i$: 
\begin{subequations}
\begin{align}
&\bar{\eta}_{mn}^i=-\frac{1}{2}\epsilon_{mnpq}\bar{\eta}_{pq}^i, \\
&\bar{\eta}_{mn}^{i}~\bar{\eta}_{pq}^{i} =\delta_{mp}\delta_{nq} -\delta_{mq}\delta_{np} -  \epsilon_{mnpq}.
\end{align}
\end{subequations}
$\eta_{mn}^{i}$ and $\bar{\eta}_{mn}^{j}$ satisfy 
\be
\eta_{mn}^{i}~\bar{\eta}_{mn}^{j}=0. \label{orthoetaetabarsu2}
\ee

\subsubsection{The $SU(4)$ generalized 't Hooft symbol}

The $SO(6)$ gamma matrices are represented as 
\be
\Gamma_{m=1,2,\cdots, 6} =\begin{pmatrix}
0 & \bar{g}_m \\
g_m & 0 
\end{pmatrix},
\ee
with 
\be
g_{m}=\{g_{i=1,2,\cdots, 5}, g_6\}=\{-i\gamma_i, 1_4\}, ~~~~\bar{g}_{m}=\{\bar{g}_{i=1,2,\cdots, 5}, \bar{g}_6\}=\{+i\gamma_i, 1_4\}. 
\ee
Here, $\gamma_{i=1,2,3,4,5}$ are the $SO(5)$ gamma matrices; $\gamma_{i=1,2,3,4}$ (\ref{so4gamma}) and  
$\gamma_5=\begin{pmatrix}
1_2 & 0 \\
0 & -1_2
\end{pmatrix}$.  
The  $SO(6)$ matrix generators, 
$\Sigma_{mn}=-i\frac{1}{4}[\Gamma_m, \Gamma_n]$, take the form of 
\be
\Sigma_{mn} = \begin{pmatrix}
\sigma_{mn} & 0 \\
0 &  \bar{\sigma}_{mn}
\end{pmatrix},   
 \label{expansso6bysu4}
\ee
where  $\sigma_{mn}$ and $\bar{\sigma}_{mn}$ are the $Spin(6)$  matrix generators: 
\be
\sigma_{ij}=\bar{\sigma}_{ij}=-i\frac{1}{4}[\gamma_i, \gamma_j], ~~~~~\sigma_{i 6}=-\bar{\sigma}_{i6}=\frac{1}{2}\gamma_i. \label{decompmso6gen} 
\ee
$\sigma_{mn}$ and $\bar{\sigma}_{mn}$  satisfy   the generalized self-dual and  anti-self-dual equations,    
\be
\sigma_{mn} =\frac{1}{12}\epsilon_{mnpqrs}\sigma_{pq}\sigma_{rs}, ~~~\bar{\sigma}_{mn} =-\frac{1}{12}\epsilon_{mnpqrs}\bar{\sigma}_{pq}\bar{\sigma}_{rs},  
\ee 
and 
\begin{align}
&{\sigma}_{mn} {\sigma}_{pq} =\frac{1}{4}(\delta_{mp}\delta_{nq}-\delta_{mq}\delta_{nq})1_4+i\frac{1}{2}(\delta_{mp} {\sigma}_{nq}-\delta_{mq}{\sigma}_{np}+\delta_{nq}{\sigma}_{mp}-\delta_{np}{\sigma}_{mq}) +\frac{1}{4}\epsilon_{mnpqrs}{\sigma}_{rs}, \nn\\
&\bar{\sigma}_{mn}\bar{\sigma}_{pq} =\frac{1}{4}(\delta_{mp}\delta_{nq}-\delta_{mq}\delta_{nq})1_4+i\frac{1}{2}(\delta_{mp}\bar{\sigma}_{nq}-\delta_{mq}\bar{\sigma}_{np}+\delta_{nq}\bar{\sigma}_{mp}-\delta_{np}\bar{\sigma}_{mq}) -\frac{1}{4}\epsilon_{mnpqrs}\bar{\sigma}_{rs}, 
\nn\\
&\sigma_{mn}\bar{\sigma}_{mn} = \bar{\sigma}_{mn} {\sigma}_{mn} =2\frac{1}{4}(10-5)1_4=\frac{5}{2}1_4 .   \label{sigsigbarsu4relmat}
\end{align}
Since $Spin(6)\simeq SU(4)$, $\sigma_{mn}$ and $\bar{\sigma}_{mn}$ can be expressed as a linear combination of  the $SU(4)$ Gell-Mann matrices \cite{Greiner-Muller-book-1981} $\lambda^A$ $(A=1,2,\cdots, 15)$: 
\be
\sigma_{mn}=\frac{1}{2}
\eta_{mn}^A\lambda_A,~~~~~\bar{\sigma}_{mn}=\frac{1}{2}\bar{\eta}_{mn}^A \lambda_A. 
\label{expsigmalambda}
\ee
Here, we introduced $\eta_{mn}^A$ and $\bar{\eta}_{mn}^A$ as the expansion coefficients which we refer to  as the $SU(4)$ generalized 't Hooft symbols.   
(\ref{decompmso6gen}) implies  
\be
\eta_{ij}^A =\bar{\eta}_{ij}^A ~~~~~\eta_{i 6}^A =-\eta_{6 i}^A =-\bar{\eta}_{i6}^A =\bar{\eta}^A_{6 i}. 
\ee
The  $SU(4)$ Gell-Mann matrices are inversely represented as   
\be
\lambda_A =\frac{1}{4}\eta_{mn}^A\sigma_{mn} =\frac{1}{4}\bar{\eta}_{mn}^A\bar{\sigma}_{mn}.
\ee
The $SU(4)$ Gell-Mann matrices have the following properties 
\be
[\lambda_A, \lambda_B] =2if^{ABC}\lambda_C, ~~~~\{\lambda_A, \lambda_B\} =\delta^{AB}1_4  +2d^{ABC}\lambda_C, 
\ee
or 
\be
\lambda_A \lambda_B =\frac{1}{2}\delta_{AB} 1_4 +i(f_{ABC}-id_{ABC})\lambda_C,  \label{completerellambdas}
\ee
where $f_{ABC}$ are the structure constants (totally antisymmetric tensors) and   $d_{ABC}$ are the totally symmetric tensors \cite{Greiner-Muller-book-1981}:  
\begin{align}
&f_{ABC}=-i\frac{1}{12}\text{Atr}(\lambda_{A}\lambda_B\lambda_{C})
=-i\frac{1}{4}\text{tr}([\lambda_A,\lambda_B] \lambda_C),  \nn\\
&d_{ABC} 
= \frac{1}{12}\text{Str}(\lambda_{A}\lambda_B\lambda_{C})
=\frac{1}{4}\text{tr}(\{\lambda_A,\lambda_B\} \lambda_C).  
\end{align}
Substituting (\ref{expsigmalambda}) into the equations of the $Spin(6)$ matrix generators, 
one may find properties of   the $SU(4)$ generalized 't Hooft symbol:  
\begin{subequations}
\begin{align}
&\eta_{mn}^A =\frac{1}{24}\epsilon_{mnpqrs}d_{ABC}~\eta_{pq}^B~\eta_{rs}^C,  \label{selfduso6} \\
&{\eta}_{mn}^{A} {\eta}_{pq}^{A} =2(\delta_{mp}\delta_{nq} -\delta_{mq}\delta_{np} ), \\
&(f_{ABC}-id_{ABC}){\eta}_{mn}^{B}{\eta}_{pq}^{C} = (\delta_{mp} {\eta}_{nq}^{A} - \delta_{mq} {\eta}_{np}^{A} - \delta_{nq} {\eta}_{mp}^{A} - \delta_{np} {\eta}_{mq}^{A})-i \frac{1}{2}\epsilon_{mnpqrs}\bar{\eta}_{rs}^{A} , 
\label{epconcetasu4}
\end{align} 
\end{subequations}
and 
\be
{\eta}_{mn}^A {\eta}_{mn}^B=4\delta^{AB}, ~~~~ {\eta}_{mn}^A {\eta}_{np}^B {\eta}_{pm}^C=4f^{ABC}, ~~~~\epsilon_{mnpqrs} {\eta}_{mn}^A {\eta}_{pq}^B{\eta}_{rs}^C=32 d^{ABC}.  
\ee
Similar  relations also hold for 
$\bar{\eta}_{mn}^i$ except for (\ref{selfduso6}) and  (\ref{epconcetasu4}):  
\begin{subequations}
\begin{align}
&\bar{\eta}_{mn}^A =-\frac{1}{24}\epsilon_{mnpqrs}d_{ABC}~\bar{\eta}_{pq}^B~\bar{\eta}_{rs}^C, \\
&(f_{ABC}-id_{ABC})\bar{\eta}_{mn}^{B}\bar{\eta}_{pq}^{C} = (\delta_{mp}\bar{\eta}_{nq}^{A} - \delta_{mq}\bar{\eta}_{np}^{A} - \delta_{nq}\bar{\eta}_{mp}^{A} - \delta_{np}\bar{\eta}_{mq}^{A})+i \frac{1}{2}\epsilon_{mnpqrs}\bar{\eta}_{rs}^{A}.
\end{align}
\end{subequations}
The last equation of (\ref{sigsigbarsu4relmat}) yields   
\be
\eta_{mn}^{A}~\bar{\eta}_{mn}^{A} =20,~~~~~d_{ABC}\eta_{mn}^{B}~\bar{\eta}_{mn}^{C}=0. \label{sumetaetabarsu4}
\ee

\section{Tensor gauge field theory}\label{appendix:tensorfieldtheory}

Here, we review tensor gauge field theories in  even dimensions mainly based on \cite{Tchrakian-1980,  Tchrakian-1985, OSe-Tchrakian-1987} . 

\subsection{Basic properties of the tensor field}

From the following property of the anti-commutator 
\be
M_{[1}M_2 M_3 M_4 \cdots M_{2l]} =\frac{1}{2^2 (2l-2)!} \epsilon_{\mu_1\mu_2\cdots\mu_{2l}} \{ M_{[\mu_1}M_{\mu_2]}, M_{[\mu_3}M_{\mu_4}\cdots M_{\mu_{2l}]}   \} , 
\ee
we have 
\begin{align}
F_{123\cdots ,2l} &\equiv \frac{1}{(2l)!}F_{[12 }F_{34}\cdots F_{2l-1, 2l]} \nn\\
&=\frac{1}{2(2l)!} \epsilon_{\mu_1\mu_2\mu_3\cdots \mu_{2l}} \{F_{\mu_1\mu_2}, F_{\mu_3\mu_4\cdots\mu_{2l}}\}\nn\\
&=\frac{1}{2(2l-1)} (\{ F_{12}, F_{34\cdots 2l}\} -\{ F_{13}, F_{24 \cdots 2l}\} +\cdots +\{ F_{1,2l}, F_{2 3 \cdots ,2l-1}\} ).   
\label{ftensotrdecomff}
\end{align}
A covariant fashion of (\ref{ftensotrdecomff}) yields 
\begin{align} 
F_{\mu_1\mu_2\cdots\mu_{2l}} &\equiv \frac{1}{(2l)!}F_{[\mu_1\mu_2} F_{\mu_3\mu_4}\cdots F_{\mu_{2l-1}\mu_{2l}]}\nn\\
&= \frac{1}   {2(2l-1)} \sum_{i=2}^{2l} (-1)^i \{F_{\mu_1 \mu_i} , F_{\mu_2 \mu_3 \cdots \mu_{i-1}\mu_{i+1} \cdots \mu_{2l}}\} \nn\\
&=\frac{1}   {2(2l-1)} (\{ F_{\mu_1 \mu_2}, F_{\mu_3 \mu_4\cdots \mu_{2l}}\} -\{ F_{\mu_1\mu_3}, F_{\mu_2\mu_4 \cdots \mu_{2l}}\} +\cdots +\{ F_{\mu_1,\mu_{2l}}, F_{\mu_2 \mu_3 \cdots ,\mu_{2l-1}}\}) .  
\label{defdecomftensor} 
\end{align}
For instance, 
\begin{align}
&F_{\mu\nu} =\frac{1}{2!}F_{[\mu\nu]}, \nn\\
&F_{\mu\nu\rho\sigma}=\frac{1}{4!} F_{[\mu\nu}F_{\rho\sigma]}
=\frac{1}{6} (\{  F_{\mu\nu}, F_{\rho\sigma} \} - \{  F_{\mu\rho}, F_{\nu\sigma} \}+\{  F_{\mu\sigma}, F_{\nu\rho} \}),\nn \\
&F_{\mu\nu\rho\sigma\kappa\tau}=\frac{1}{6!} F_{[\mu\nu}F_{\rho\sigma}F_{\kappa\tau]}=\frac{1}{10} (\{F_{\mu\nu},F_{\rho\sigma\kappa\tau}\} -\{F_{\mu\rho},F_{\nu\sigma\kappa\tau}\}+\{F_{\mu\sigma},F_{\nu\rho\kappa\tau}\}-\{F_{\mu\kappa},F_{\nu\rho\sigma\tau}\}+\{F_{\mu\tau},F_{\nu\rho\sigma\kappa}\}). \label{examanticommu}
\end{align} 
One may observe that the higher rank tensor fields are hierarchically constituted of the lower rank tensor fields.   
The squares of the four-rank and six-rank tensor field strengths are respectively given by\footnote{(\ref{fourstdecffff}) was utilized in  8D tensor gauge theory of \cite{Nakamula-Sasaki-Takesue-2017} to realize  a 7(+1)D Skyrmion  from the Atiyah-Manton construction. }  
\begin{subequations}
\begin{align}
&\tr({F_{\mu\nu\rho\sigma}}^2) = \frac{1}{6}\tr(({F_{\mu\nu}}^2)^2)  -\frac{2}{3}\tr(F_{\mu\nu}F_{\rho\sigma}F_{\mu\rho}F_{\nu\sigma})+\frac{1}{6}\tr((F_{\mu\nu}F_{\rho\sigma})^2 ),  \label{fourstdecffff} \\
&\tr({F_{\mu\nu\rho\sigma\kappa\tau}}^2) = \frac{1}{15}\tr(  (F_{\mu\nu}F_{\rho\sigma\kappa\tau})^2) -\frac{116}{225}\tr( F_{\mu\nu}F_{\rho\sigma\kappa\tau}F_{\mu\rho}F_{\nu\sigma\kappa\tau}) + \frac{94}{225} \tr( F_{\mu\nu}F_{\rho\sigma\kappa\tau}F_{\rho\sigma}F_{\mu\nu\kappa\tau}). 
\end{align}
\end{subequations} 

\subsection{Gauge Symmetry and covariant derivatives}

Under the gauge transformation 
\begin{subequations}
\begin{align}
&A_{\mu} ~~\rightarrow~~g(x)^{\dagger} A_{\mu} g(x)- ig(x)^{\dagger}~\partial_{\mu}~g(x),  ~~(g(x)^{\dagger}g(x)=1) \label{transa}\\
&F_{\mu\nu}=\partial_{\mu}A_{\nu} -\partial_{\nu}A_{\mu}+i[A_{\mu}, A_{\nu}]~~\rightarrow~~g(x)^{\dagger} F_{\mu\nu} g(x), ~
\end{align}
\end{subequations}
the tensor field strength (\ref{defdecomftensor}) is transformed as 
\be
F_{\mu_1\mu_2\cdots \mu_{2l}}~~\rightarrow~~g(x)^{\dagger}~ F_{\mu_1\mu_2\cdots \mu_{2l}}~ g(x).  
\label{gaugetrafietens}
\ee
The covariant derivative of the tensor field strength is introduced so as to satisfy
\be
D_{\mu}F_{\mu_1\mu_2\cdots\mu_{2l}} ~~\rightarrow~~g(x)^{\dagger} D_{\mu}F_{\mu_1\mu_2\cdots\mu_{2l}} g(x),   \label{dgaugetrafietens}
\ee
and such  covariant derivative is simply constructed as 
\be
D_{\mu} F_{\mu_1\mu_2\cdots\mu_{2l}} \equiv \partial_{\mu}F_{\mu_1\mu_2\cdots\mu_{2l}} +i[A_{\mu}, F_{\mu_1\mu_2\cdots\mu_{2l}}].  
\label{derivfieltens}
\ee
Note that the covariant derivative linearly acts to the original constituent  2-rank field strength of the tensor field strength.  For instance,  
\be
D_{\mu}F_{\nu\rho\sigma\tau} =\frac{1}{4!}(D_{\mu}F_{[\nu\rho}\cdot F_{\sigma\tau]} + F_{[\nu\rho}\cdot D_{\mu} F_{\sigma\tau]}), \label{linearcovderi}
\ee
where  index $\mu$ in the second term is not included in the antisymmetrization. 

\subsection{Bianchi Identity and  equations of motion }

The original Bianchi identity 
\be
D_{[\mu}F_{\rho\sigma]}=0,  \label{origibianchiiden}
\ee
is readily verified by the definition of the field strength,   $F_{\mu\nu}=\partial_{\mu}A_{\nu}-\partial_{\nu}A_{\mu}+i[A_{\mu}, A_{\nu}]$.  For tensor field strength,  
(\ref{origibianchiiden}) is generalized as 
\be
D_{[\mu} F_{\mu_1\mu_2\cdots \mu_{2l}]}=0. \label{bianchitensorf}
\ee
One may easily  verify (\ref{bianchitensorf}) using the linearity of the covariant derivative  (\ref{linearcovderi}) and the original Bianchi identity (\ref{origibianchiiden}).

We introduce  (Euclidean) tensor field theory action as 
\be
S=\frac{1}{4}\int d^{d}x ~\tr~({F_{\mu_1\mu_2\cdots\mu_{2l}}}^2 ).   
\ee
Since tensor field strength is originally made of the field strength,   we should take a variation of $S$ with respect to $A_{\mu}$ to derive   equations of motion:  
\be
\frac{\delta}{\delta A^{\nu}}S = -D_{\mu}G_{\mu\nu}=0, \label{tenseofbyamu}
\ee
where 
\begin{align}
G_{\mu_1\mu_2}& \equiv \sum_{p=1}^k F_{\mu_3\mu_4\cdots \mu_{2p}}F_{\mu_1\mu_2\cdots\mu_{2l}} 
F_{\mu_{2p+1}\mu_{2p+1}\cdots \mu_{2l}}\nn\\
&=F_{\mu_1\cdots\mu_{2l}} F_{\mu_3\cdots \mu_{2l}} + F_{\mu_3 \mu_4}  F_{\mu_1\cdots\mu_{2l}} F_{\mu_5\cdots \mu_{2l}}+   F_{\mu_{3}\mu_4\mu_5 \mu_{6}} F_{\mu_1\cdots\mu_{2l}} F_{\mu_7\cdots \mu_{2l}} + \cdots + F_{\mu_3\cdots \mu_{2l}} F_{\mu_1\cdots\mu_{2l}} . 
\end{align}
For instance, 
\begin{align}
&l=1~:~G_{\mu\nu} =F_{\mu\nu}, \nn\\
&l=2~:~G_{\mu\nu} =F_{\mu\nu\rho\sigma}F_{\rho\sigma}+F_{\rho\sigma}F_{\mu\nu\rho\sigma}  =\{F_{\mu\nu\rho\sigma}, F_{\rho\sigma}\}, \nn\\
&l=3~:~G_{\mu\nu} = F_{\mu\nu\rho\sigma\kappa\tau}F_{\rho\sigma\kappa\tau}  +F_{\rho\sigma} F_{\mu\nu\rho\sigma\kappa\tau} F_{\kappa\tau} +  F_{\rho\sigma\kappa\tau}F_{\mu\nu\rho\sigma\kappa\tau} . 
\end{align}
From the Bianchi identity (\ref{bianchitensorf}) and the linearity of the covariant derivative (\ref{linearcovderi}),  we have 
\begin{align}
D_{\mu_1}G_{\mu_1\mu_2} &= \sum_{p=1}^k F_{\mu_3\mu_4\cdots \mu_{2p}}(D_{\mu_1}F_{\mu_1\mu_2\cdots\mu_{2l}} )F_{\mu_{2p+1}\mu_{2p+1}\cdots \mu_{2l}}\nn\\
&=(D_{\mu_1}F_{\mu_1\cdots\mu_{2l}}) F_{\mu_3\cdots \mu_{2l}} + F_{\mu_3 \mu_4}  (D_{\mu_1}F_{\mu_1\cdots\mu_{2l}}) F_{\mu_5\cdots \mu_{2l}}\nn\\
&~+   F_{\mu_{3}\mu_4\mu_5 \mu_{6}} (D_{\mu_1}F_{\mu_1\cdots\mu_{2l}}) F_{\mu_7\cdots \mu_{2l}} + \cdots + F_{\mu_3\cdots \mu_{2l}} (D_{\mu_1}F_{\mu_1\cdots\mu_{2l}} ), \label{dmutensorg}
\end{align}
which implies 
\be
D_{\mu_1}F_{\mu_1\mu_2 \mu_3\cdots\mu_{2l}}  =0 ~~\rightarrow~~~D_{\mu_1}G_{\mu_1\mu_2}=0. \label{eomtensgmunu}
\ee

\subsection{Self-dual equations}

The tensor field Bianchi identity (\ref{bianchitensorf}) can be expressed as  
\be
D_{\mu_1}\tilde{F}_{\mu_1\mu_2\cdots \mu_{2l}} =0,  
\ee
where 
\be
\tilde{F}_{\mu_1\mu_2\cdots \mu_{2l}} \equiv \frac{1}{(2k-2l)!} ~\epsilon_{\mu_1\mu_2\cdots\mu_{2k}}F_{\mu_1\mu_2\cdots \mu_{2k-2l}}  .   \label{tensordual2l}
\ee
For $l=k/2$ ($k$: even), the self-dual equation is given by 
\be
\tilde{F}_{\mu_1\mu_2 \cdots \mu_{2l}}={F}_{\mu_1\mu_2 \cdots \mu_{2l}}. \label{selfdualeq}
\ee
When (\ref{selfdualeq}) holds, 
its dual equation automatically follows:  
\be
\tilde{F}_{\mu_1\mu_2 \cdots \mu_{2k-2l}}={F}_{\mu_1\mu_2 \cdots \mu_{2k-2l}}, 
\ee
and then there are $[k/2]$ independent self-dual equations in $2k$D. 
In low dimensions,  the independent self-dual equations  are 
\begin{align}
&k=2~:~\tilde{F}_{\mu\nu}=F_{\mu\nu}, \nn\\
&k=3~:~\tilde{F}_{\mu\nu}=F_{\mu\nu}, \nn\\
&k=4~:~\tilde{F}_{\mu\nu}=F_{\mu\nu}, ~~\tilde{F}_{\mu\nu\rho\sigma}=F_{\mu\nu\rho\sigma}. 
\end{align}
The self-dual tensor field  satisfies  
\be
D_{\mu_1}{F}_{\mu_1\mu_2\cdots \mu_{2l}}=D_{\mu_1}\tilde{F}_{\mu_1\mu_2\cdots \mu_{2l}} =0. 
\ee
From (\ref{eomtensgmunu}),  one may find that the self-dual tensor field  realizes a solution of  the equations of motion (\ref{tenseofbyamu}).

\section{Hidden local symmetry}\label{append:hls}

Hidden local symmetries of non-linear sigma models and Skyrme model are  discussed in \cite{Bando-Kugo-Yamawaki-1988, Bando-Kugo-Uehara-Yamawaki-Yanagida-1985, Ma-Harada-2016}. Here, we apply the discussions to the present $O(d+1)$ non-linear sigma models. 

\subsection{$O(2k+1)$ non-linear sigma model }

Let us consider the non-linear realization  of $O(2k+1)$ group associated with the symmetry breaking:  
\be
O(2k+1) ~~\rightarrow~~O(2k). 
\ee
We take the broken generators as 
\be
\Sigma_{m,2k+1} =i\frac{1}{2}\begin{pmatrix}
0 & \bar{g}_m \\
-g_m & 0
\end{pmatrix}, ~~ ~~(m=1,2,\cdots, 2k)
\ee
with $g_m$ (\ref{defgm}) and $\bar{g}_m$ (\ref{defgbarm}).  
In the unitary gauge \cite{Weinberg-1971, Weinberg-1973},\footnote{
The gauge is called the unitary gauge because in the  gauge  all of the fields are physical and the unitarity of  $S$-matrix is apparent. } 
 the non-linear realization  $\xi(n)$ is expressed as  
\be
\xi(n) =e^{i\theta \sum_{m=1}^{2k} \hat{n}_m \Sigma_{m,2k+1}} =\cos\biggl(\frac{\theta}{2}\biggr)1_{2^k} +2i \sin\biggl(\frac{\theta}{2}\biggr) \sum_{m=1}^{2k} \hat{n}_m\Sigma_{m,2k+1}, 
\ee 
where  $\theta$ and $\hat{n}_m$ $(\sum_{m=1}^{2k}\hat{n}_m\hat{n}_m=1)$ denote the azimuthal angle and normalized $S^{2k-1}$-latitude  of the coset $S^{2k}~\simeq~SO(2k+1)/SO(2k)$. 
The $O(2k+1)$ global transformation acts to $\xi(n)$ as 
\be
\xi(n) ~~\rightarrow~~g\cdot \xi(n) , ~~~~~(g \in O(2k+1))  \label{globalso2kfor}
\ee
while $O(2k)$ local transformation acts to $\xi(n)$ as  
\be
\xi(n) ~~\rightarrow~~ \xi(n) \cdot h.~~~~~(h\in O(2k)) \label{localso2kfor}
\ee
With the $O(2k)$ generators 
\be
\Sigma_{mn} =\begin{pmatrix}
\sigma_{mn} & 0 \\
0 & \bar{\sigma}_{mn}
\end{pmatrix}, 
\ee
the $O(2k)$ group element $h$ is expressed as  
\be
h=e^{i\sum_{m<n=1}^{2k}\omega_{mn}\Sigma_{mn}}=\begin{pmatrix}
h_L & 0 \\
0 & h_R
\end{pmatrix} =\begin{pmatrix}
e^{i\sum_{m<n=1}^{2k}\omega_{mn}\sigma_{mn}} & 0 \\
0 & e^{i\sum_{m<n=1}^{2k}\omega_{mn}\bar{\sigma}_{mn}}
\end{pmatrix} , \label{hhlhr}
\ee
where $h_L$ and $h_R$ are $2^{k-1}\times 2^{k-1}$ matrix generators of the  $Spin(2k)$ group.   
Therefore, there are ``two kinds of'' gauge transformations, $L$ and $R$. We decompose the non-linear realization $\xi(n)$  as 
\be
\xi(n)=\begin{pmatrix}
\Psi_L & \Psi_R 
\end{pmatrix} 
\ee
where $\Psi_L(n)$ and $\Psi_R(n)$ are $2^k\times 2^{k-1}$ rectangular matrices:\footnote{
The gauge invariant quantity is constructed as the projection matrix \cite{Hasebe-2014-1} 
\be
\Psi_L \Psi^{\dagger}_L =\frac{1}{2}(1_{2^{k-1}}+n_a\gamma_a),~~~~~  \Psi_R \Psi^{\dagger}_R =\frac{1}{2}(1_{2^{k-1}}-n_a\gamma_a).  
\ee
$\Psi_L$ and $\Psi_R$ realize a generalization of the Hopf maps: 
\be
n_a 1_{2^{k-1}} = \Psi^{\dagger}_L\gamma_a \Psi_L=- \Psi^{\dagger}_R\gamma_a \Psi_R.  
\ee 
}  
\be
\Psi_L=
\frac{1}{\sqrt{2(1+n_{2k+1})}}
\begin{pmatrix}
(1+n_{2k+1})1_{2^{k-1}} \\
n_m g_m 
\end{pmatrix}, ~~~~~
\Psi_R=
\frac{1}{\sqrt{2(1+n_{2k+1})}}
\begin{pmatrix}
 -n_m\bar{g}_m \\
 (1+n_{2k+1})1_{2^{k-1}}
\end{pmatrix},  \label{psiright}
\ee
with 
\be
n_m =\hat{n}_m\sin(\theta), ~~~~n_{2k+1}=\cos(\theta).  
\ee
The global transformation (\ref{globalso2kfor}) and the gauge transformation (\ref{localso2kfor})  can be rephrased as 
\be
\Psi_L~\rightarrow~g\cdot \Psi_L, ~~~~\Psi_L~\rightarrow~\Psi_L\cdot h_L, 
\ee 
and 
\be
\Psi_R~\rightarrow~g\cdot \Psi_R, ~~~\Psi_R~\rightarrow~\Psi_R\cdot h_R. 
\ee 
Therefore, we can regard the $O(2k+1)$ NLS model as a ``sum'' of the two independent NLS models with local $Spin(2k)_L$ and $Spin(2k)_R$ symmetries. 
The $Spin(2k)_{L/R}$ gauge fields are derived as 
\be
A_{\mu}^L =-i\Psi_L^{\dagger}\partial_{\mu}\Psi_L =-\frac{1}{1+n_{2k+1}}\sigma_{mn}n_n\partial_{\mu}n_m, ~~~A_{\mu}^R =-i\Psi_R^{\dagger}\partial_{\mu}\Psi_R =-\frac{1}{1+n_{2k+1}}\bar{\sigma}_{mn}n_n\partial_{\mu}n_m. 
\ee 
$A_{\mu}^L$ exactly coincides with (\ref{afieldstrengthnls2k+1}).  
Under each of the $Spin(2k)$ local transformation, $A_L$ and $A_R$ are transformed as 
\be 
A_{L}~~\rightarrow~~h_{L}^{\dagger}A_{L} h_{L}-ih_{L}^{\dagger}dh_{L},~~~A_{R}~~\rightarrow~~h_{R}^{\dagger}A_{R} h_{R}-ih_{R}^{\dagger}dh_{R}. 
\ee 
We can treat $\Psi_L$ and $\Psi_R$ as independent $SO(2k+1)$ spinors, and their covariant derivatives are constructed as 
\be
D_{\mu}\Psi_{L}\equiv \partial_{\mu}\Psi_{L}-i\Psi_{L} A^{L}_{\mu},~~~~~D_{\mu}\Psi_{R}\equiv \partial_{\mu}\Psi_{R}-i\Psi_{R} A^{R}_{\mu}.
 \label{covderixi}
\ee 
Under the local transformation, (\ref{covderixi}) behaves as 
\be
(D_{\mu}\Psi_{L}) ~~\rightarrow~~(D_{\mu}\Psi_{L})\cdot h_{L},~~~~~(D_{\mu}\Psi_{R}) ~~\rightarrow~~(D_{\mu}\Psi_{R})\cdot h_{R}. 
 \label{covderixitrans}
\ee
Similarly, the corresponding  field strength is given by 
\be
F_L=dA_L +iA_L^2=\frac{1}{2}F_{\mu\nu}^L dx_{\mu}dx_{\nu}, ~~~~~F_R=dA_R +iA_R^2=\frac{1}{2}F_{\mu\nu}^R dx_{\mu}dx_{\nu}  
\ee
with 
\begin{subequations}
\begin{align}
F^L_{\mu\nu}& =-i(D_{[\mu}\Psi_L)^{\dagger}~ (D_{\nu]}\Psi_L)=-i(\partial_{[\mu}\Psi_L)^{\dagger}(1-\Psi_L \Psi_L^{\dagger}) (\partial_{\nu]}\Psi_L^{\dagger})  \nn\\
& =\sigma_{mn}\partial_{\mu}n_m \partial_{\nu}n_n -\frac{1}{1+n_{2k+1}} \sigma_{mn}n_n (\partial_{\mu}n_m \partial_{\nu}n_{2k+1}- \partial_{\nu}n_m \partial_{\mu}n_{2k+1} ), \\  
F^R_{\mu\nu}& =-i(D_{[\mu}\Psi^R)^{\dagger}~ (D_{\nu]}\Psi^R)  =-i(\partial_{[\mu}\Psi_R)^{\dagger}(1-\Psi_R \Psi_R^{\dagger}) (\partial_{\nu]}\Psi_R^{\dagger}) \nn\\
&= \bar{\sigma}_{mn}\partial_{\mu}n_m \partial_{\nu}n_n -\frac{1}{1+n_{2k+1}} \bar{\sigma}_{mn}n_n (\partial_{\mu}n_m \partial_{\nu}n_{2k+1}- \partial_{\nu}n_m \partial_{\mu}n_{2k+1} ). 
\end{align}
\end{subequations}
Obviously,  $F_{L/R}$ is  transformed as 
\be
F_{L}~~\rightarrow~~h^{\dagger}_L \cdot F_{L}\cdot h_L,~~~~~F_{R}~~\rightarrow~~h^{\dagger}_R \cdot F_{R}\cdot h_R. 
\ee
The $k$th Chern number is expressed as 
\begin{align}
c_k&=\frac{1}{(2\pi)^k k!} \int \tr((F^{L/R})^k) =\frac{1}{(4\pi)^k k!} \int d^{2k}x ~\epsilon_{\mu_1\mu_2\cdots \mu_{2k-1 ,2k}}\tr(F^{L/R}_{\mu_1\mu_2} \cdots F^{L/R}_{\mu_{2k-1}\mu_{2k}}) \nn\\ 
&=\frac{1}{ k!} \biggl(-i\frac{1}{2\pi}\biggr)^{k} \int d^{2k}x ~\epsilon_{\mu_1\mu_2\cdots \mu_{2k-1 ,2k}}\tr((D_{\mu_1}\Psi_{L/R})^{\dagger} (D_{\mu_2}\Psi_{L/R}) \cdots (D_{\mu_{2k-1}}\Psi_{L/R})^{\dagger}(D_{\mu_{2k}}\Psi_{L/R})) \nn\\
&=\pm \frac{1}{(2k)! A(S^{2k})}\int_{\mathbb{R}^{2k}} d^{2k}x ~\epsilon_{m_1 m_2 \cdots m_{2k+1}}\epsilon_{\mu_1\mu_2\cdots\mu_{2k}} n_{m_{2k+1}}\partial_{\mu_1}n_{m_1}\partial_{\mu_2}n_{m_2}\cdots \partial_{\mu_{2k}}n_{m_{2k}}\nn\\
&=N_{2k},  
\label{windingnumber2k}
\end{align}
which is the winding number associated with $\pi_{2k}(S^{2k})\simeq \mathbb{Z}$. 
The kinetic term of the $O(2k+1)$ NLS model  can be written as 
\be
\frac{1}{4}\sum_{a=1}^{2k+1}(\partial_{\mu}n_{a})^2 \cdot {1}_{2^{k-1}} =(D_{\mu}\Psi_L)^{\dagger}(D_{\mu}\Psi_L) =(D_{\mu}\Psi_R)^{\dagger}(D_{\mu}\Psi_R) . \label{o2k+1kineticexp} 
\ee
The RHS is invariant under the hidden local $SO(2k)$ symmetry, and so $\Psi\cdot g$ also yields the same result  (\ref{o2k+1kineticexp}). We thus verified that  $O(2k+1)$ NLS model enjoys the hidden local $SO(2k)$ symmetry.

\subsection{$O(2k)$ non-linear sigma model }\label{append:o2khls}

Let us consider the symmetry breaking
\be
O(2k) ~~\rightarrow~~O(2k-1), 
\ee
and choose the broken generators as 
\be
\Sigma_{i, 2k} =\frac{1}{2}\begin{pmatrix}
\gamma_i & 0 \\
0 & -\gamma_i
\end{pmatrix}  , 
\ee
where $\gamma_i$ $(i=1,2,\cdots, 2k-1)$ denote the $SO(2k-1)$ gamma matrices. In the unitary gauge, the non-linear realization of $O(2k)$ group is given by 
\be 
\xi(n) =e^{i\theta \sum_{m=1}^{2k-1} \hat{n}_m \Sigma_{m,2k}} =\cos\biggl(\frac{\theta}{2}\biggr)1_{2^k} +2i\sin\biggl(\frac{\theta}{2}\biggr) \sum_{i=1}^{2k-1} \hat{n}_i\Sigma_{i,2k}
=
\begin{pmatrix}
\xi_L(n) & 0 \\
0 & \xi_R(n)
\end{pmatrix},  \label{xinso2nsquare}
\ee 
where\footnote{ 
$\xi_L(n)  =\xi_R(n)^{\dagger}$ is a special relation in the unitary gauge. 
}  
\be
\xi_L(n)  =\xi_R(n)^{\dagger}  
=\cos(\frac{\theta}{2})1_{2^{k-1}}+i\sin(\frac{\theta}{2})\hat{n}_i\gamma_i= \frac{1}{\sqrt{2(1+n_{2k})}}(1_{2^{k-1}}+\sum_{m=1}^{2k} n_m\bar{g}_m)  
\ee  
with  
\be
n_i=\sin\theta ~\hat{n}_i,~~~n_{2k}=\cos\theta. ~~~~~(\sum_{m=1}^{2k}n_mn_m=1)
\ee
The $O(2k)$ global transformation acts to $\xi(n)$ as 
\be
\xi(n) ~~\rightarrow~~g\cdot \xi(n) , \label{glocal2ktrans}
\ee
where 
\be
g=e^{i\sum_{m<n=1}^{2k}\omega_{mn}\Sigma_{mn}}=\begin{pmatrix}
e^{i\sum_{m<n=1}^{2k}\omega_{mn}\sigma_{mn}} & 0 \\
0 & e^{i\sum_{m<n=1}^{2k}\omega_{mn}\sigma_{mn}}
\end{pmatrix} ~ \in ~ O(2k).
\ee
Meanwhile, $O(2k-1)$ local transformation acts to $\xi(n)$ as  
\be
\xi(n) ~~\rightarrow~~ \xi(n) \cdot h,  \label{gaugetranso2k-1}
\ee
where 
\be
h=e^{i\sum_{i<j=1}^{2k-1}\omega_{ij}\Sigma_{ij}} =
\begin{pmatrix}
{h}_D(\omega) & 0 \\
0 & {h}_D(\omega)
\end{pmatrix} ~\in ~O(2k-1)
\ee
with 
\be
{h}_D(\omega) \equiv e^{i\sum_{i<j=1}^{2k-1}\omega_{ij}\sigma_{ij}}.
\ee
Notice unlike the $SO(2k+1)$ case (\ref{hhlhr}), there is only a ``single'' local transformation denoted by ${h}_D(\omega)$. 
We combine $\xi_L$ and $\xi_R$  (\ref{xinso2nsquare}) to construct a $2^k\times 2^{k-1}$ rectangular matrix\footnote{
(\ref{consrecso2k}) realizes the chiral Hopf maps \cite{Hasebe-2017}: 
\be
n_m 1_{2^{k-1}}=\Phi^{\dagger}\gamma_m   \Phi =\frac{1}{2} (\xi_L^{\dagger}~\bar{q}_m~\xi_R + \xi_R^{\dagger}~{q}_m~\xi_L ),  
\ee
and (\ref{consrecso2k}) is gauge equivalent to $\Psi_L$ (\ref{psiright}) at $n_{2k+1}=0$: 
\be
\Psi_L|_{n_{2k+1}=0} =\Phi \cdot {h}', 
\ee
with 
\be
{h}'=\frac{1}{\sqrt{2(1+n_{2k})}} (1_{2^{k-1}}+ \sum_{m=1}^{2k} n_m {g}_m). 
\ee
} 
\be
\Phi(n) =\frac{1}{\sqrt{2}}\begin{pmatrix} 
\xi_L(n)   \\
\xi_R(n)   
\end{pmatrix}=
\frac{1}{2\sqrt{1+n_{2k}}}
\begin{pmatrix}
1_{2^{k-1}}+\sum_{m=1}^{2k}n_m\bar{g}_m \\
1_{2^{k-1}}+\sum_{m=1}^{2k}n_m {g}_m 
\end{pmatrix}.   \label{consrecso2k}
\ee
The global transformation (\ref{glocal2ktrans}) and the gauge transformation (\ref{gaugetranso2k-1}) simply act to $\Phi$ as 
\be
\Phi(n)~~\rightarrow~~g\cdot \Phi(n), ~~~~~\Phi(n)~~\rightarrow~~\Phi(n)\cdot {h}_D. \label{phihath}
\ee 
We can treat $\Phi$  as  an $SO(2k)$ Dirac spinor.   
Associated with the $Spin(2k-1)$ local transformation,  
the $Spin(2k-1)$ gauge field is obtained as 
\be
A=-i\Phi^{\dagger}d\Phi =-i\frac{1}{2}(\xi_L^{\dagger}d\xi_L + \xi_R^{\dagger}d\xi_R)=-\frac{1}{1+n_{2k}}\sigma_{ij}n_j\partial_{\alpha}n_i dx_{\alpha}. 
\ee
Under the $Spin(2k)$ local transformation, 
$A$ is transformed as 
\be
A~~\rightarrow~~{h}_D^{\dagger}A {h}_D-i{h}_D^{\dagger}d{h}_D. 
\ee
Their covariant derivative is  given by 
\be 
D_{\alpha}\Phi\equiv \partial_{\alpha}\Phi-i\Phi A_{\alpha}.  
\ee
Under the local transformation, (\ref{covderixi}) behaves as 
\be 
D_{\alpha}\Phi ~~\rightarrow~~(D_{\alpha}\Phi)\cdot {h}_D. 
\ee
The corresponding field strength is constructed as 
\be
F=dA+iA^2=\frac{1}{2}F_{\alpha\beta}dx_{\alpha}dx_{\beta}
\ee
where
\begin{align}
F_{\alpha\beta} &=-i(D_{[\alpha}\Phi)^{\dagger}(D_{\beta]}\Phi) = -i (\partial_{[\alpha}\Phi)^{\dagger}(1-\Phi\Phi^{\dagger})(\partial_{\beta]}\Phi)\nn\\
&=\sigma_{ij}\partial_{\alpha}n_i\partial_{\beta}n_j-\frac{1}{1+n_{2k}}\sigma_{ij}n_j(\partial_{\alpha}n_i\partial_{\beta}n_{2k}-\partial_{\beta}n_i\partial_{\alpha}n_{2k}). 
\end{align}
The kinetic term of the $O(2k)$ NLS model can be written as 
\be
\frac{1}{4}\sum_{m=1}^{2k}(\partial_{\alpha}n_{m})^2 \cdot {1}_{2^{k-1}}=\frac{1}{2}((D_{\alpha}\xi_L)^{\dagger}(D_{\alpha}\xi_L) +(D_{\alpha}\xi_R)^{\dagger}(D_{\alpha}\xi_R)  )=(D_{\alpha}\Phi)^{\dagger}(D_{\alpha}\Phi). 
  \label{unitarygaugeresu}
\ee
From $\Phi$, we can readily construct an invariant  quantity under the local $O(2k-1)$ transformation: 
\be
\Phi(n)~\Phi(n)^{\dagger}= \frac{1}{2}\begin{pmatrix}
1_{2^{k-1}} &  U^{\dagger} \\
  U   & 1_{2^{k-1}}
\end{pmatrix}, 
\ee
 where 
\be
U=\xi_R \cdot \xi_L^{\dagger}=\sum_{m=1}^{2k}n_m \bar{g}_m. 
\ee
With $U$, we introduce  
\begin{subequations}
\begin{align}
&{W}^L_{\alpha}=-iU^{\dagger} \partial_{\alpha}U=-i\xi_L ~(D_{\alpha}\xi_L)^{\dagger}-i\xi_L ~\xi_R^{\dagger} ~(D_{\alpha}\xi_R)~\xi_L^{\dagger}=-2\bar{\sigma}_{mn}n_n\partial_{\alpha}n_m, \\
&W_{\alpha}^R=-iU \partial_{\alpha}U^{\dagger}= -i\xi_R ~(D_{\alpha}\xi_R)^{\dagger}-i\xi_R ~\xi_L^{\dagger} ~(D_{\alpha}\xi_L)~\xi_R^{\dagger}=-2{\sigma}_{mn}n_n\partial_{\alpha}n_m.  
\end{align}
\end{subequations}
$U$ is  identical to the transition function (\ref{noneqspin2k}), and so ${W}^L_{\alpha}$ turns out to be  $\mathcal{A}_{\alpha}$ (\ref{puregauge2k-1}).  
The winding number associated with $\pi_{2k-1}(S^{2k-1})\simeq \mathbb{Z}$ is represented as 
\begin{align}
N_{2k-1}&=\pm\frac{1}{(2k-1)! A(S^{2k-1})}\int_{\mathbb{R}^{2k-1}} d^{2k-1}x ~\epsilon_{m_1 m_2 \cdots m_{2k}}\epsilon_{\alpha_1\alpha_2\cdots\alpha_{2k-1}} n_{m_{2k}}\partial_{\alpha_1}n_{m_1}\partial_{\alpha_2}n_{m_2}\cdots \partial_{\alpha_{2k-1}}n_{m_{2k-1}}\nn\\
&=(-i)^{k-1}\frac{1}{(2\pi)^k}\frac{(k-1)!}{(2k-1)!}\int_{\mathbb{R}^{2k-1}} d^{2k-1}x ~\epsilon_{\alpha_1\alpha_2\cdots \alpha_{2k-1}} \tr(W_{\alpha_1}^{L/R}W_{\alpha_2}^{L/R}\cdots W_{\alpha_{2k-1}}^{L/R}).  
\end{align}




\end{document}